\documentclass[prx,superscriptaddress,floatfix,letterpaper,twocolumn,aps,showpacs]{revtex4-1}

\usepackage[utf8]{inputenc}
\usepackage[usenames,dvipsnames,svgnames,table]{xcolor}
\usepackage{natbib,amsmath,amsbsy,amssymb,scrextend,graphicx,color,placeins,array,siunitx,bm}
\usepackage{mathrsfs}  
\setcitestyle{square, comma, numbers, nccmath}
\usepackage{hyperref}
\usepackage{lipsum}

\renewcommand{\cite}[1]{[\onlinecite{#1}]}

\newcommand{\dep}{\mathrm{dp}}

\newcommand{\pin}{\mathrm{pin}}
\newcommand{\C}{\bar{C}}

\renewcommand{\vec}{\mathbf}
\newcommand{\eff}{\mathrm{eff}}

\newcommand{\br}{\vec{r}}

\newcommand{\hr}{\hat{r}}
\newcommand{\hbr}{\hat{\vec{r}}}

\newcommand{\vD}{\vec{\Delta}}
\renewcommand{\vr}{\vec{r}}

\newcommand{\+}{{\scriptscriptstyle +}}
\renewcommand{\-}{{\scriptscriptstyle -}}
\newcommand{\ppm}{{\scriptscriptstyle \pm}}

\renewcommand{\t}{\mathrm}
\newcommand{\vrh}{\bm{\rho}}
\newcommand{\pair}{\mathrm{pair}}
\newcommand{\BZ}{{\rm\scriptscriptstyle BZ}}

\begin{document}
\title{The role of rare events in the pinning problem}

\author{M.\ Buchacek}
\affiliation{Institute for Theoretical Physics, ETH Zurich, 8093 Zurich, Switzerland}
\author{V.B.\ Geshkenbein}
\affiliation{Institute for Theoretical Physics, ETH Zurich, 8093 Zurich, Switzerland}
\author{G.\ Blatter}
\affiliation{Institute for Theoretical Physics, ETH Zurich, 8093 Zurich, Switzerland}
\date{\today}

\begin{abstract}
Type II superconductors exhibit a fascinating phenomenology that is determined
by the dynamical properties of the vortex matter hosted by the material.  A
crucial element in this phenomenology is vortex pinning by material defects,
e.g., immobilizing vortices at small drives and thereby guaranteeing
dissipation-free current flow. Pinning models for vortices and other
topological defects, such as domain walls in magnets or dislocations in
crystals, come in two standard variants: i) weak collective pinning, where
individual weak defects are unable to pin, while the random accumulation of
many force centers within a collective pinning volume combines into an
effective pin, and ii) strong pinning, where strong defects produce large
vortex displacements and bistabilities that lead to pinning on the level of
individual defects. The transition between strong and weak pinning is
quantified by the Labusch criterion $\kappa \approx f_p/\bar{C}\xi = 1$, where
$f_p$ and $\bar{C}$ are the force of one defect and the effective elasticity
of the vortex lattice, respectively ($\xi$ is the coherence length).  Here, we
show that a third generic type of pinning becomes dominant when the pinning
force $f_p$ enters the weak regime, the pinning by rare events.  We find that
within an intermediate regime $1/2 < \kappa < 1$, compact pairs of weak
defects define strong pinning clusters that extend the mechanism of strong
pinning into the weak regime. We present a detailed analysis of this
cluster-pinning mechanism and show that its pinning-force density
parametrically dominates over the weak pinning result. The present work is a
first attempt to include correlations between defects into the discussion of
strong pinning.
\end{abstract}

\maketitle 

\section{Introduction}\label{sect:introduction}

Broken-symmetry phases, as they appear in super- conducting-, magnetic-, or
density wave systems, exhibit physical properties on top of those originating
from the underlying material.  Typically, these ordered phases develop
topological excitations (or defects) that govern the material properties,
e.g., vortices in superconductors \cite{Abrikosov1957} or domain-walls in
magnets \cite{Bloch1932, LandauLifshitz1935}.  Remarkably, it is the
interaction between the material's and the topological defects that determines
the static and dynamical properties of the latter, with pinning immobilizing
vortices in superconductors guaranteeing the material's dissipation-free
current transport \cite{Labusch1969,Larkin1979} and fixing domain-walls in the
magnet determining its coercive field \cite{Kittel1949}. On the fundamental
side, pinning of topological defects constitutes a rich branch of disordered
statistical physics with challenging phase-space and ergodicity properties,
including the phenomenon of glassiness \cite{Blatter1994,
NattermannScheidl2000}.

\begin{figure}[hb!]
\includegraphics[scale=1]{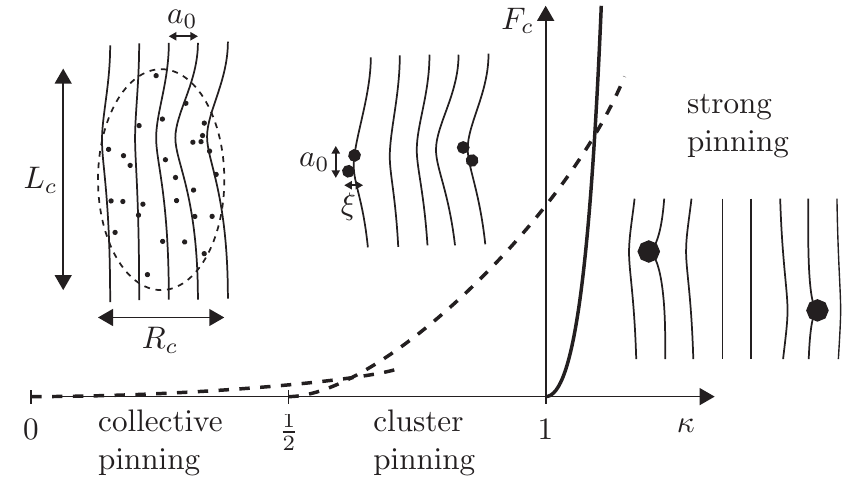}
\caption{Pinning mechanisms for flux lattices in type-II superconductors in
the regime of low defect density $n_p a_0\xi^2 \ll 1$. Shown is the critical
force density $F_c$ as a function of pinning strength $\kappa\propto
f_p/\C\xi$.  For small $\kappa \to 0$, pinning arises due to the collective
action of a large number of defects within the Larkin volume $V_c\sim R_c^2
L_c$, resulting in the collective pinning-force density $F_\mathrm{coll}\sim
(\xi/\lambda)^2 \kappa^3 f_p n_p (n_p a_0 \xi^2)$. For intermediate values
$1/2 < \kappa < 1$, pairs of defects in close proximity form strong pinning
clusters that produce the cluster pinning-force density $F_\mathrm{clust}\sim
(\xi/a_0)^2 (\kappa - \tfrac{1}{2})^4 f_p n_p (n_p a_0 \xi^2)$; the latter
dominates over the collective pinning result for a Labusch parameter $\kappa >
1/2 + \mathcal{O}[(a_0/\lambda)^{1/2}]$. For $\kappa \gtrsim 1$, pinning is
strong, with the pinning-force density due to individual defects rising as
$(\xi/a_0)^2 (\kappa - 1)^2 f_p n_p$; the latter dominates over the cluster
pinning when increasing $\kappa$ beyond unity by the amount $(n_p a_0
\xi^2)^{1/2}$.} \label{fig:pinning_overview}
\end{figure}

Traditionally, pinning in such systems was thought of as due to large
ensembles of weak defects; the ensuing collective pinning theory
\cite{Larkin1974, Larkin1979, Schmid1973, Larkin1970} has become a common
framework for the description of pinning of superconducting vortices
\cite{Blatter1994, Giamarchi1994, Giamarchi1995, Korshunov1993}, magnetic
domain-walls \cite{Kleemann2007, Gorchon2014, Jeudy2016}, charge density waves
(CDWs) \cite{Lee1979, Brazovskii2004} and other types of elastic media
\cite{Chauve2000}.  At the same time, an alternative viewpoint describing
pinning due to a low density of strong centres was proposed early on, see
Refs.\ \cite{Labusch1969} and \cite{Larkin1979}; recently, this strong pinning
scenario has attracted increasing attention, particularly in studies of charge
density waves \cite{Brazovskii1996, Brazovskii1999, Brazovskii2004} and of
magnetic flux-line lattices \cite{Thomann2017, Willa2018a, Willa2018b,
Buchacek2018, Buchacek2019}.  Although some effort has been made to
qualitatively understand the crossover between the two regimes
\cite{Blatter2004, Koopmann2004}, a quantitative model describing this
transition has not been developed so far.  In this paper, we describe a new
regime at the crossover between the two theories. We show that in a
considerable part of the weak region, pinning is dominated by defect clusters
co-operating on short distances and forming strong pinning centers that are
described with the tools of strong pinning theory.  The dominance of these
strongly-pinning small pairs over the weak collective ensembles can be traced
back to the dispersive nature of the vortex elasticity.  Pinning by rare
events then interpolates between the strong pinning of individual defects and
the random sum of weak pinning forces due to the many defects within the
Larkin domains of collective pinning theory, as illustrated in
Fig.~\ref{fig:pinning_overview}.

The central problem arising in studies of pinned systems is the determination
of the maximal driving (or critical) force density $F_c$ below which the
system remains immobilized. This critical force is determined by the
competition between the pinning centers characterized by their density $n_p$
and individual forces $f_p$ and the elastic properties of the manifold. In the
present study, we focus on the vortex lattice \cite{Abrikosov1957} formed by
flux lines or vortices, each carrying a superconducting flux quantum $\Phi_0$
and characterized by a line energy $\varepsilon_0 = (\Phi_0/4\pi\lambda)^2$
($\lambda$ denotes the London penetration depth). The effective elasticity
$\C\sim \varepsilon_0/a_0$, with $a_0$ the distance between vortices, captures
the full elastic properties of the vortex lattice that combines the line
tension and the interaction between vortices. The competition between pinning
and elastic forces then can be quantified by the dimensionless Labusch
parameter \cite{Labusch1969} $\kappa \sim f_p/\xi\bar{C}$, where $\xi$ denotes
the coherence length (or vortex diameter) in the superconductor. When $\kappa$
increases beyond unity, individual pins change from weak to strong. The three
scenarios, strong-, weak-collective-, and cluster-pinning as illustrated in
Fig.\ \ref{fig:pinning_overview} then provide different mechanisms and scaling
laws for the critical force density $F_c$.

The {\it strong pinning paradigm} rests on the assumption of a low defect
density $n_p$, such that $\kappa n_p a_0 \xi^2 < 1$ \cite{Blatter2004}, and
strong defects, i.e., $\kappa >1$. In this setting, material defects act
independently, resulting in a critical force density $F_c \propto n_p$ that is
linear in the density $n_p$ of pinning centers. The task of calculating $F_c$
then simplifies considerably and even allows for a quantitative treatment: as
defects act independently, the calculation of their contribution to the
critical force density $F_c$ boils down to an effective single-particle
problem where a strong defect interacts with an elastic manifold.  The
competition between potential and elastic forces does, however, add quite some
complexity to the problem, with strong pinning inducing plastic deformations
and bi-stable (pinned and free) states of the elastic manifold
\cite{Larkin1979, Brazovskii1996, Brazovskii2004, Blatter2008}. The
non-symmetric occupation of these bi-stable solutions then generates a finite
pinning force, with the critical force density derived from the maximally
asymmetric occupation of metastable states given by $F_c \sim
(S_\mathrm{trap}/a_0^2)\, n_p f_p\sim (\kappa \xi^2/a_0^2)\, n_p f_p$; here
$S_\mathrm{trap}/a_0^2$ defines the fraction of vortices falling into the
defect trapping area $S_\mathrm{trap}\sim \kappa\xi^2$ with longitudinal and
transverse  dimensions $\sim\kappa\xi$ and $\sim\xi$
\cite{IvlevOvchinnikov1991, Blatter2008}.

{\it Weak collective pinning} instead, relies on the joint action of many
defects, as individual weak pins with $\kappa < 1$ cannot hold the manifold.
In the weak-collective pinning scenario, distant defects act with {\it random
forces} on the manifold and their (random) addition within the Larkin volume
$V_c \sim \lambda^3 (\lambda/a_0)/(\kappa^2 n_p a_0 \xi^2)^3$ (that contains a
large number of pins) produces a critical force density $F_c \sim
[(\xi^2/a_0^2) n_p f_p^2 V_c]^{1/2}/V_c \sim (\xi^2/\lambda^2) \kappa^3 (n_p
a_0\xi^2)\, n_p f_p$, where the factor $\xi^2/a_0^2$ accounts for the fraction
of defects within $V_c$ that overlap with the vortex cores.  In fact, the
collective force randomly accumulated in the Larkin volume $V_c$ produces an
effectively strong pin \cite{Blatter2004} that satisfies the Labusch criterion
$\kappa(V_c) = 1$.

In the present paper, we study the crossover between the strong- and
weak-collective-pinning mechanisms near $\kappa \sim 1$; this study leads us
to the mechanism of {\it pinning by rare events}. Pairs of defects that
reinforce one another appear with relative probability $n_p^2$ and thus
potentially compete with the force generated in the weak-pinning scenario. In
identifying suitable pairs, we find that closeby defects within the action
volume $\xi^2 a_0$ of one defect define the relevant clusters; the density of
such clusters then is given by $(n_p a_0 \xi^2)\,n_p$. Defects in one cluster act
cooperatively rather then competitively. For defects with a pinning strength
$1/2 < \kappa < 1$, such neighboring pairs jointly produce a strong defect
with $2\kappa > 1$. Applying the strong-pinning formalism to these strong
cluster-defects then produces a critical force density ${F_c \sim
(\xi^2/a_0^2) (n_p a_0 \xi^2)\, n_p f_p}$ that is larger than the
weak-collective force density by a factor $(\lambda/a_0)^2$. This factor is a
consequence of the dispersive nature of the tilt elasticity $c_{44}(\vec{k})$:
while (non-dispersive) collective pinning involves the large Larkin scale $R_c
> \lambda$, cluster-pinning appears on short distances below $a_0$ and hence
involves the line rather than the bulk elasticity. Hence, we find a new
transition region in the pinning strength $\kappa$ where rare events,
neighboring defects forming a strong-pinning cluster, determine the critical
force density $F_c$. 

The relevance of rare events has been pointed out before in the context of
charge density wave pinning \cite{Fisher1985}, where an analysis in $D > 4$
dimensions demonstrated the irrelevance of weak collective pinning. Instead, a
finite but exponentially small (in the disorder strength) pinning-force
density was found that originates from rare regions with anomalously coherent
pinning. In our case, we deal with a $D = 3$ dimensional vortex lattice, where
both types of pinning, weak collective and rare events contribute
simultaneously, with the rare events identified as small defect pairs.

The paper is organized as follows: in Sect.~\ref{sect:VL}, we discuss the
formalism used in the description of vortex pinning for the generic case of an
isotropic material and briefly present the main steps in the derivation of the
pinning-force density $F_c$ in the strong- and weak-pinning scenarios and for
the newly-introduced framework of pinning by close pairs of defects. In
Sect.~\ref{sect:two_defect_problem}, we first introduce the general two-defect
problem for pairs of any size. In the overview section \ref{sect:overview}, we
identify the strongly-pinning pairs and discuss their contribution to the
pinning-force density as a function of the spatial separation between the
defects constituting the pair.  We show that pairs of distant defects provide
a smaller contribution, justifying the assumption of dominant pinning by rare
clusters of close defect pairs. We proceed with a detailed analytical
derivation of our results, involving an in-depth discussion of the effective
anisotropic pinning potential of defect pairs (Sect.\
\ref{sect:effective_potential}), of the effective Labusch parameter of defect
pairs in Sect.\ \ref{sect:kappa_eff}, and the average pinning force of
defect pairs in Sect.\ \ref{sect:average_f_pin} including a comparison to
numerical results. Finally, in Sect.~\ref{sect:conclusion}, we summarize our
results and place them into context, including also some further directions of
research.

\section{Vortex lattice pinning}\label{sect:VL}
The pinning of a vortex lattice is an example of the $(D + n)$-dimensional
random manifold problem; the latter describes a $D$-dimensional elastic
manifold parametrized by $\vrh\in \mathbb{R}^D$ that is distorted with an
$n$-dimensional displacement field $\vec{u}(\vrh) \in \mathbb{R}^n$ due to a
pinning potential $\varepsilon_\pin(\vrh,\vec{u})$. Assuming small
distortions, the generic Hamiltonian
\begin{equation}\label{eq:H_elastic}
   H = \int d^D\! \vrh \, \Bigl[ \frac{c}{2} (\nabla \vec{u})^2 
   + \varepsilon_\pin(\vrh,\vec{u})\Bigr]
\end{equation}
describes this type of systems.  Minimizing Eq.~\eqref{eq:H_elastic} yields
the equation for the displacement field in the form,
\begin{align}\label{eq:u_general}
   \vec{u}(\vrh) = \int d^D \! \vrh' \, G(\vrh-\vrh')
   \Bigl[-\nabla_\vec{u}\varepsilon_\pin(\vrh',\vec{u}(\vrh'))\Bigr],
\end{align}
with the Green's function $G(\vrh)$; in reciprocal space, $G(\vec{k}) =
1/c\vec{k}^2$.  In the following, we first discuss the relevant properties of
the real-space Green's function $G(\vrh)$ for our vortex problem and then turn
to the peculiarities of the disorder potential $\varepsilon_\pin (\vrh,
\vec{u})$ for the weak- and strong pinning situations.

\subsection{Green's function}\label{sect:Greens_fun}
The vortex pinning problem considered here belongs to the class $D = 3$, $n =
2$ and the complex structure of the vortex lattice brings a number of
modifications to the simple pinning model in Eq.~\eqref{eq:H_elastic}. The
Green's function for the vortex lattice (aligned along the $z$-axis) is in
fact non-diagonal and features anisotropic and dispersive elastic moduli;
focusing the discussion to isotropic superconductors and writing ${\vec{k} =
(\vec{K},k_z)}$ with the transverse ($\vec{K}$) and longitudinal ($k_z$)
components of the reciprocal vector, it assumes the form \cite{Blatter1994}
\begin{align}\label{eq:G_VL}
   G_{\alpha\beta}(\vec{k}) = \frac{\mathcal{P}^{\parallel}_{\alpha\beta}(\vec{K})}
   {c_{11}(\vec{k})K^2 \! + \! c_{44}(\vec{k})k_z^2} 
   + \frac{\mathcal{P}^{\perp}_{\alpha\beta}(\vec{K})}{c_{66}K^2 \! 
   + \! c_{44}(\vec{k})k_z^2}
\end{align}
with indices $\alpha,\beta \in {1,2}$ and the projection operators
$\mathcal{P}^{\parallel}_{\alpha\beta}(\vec{K}) = K_\alpha K_\beta/K^2$ and
$\mathcal{P}^{\perp}_{\alpha\beta}(\vec{K}) = \delta_{\alpha\beta} - K_\alpha
K_\beta/K^2$. The compression and tilt moduli $c_{11}(\vec{k}) \approx
c_{44}(\vec{k}) \approx (B^2/4\pi)(1+\lambda^2 k^2)^{-1}$ exhibit strong
dispersion due to the long-range interaction between vortices; $c_{66} = B
\Phi_0 / (8\pi\lambda)^2$ is the non-dispersive shear modulus ($\vec{B}
\parallel \hat{\vec{z}}$ is the magnetic field induced in the bulk of the
superconductor). The corresponding real-space Green's function is obtained via
standard Fourier transformation,
\begin{align}\label{eq:G_real_space}
   G_{\alpha\beta}(\vrh) = \int\limits_{K<K_{\rm\scriptscriptstyle BZ}} 
   \frac{d^2\vec{K}\, d k_z}{(2\pi)^3} 
   G_{\alpha\beta}(\vec{k})\,e^{i\vec{k}\cdot \vrh},
\end{align}
with the integration over $K$ restricted to the Brillouin zone of the vortex
lattice, $K_{\rm\scriptscriptstyle BZ} \approx \sqrt{4\pi}/a_0$. Of key
importance will be the on-site Green's function $G_{\alpha\beta}(\vrh = 0) =
G(\vec{0})\, \delta_{\alpha\beta}$. The integration in
Eq.~\eqref{eq:G_real_space} then is dominated by transverse momenta near the
Brillouin zone boundary $K\sim K_\BZ$, and estimating the relevant
longitudinal momentum by comparing the shear and tilt elastic energies
$c_{66}K^2\sim c_{44}(K_\BZ)k_z^2$, we obtain the scaling result $G(\vec{0})
,\sim 1/[a_0 \sqrt{c_{44}(K_\BZ)c_{66}}]$. The precise integration in
Eq.~\eqref{eq:G_real_space} gives the result \cite{Willa2016, Thomann2017,
Willa2018b}
\begin{align}\label{eq:G_on_site}
   G(\vec{0})^{-1} = \zeta (a_0^2/\lambda)\sqrt{c_{44}(\vec{0})c_{66}},
\end{align}
with a numerical factor $\zeta$ that depends on the chosen approximation for
the elastic moduli.

To evaluate the spatial variations of the Green's function, we consider a
simplified model of the vortex lattice elasticity: we drop the first term in
Eq.~\eqref{eq:G_VL} involving the large compression modulus
$c_{11}(\vec{k})>c_{66}$  and replace the projection operator in the remaining
term by $\delta_{\alpha\beta}$, such that $G_{\alpha\beta}(\vrh) =
G(\vrh)\,\delta_{\alpha\beta}$. Our diagonal reponse function $G[\vrh =
(\vec{R},z)]$ is characterized by a sharp and structured peak around the
origin and a smooth decay $\propto 1/\tilde{\rho}$ at large distances
$\tilde\rho > \lambda$, where $\tilde{\vrh} = (\vec{R}, \sqrt{c_{66}/
c_{44}(\vec{0})}\,z)$ is the properly scaled distance due to the anisotropic
elasticity of the vortex lattice.  Going beyond the diagonal
approximation does not change our strong pair-pinning results obtained below.
Note that the function $G(\vrh)$ provides us with the displacement field
$u(\vrh) = G(\vrh) F$ due to a $\delta$-force $F \delta(\vrh)$ at the origin.

\begin{figure}
\centering
\includegraphics[scale=1]{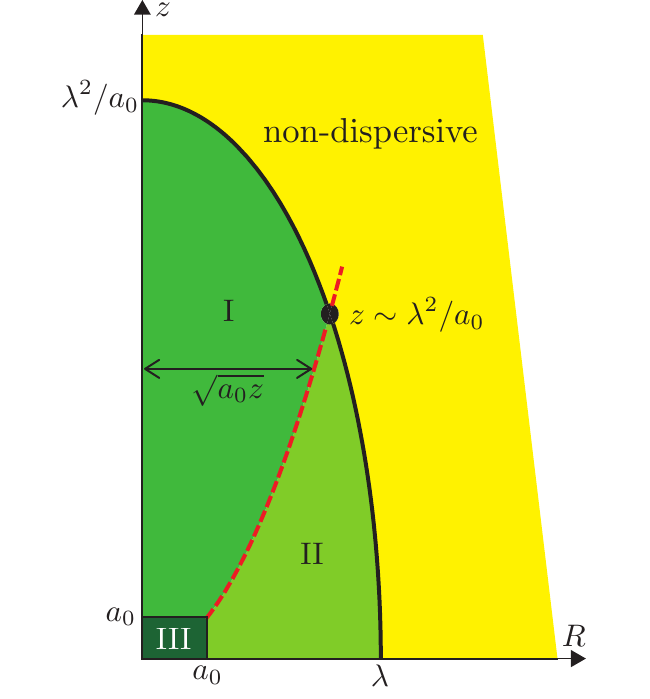}
\caption{Different domains analyzed in the evaluation of the scaled Green's
function $g(\vec{R},z) = G(\vec{R},z)/G(\vec{0})$.  In the non-dispersive
region (yellow) outside the ellipse $\tilde{\rho}^2 = R^2 +
(a_0^2/16\pi\lambda^2)\,z^2 \approx \lambda^2$ the Green's function decays
$\propto 1/\tilde{\rho}$, see Eq.~\eqref{eq:G_non_disp}.  The value of the
Green's function on the ellipse boundary is $g(\tilde\rho\sim \lambda)\sim
a_0^2/\lambda^2$. Inside the ellipse (green), we find several regions
characterized by different scaling results.  For $a_0\lesssim z \lesssim
\lambda^2/a_0$, the Green's function reads $g(\vec{R},z)\sim
(a_0/z)e^{-\sqrt{\pi}R^2/a_0z} + a_0^2/\lambda^2$; starting out at
$g(\vec{0},z)\sim a_0/z$, it decays exponentially fast along $R$ on the scale
$R\sim\sqrt{a_0 z}$ (region I, green) before saturating (ignoring slow
logarithmic variations) at $g \sim a_0^2/\lambda^2$ [region II, light green].
For small longitudinal coordinates $z\lesssim a_0$, the Green's function
evaluated on the $z$-axis is $g(\vec{0},z)\approx 1-z/a_h$, with $a_h\sim
a_0[\ln(a_0/\xi)]^{1/2}$ the healing length, and its decay along the
transverse coordinate $R$ is governed by the same scale $\sim a_h$ (region
III, dark green). For $z\lesssim a_0, R\gtrsim a_0$, the Green's function
again saturates at the value $g(\vec{R},z)\sim a_0^2/\lambda^2$. Increasing
$z$ at fixed $R$ within the interval $a_0 < R < \lambda$, the ratio $g$ first
increases and goes over a maximum when $z$ reaches the value $R^2/a_0$ (red
dashes); this feature produces a distinct ridge in the peak region of
$g$.}\label{fig:G_schematic}
\end{figure}

We first evaluate the Green's function in the non-dispersive regime (large
distances $\rho$), with the dominant contributions to the integration in
Eq.~\eqref{eq:G_real_space} originating from small momenta $\lambda^2 k^2
\lesssim 1$, such that $c_{44}(\vec{k})\approx c_{44}(\vec{0})$. The
anisotropy of the Green's function in Eq.~\eqref{eq:G_VL} generates different
decays along the directions longitudinal and transverse to the induced
magnetic field, that is for $\vrh = (\vec{0}, z)$ and $\vrh = (\vec{R},0)$. To
simplify the calculation, we remove this anisotropy by introducing the
rescaled momentum vector
$\vec{q}=(\vec{K},\sqrt{c_{44}(\vec{0})/c_{66}}\,k_z)$ with
$c_{44}(\vec{0})/c_{66} = 16\pi\lambda^2/a_0^2$, which leads to
\begin{align}\label{eq:G_non_disp_rescaled}
   G(\tilde{\vrh})
   \approx \frac{1}{\sqrt{c_{44}(\vec{0})c_{66}}}\int \frac{d^3\vec{q}}{(2\pi)^3}
   \frac{e^{i \vec{q}\cdot \tilde{\vrh}}}{q^2},
\end{align}
with $\sqrt{c_{44}(\vec{0})\, c_{66}} = (B^2/16\pi\sqrt{\pi})(a_0/\lambda)$ and
the rescaled distance $\tilde{\vrh} = (\vec{R}, \sqrt{c_{66}/
c_{44}(\vec{0})}\,z)$.  Integrating over the momenta $\vec{q}$ yields
$G(\tilde\rho)=1/[4\pi\sqrt{c_{44}(\vec{0})c_{66}}\,\tilde\rho\,]$ and the
reverse transformation $\tilde{\vrh}\to\vrh = (\vec{R},z)$ provides us with
the result
\begin{align}\label{eq:G_non_disp}
   G(R,z) \approx \frac{1/4\pi\sqrt{c_{44}(\vec{0})c_{66}}} 
   {\sqrt{R^2 + (a_0^2/16\pi\lambda^2)\, z^2}}.
\end{align}
Eq.~\eqref{eq:G_non_disp} describes the situation where the dispersion in the
tilt modulus can be neglected, which is the case at large distances $R^2 +
(a_0^2/16\pi\lambda^2)\,z^2 \gtrsim \lambda^2$, see the yellow region in Fig.\
\ref{fig:G_schematic}; on the inner boundary (an ellipsoid with extensions
$R\sim \lambda$ and $z\sim \lambda^2/a_0$), the Green's function assumes a
constant value $G \sim (a_0^2/\lambda^2)G(\vec{0})$ and decays $\propto
1/\tilde{\rho}$ further out, see Eq.\ \eqref{eq:G_non_disp}.  Indeed, in order
to drop the dispersion in $c_{44}$, we require the $q$-integral in
\eqref{eq:G_non_disp_rescaled} to be cut by a large distance $\tilde{\rho}$
(rather than the Brillouin zone), $q \lesssim 1/\tilde{\rho}$, at values where
$q \lambda < 1$ (rendering dispersion irrelevant), implying that $\tilde{\rho}
> \lambda$.
\begin{figure}[ht!]
\centering \includegraphics[height=70.94mm]{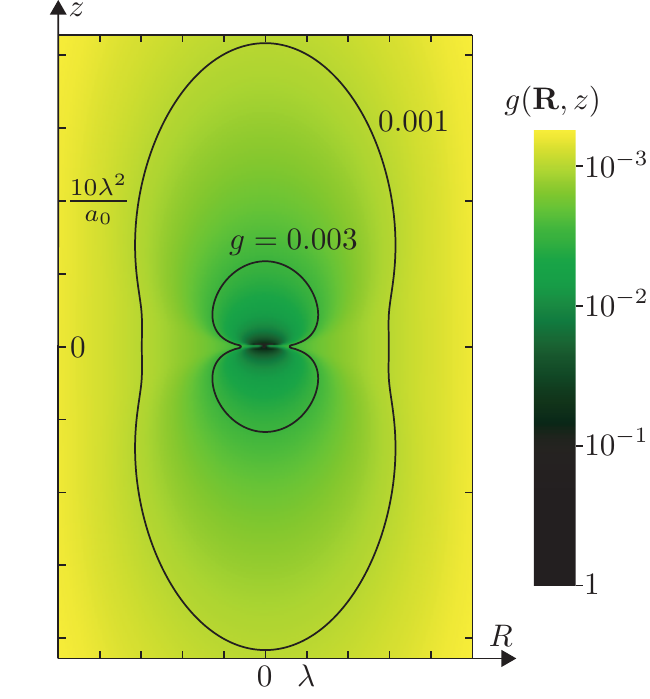}
\caption{Color plot of the rescaled Green's function $g(x=R,0,z) = G(R, 0, z)
/G(\vec{0})$ evaluated for $\lambda = 10\,a_0$; axes are not to scale.  The
dark-green peak in the center saturates to unity over a region $\sim a_0^2$;
at large distances $\tilde\rho  = \sqrt{R^2 + (a_0^2/16\pi\lambda^2) z^2} >
\lambda$, a smooth decay $\propto 1/\tilde\rho$ is observed (yellow).  The
peak at small distances (green) exhibits a dumbbell shape and gives way to a
smoothly decaying background of elliptical shape at large distances (yellow);
the two contours with $g = 0.003$ and $g = 0.001$ illustrate this change of
shape from a dumbbell- to an elliptical form. Fig.\ \ref{fig:g_peak} in
Appendix \ref{APP:g_disp} shows the detailed contour plot near the center of
the structured peak, including the position of the ridge.}
\label{fig:g_Rz}
\end{figure}

The evaluation of the Green's function at locations inside the ellipsoid
requires proper integration both over small ($k\lesssim \lambda^{-1}$) and
large ($k\gtrsim \lambda^{-1}$) momenta; the full calculation is presented in
Appendix \ref{APP:g_disp}. For longitudinal distances $z\gtrsim a_0$ and
arbitrary $R$, we find the interpolation formula
\begin{align}\label{eq:G_interp}
   G(\vec{R},z)&\approx \frac{\lambda/\sqrt{4\pi}}{a_0 z 
   \sqrt{c_{44}(\vec{0})c_{66}}} e^{-\sqrt{\pi} R^2/a_0 z}\\ \nonumber
   &+\frac{1/16\pi}{\lambda \sqrt{c_{44}(\vec{0})c_{66}}}\Bigl[1\!-\!2\gamma \!
   +\! \ln \frac{16\lambda^2}{R^2 + a_0 z / e^\gamma \sqrt{\pi}}\Bigr]
\end{align}
with $\gamma\approx 0.577$ the Euler-Mascheroni constant. This result provides
us with various scaling regimes for the Green's function, as illustrated in
Fig.~\ref{fig:G_schematic}. First, fixing $R = 0$ and going away from the
origin along the longitudinal direction, the rescaled Green's function decays
as $G(\vec{0},z)\sim (a_0/z)G(\vec{0})$; the result in Eq.~\eqref{eq:G_interp}
matches the non-dispersive expression Eq.~\eqref{eq:G_non_disp} at the
crossover $z\sim \lambda^2/a_0$.

Increasing $R$ for $z < \lambda^2/a_0$, the Green's function is dominated
by the first term in Eq.\ \eqref{eq:G_interp} that describes a Gaussian with
height $G(\vec{R}=\vec{0},z)\sim (a_0/z)\, G(\vec{0})$ and of width $R\sim
\sqrt{a_0 z}$; with decreasing $z$ this Gaussian peak becomes higher and
narrows down to produce a dumbbell shape peak, see region I in the schematic
Fig.\ \ref{fig:G_schematic} and the neck in the contour $g = 0.003$ in Fig.\
\ref{fig:g_Rz}. Increasing $R$ beyond $\sim \sqrt{a_0 z}$, we enter region II
in Fig.\ \ref{fig:G_schematic} where the second term in Eq.\
\eqref{eq:G_interp} dominates, interpolating smoothly between the peak and the
non-dispersive result Eq.\ \eqref{eq:G_non_disp}. This smooth interpolation
through region II is of order $G(\vec{R},z) \sim (a_0/\lambda)^2 G(\vec{0})$,
with logarithmic corrections that become large at small values of $z$ where
the narrow dumbbell peak at the origin decays more rapidly.  Note that beyond
the point $z\sim \lambda^2/a_0$ where the decay length $R\sim \sqrt{a_0 z}$
meets the ellipsoidal shell, the Green's function for $\vec{R} = \vec{0}$
already assumes a value $G(\vec{0},z\sim\lambda^2/a_0)\sim (a_0^2/\lambda^2)\,
G(\vec{0})$ and no substantial variations with $R$ are seen within the region
I.

Increasing instead the longitudinal distance $z$ at fixed $R < \lambda$,
the Green's function first remains flat (region II), then steeply increases
$\propto e^{-\sqrt{\pi} R^2/a_0 z}$ upon entering the peak region I at $z \sim
\pm R^2/a_0$, and then decreases smoothly $\propto 1/z$, thus defining a
maximum at $z \sim \pm R^2/a_0$. The resulting ridges located at the edges of
the Gaussian peak are another manifestation of the dumbbell structure of the
peak in $G(\vec{R},z)$, see Figs.\ \ref{fig:g_Rz} and \ref{fig:g_peak}.

The discussion has to be further refined in the regime of small $z\lesssim
a_0$. As $z\to 0$, the first term in Eq.~\eqref{eq:G_interp} diverges for $R =
0$ and vanishes for $R> 0$, formally approaching the 2D delta-function
$\propto \delta^2(\vec{R}/a_0)$. In reality, accounting for the
$\vec{q}$-cutoff at the Brillouin zone boundary in Eq.~\eqref{eq:G_real_space}
provides us with the finite result for the on-site Green's function $G(\vec{R} =
\vec{0},z = 0)$. An expansion in the longitudinal direction for small $z\lesssim
a_0$ then gives \cite{Koshelev_priv_comm}
\begin{align}\label{eq:G_vertical_small_z}
   G(\vec{0},z) \approx (1 - |z|/a_h)\, G(\vec{0}),
\end{align}
with the healing length $a_h\sim a_0[\ln (a_0/\xi)]^{1/2}$. 

The decay length in the transverse direction at small $z$ is affected by the
single-vortex elasticity that becomes relevant near the Brillouin zone
boundary \cite{Blatter1994, Brandt1977b}. Replacing the tilt modulus by
$c_{44}(\vec{k})\to c_{44}(\vec{k}) + (\varepsilon_0/a_0^2) \ln (a_0/\xi)$
then entails a saturation of the decay scale $R\sim \sqrt{a_0 z}$ in
Eq.~\eqref{eq:G_interp} at $R\sim a_h\sim a_0$ (we ignore a factor $\ln
(a_0/\xi)$ in the scaling estimates) for $z\lesssim a_0$ (region III).  For
$R\gtrsim a_h$, we again cross over to the region II where the Green's
function assumes the constant value $\sim a_0^2/\lambda^2$, up to slow
logarithmic corrections. 

The above analysis has been carried out for a simplified diagonal
expression $G_{\alpha\beta} = G\, \delta_{\alpha\beta}$. In a further step,
one may replace the identity matrix $\delta_{\alpha\beta}$ by the full
transverse projector $P_{\alpha\beta}^\perp (\vec{K})$, see Eq.\
\eqref{eq:G_VL}.  Focusing on the non-dispersive regime, the
$\vec{q}$-integral in Eq.\ \eqref{eq:G_non_disp_rescaled} picks up an
additional angular dependence that depends on the geometry of the problem.
For the component $G_{xx}$ evaluated in the $xz$-plane, we obtain the
asymptotic dependence
\begin{align}\label{eq:Gxx}
   G_{xx}(R,0,z) \approx \frac{1}{4\pi \sqrt{c_{44}(\vec{0})c_{66}}}
   \frac{\sqrt{R^2 + \tilde{z}^2} -\tilde{z}}{R^2},
\end{align}
where $\tilde{z} = (a_0/4\sqrt{\pi} \lambda) z$ is the scaled longitudinal
length. The result \eqref{eq:Gxx} then exhibits a modified anisotropy at large
distances: the simple scaling $G \propto 1/\sqrt{R^2 + \tilde{z}^2}$ in the
expression \eqref{eq:G_non_disp} is replaced with $G_{xx} \propto
1/2\tilde{z}$ when $\tilde{z} \gg R$ and $G_{xx} \propto 1/R$ at large $R \gg
\tilde{z}$.  Finally, while $G_{xy} = 0$, we find that $G_{yy}(R,0,z) =
(\tilde{z}/\sqrt{R^2 + \tilde{z}^2})\,G_{xx}$.  Note that $G_{xx} + G_{yy} =
G$, as expected.

Having analyzed the elastic component in the pinning problem, we now turn
to the discussion of the pinning potential $\varepsilon_\pin(\vrh,\vec{u})$ in
Eq.~\eqref{eq:H_elastic} for the cases of strong pinning, weak collective
pinning, and the pinning by rare clusters.  Note that the smallest transverse
scale $R$ in the context of elasticity is the separation $a_0$ between
vortices, while separations between defects as discussed below are considered
small when $R$ reaches the effective size $\xi$ of defects. Hence, small
lengths $R$ take a different meaning when talking about the vortex lattice
(elasticity) or the pinning landscape.

\subsection{Strong pinning}\label{sect:sp}

We consider a lattice of flux lines or vortices aligned with the $z$-axis and
described by the unperturbed vortex core positions $\vec{R}_\mu \in
\mathbb{R}^2$.  The pinning force acts on the vortex cores and the pinning
energy can be expressed in the form [with ${\vrh = (\vec{R}, z)]}$
\begin{align}\label{eq:pinning_potential}
   \varepsilon_\pin(\vrh,\vec{u}) = \sum_\mu \delta^{(2)}(\vec{R} - \vec{R}_\mu)\,
   \varepsilon_\pin^\mu[z,\vec{u}_\mu(z)]
\end{align}
with $\varepsilon_\pin^\mu[z,\vec{u}_\mu(z)]$ the random pinning potential
acting on the $\mu$-th vortex line,
\begin{align}\label{eq:line_pinning_potential}
   \varepsilon_\pin^\mu[z,\vec{u}_\mu(z)] \!
    = \!\! \int \! d^2 \vec{R}\, U_\pin(\vec{R},z)\, p[\vec{R} 
    - \vec{R}_\mu \! - \vec{u}_\mu(z)].
\end{align}
Here, $U_\pin(\vec{R},z)$ denotes the disorder potential generated by the
material defects; assuming pinning due to point-like defects located at
$\vec{r}_i = (\vec{R}_i, z_i)$, each with identical pinning energy $e_p$, the
disorder potential takes the form
\begin{align}\label{eq:disorder}
   U_\pin(\vec{R},z) = -\sum_{i} e_p\, \delta^2(\vec{R} - \vec{R}_i)\, \delta(z - z_i).
\end{align}
The factor $p(\vec{R})$ in Eq.~\eqref{eq:line_pinning_potential} describes the
vortex form factor, e.g., for a $\delta T_c$-type pinning mechanism
\cite{Blatter1994}, it reads $p(R) = 1 - |\psi(R)|^2$, with
$\psi(R)$ the superconducting order parameter of the single-vortex
solution to the Ginzburg-Landau equations. The simple \textit{Ansatz}
\cite{Schmid1966, Clem1975} $|\psi(R)| = R/(R^2 + 2\xi^2)^{1/2}$ provides us
with Lorentzian shape for the form factor, $p(R) = 1/(1+R^2/2\xi^2)$.

Combining Eqs.~\eqref{eq:line_pinning_potential} and \eqref{eq:disorder}, we
express the random pinning potential as
\begin{align}
   \varepsilon_\pin^\mu[z, \vec{u}_\mu(z)] 
   = \sum_i e_p[\vec{R}_i - \vec{R}_{\mu} - \vec{u}_\mu(z)]\delta (z-z_i),
\end{align}
with $e_p(\vec{R}) = -e_p\,p(\vec{R})$ the pinning potential due to a single
defect; note that $e_p(\vec{R})$ is maximally negative for $\vec{R} = 0$,
i.e., pinning is maximal when the defect position $\vec{R}_i$ coincides with
the perturbed vortex position $\vec{R}_\mu + \vec{u}_\mu(z)$. Substituting
this result to Eqs.~\eqref{eq:u_general} and \eqref{eq:pinning_potential}, we
arrive at the equation for the displacement of the $\nu$-th vortex in the form
\begin{align}\label{eq:u_VL}
   \vec{u}_\nu(z) & \equiv \vec{u}(\vec{R}_\nu, z) \\ \nonumber
   & = \sum_{\mu, i}G(\vec{R}_\nu - \vec{R}_\mu, z - z_i)
   \vec{f}_{p}[\vec{R}_\mu + \vec{u}_\mu(z_i) - \vec{R}_i]
\end{align}
with the pinning force
\begin{align}\label{eq:f_p_fun}
   \vec{f}_{p}(\vec{R}) = -\nabla_{\vec{R}}e_{p}(\vec{R}) =
   -\frac{e_p}{\xi}\frac{\vec{R}/\xi}{(1 + R^2/2\xi^2)^2}
\end{align}
acting in the direction transverse to the field. The last relation above
applies for the Lorentzian-shaped potential.

In Eq.~\eqref{eq:u_VL}, we sum over all interactions between defects and
vortices. In practice, we assume that no more than a single vortex can be
pinned by an impurity and neglect interactions of vortices with defects far
away from the vortex core, $|\vec{R}_\mu + \vec{u}_\mu(z_i) - \vec{R}_i| \gg
\xi$. The sum over the vortex index $\mu$ is then restricted to a single index
$\mu(i)$ denoting the vortex closest to the impurity $i$. The relation
\eqref{eq:u_VL} then allows to evaluate the displacement $\vec{u}_\mu(i)$ of
the vortex $\mu$ pinned to the defect $i$ at the position $z_i$; this is
nothing but the vortex tip displacement of the $\mu(i)$-th vortex,
\begin{align}\label{eq:u_i_VL}
\begin{split}
   \vec{u}_i \equiv \vec{u}_{\mu(i)}(z_i) = \sum_{j}&G(\vec{R}_{\mu(i)} 
   - \vec{R}_{\mu(j)}, z_i - z_j)\\
   &\times \vec{f}_{p}(\vec{R}_{\mu(j)} + \vec{u}_j - \vec{R}_j).
\end{split}
\end{align}
The set of equations \eqref{eq:u_i_VL} represents a system of $N$ coupled
non-linear equations for the displacements $\vec{u}_i$, with $N$ the total
number of defects.

Within the strong pinning paradigm, we assume that defects act independently,
allowing for a further simplification of Eq.~\eqref{eq:u_i_VL} where the
displacement $\vec{u}_i$ is ascribed exclusively to the action of the defect
$i$; the summation over $j$ in Eq.~\eqref{eq:u_i_VL} then reduces to the term
$j = i$, i.e., we neglect the force exerted by distant defects $j\neq i$ on
vortices $\mu(j)$ that contributes to the displacement $\vec{u}_i$ via the
non-local Green's function $G(\vec{R}_{\mu(i)} - \vec{R}_{\mu(j)}, z_i -
z_j)$. It is exactly this simplification that will be dropped later on when
considering strong pinning by pairs.  The system of
equations~\eqref{eq:u_i_VL} then reduces to $N$ independent equations
\begin{align}\label{eq:u_on_site}
   \vec{u}_i \approx \vec{f}_p (\vec{R}_{\mu(i)} + \vec{u}_i - \vec{R}_i)/\C
\end{align}
with the effective vortex-lattice elasticity defined by $\C = 1/G(\vec{0})$,
see Eq.~\eqref{eq:G_on_site}.

The resulting pinning-force density is obtained by summing the forces from all
pinning sites. Note that the solution $\vec{u}_i$ in Eq.~\eqref{eq:u_on_site}
depends only on the distance of the vortex from the pinning defect $\vec{x}_i
= \vec{R}_{\mu(i)} - \vec{R}_i$. The average pinning force density is thus
$\vec{F}_\pin = n_p\langle \vec{f}_p[\vec{x} + \vec{u}(\vec{x})]
\rangle_{\vec{x}}$, where $n_p$ denotes the density of impurities and the
average is taken with respect to the possible position vectors $\vec{x}$;
assuming a uniform distribution of relative distances $\vec{x}$, the average
then corresponds to a simple integration over $\vec{x}$.  

It turns out that the pinning force can be expressed as the gradient of the
total pinning energy, $\vec{f}_p[\vec{x} + \vec{u}(\vec{x})] = -\nabla_\vec{x}
e_\pin(\vec{x})$, with $e_\pin(\vec{x})$ involving pinning and elastic terms,
\begin{align}\label{eq:e_pin_def}
   e_\pin(\vec{x}) = e_p[\vec{x} + \vec{u}(\vec{x})] + \tfrac{1}{2}\C
   \vec{u}(\vec{x})^2.
\end{align}

If the solution $\vec{u}(\vec{x})$ to the on-site equation
\eqref{eq:u_on_site} is unique, implying a continuous evolution with
$\vec{x}$, the average pinning force density vanishes, as follows from a
simple integration of $\vec{f}_p[\vec{x}+\vec{u}(\vec{x})]$ over $\vec{x}$,
\begin{align}\label{eq:F_pin_weak}
   \vec{F}_\pin = - n_p \!\! \int \frac{d^2x}{a_0^2} \> \nabla_\vec{x} 
   e_\pin(\vec{x}) =0.
\end{align}

\begin{figure}
\begin{center}
\includegraphics[scale=1]{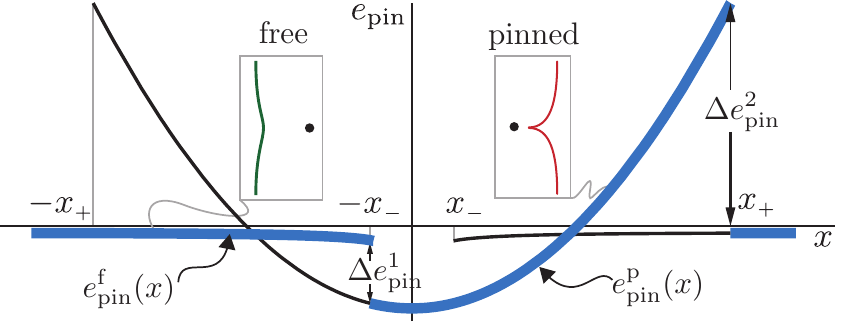}
\end{center}
\caption{Sketch of a typical large-$\kappa$ energy profile $e_{\pin}(x)$ with
multiple branches involving an approximately parabolic pinned branch and
nearly flat free branches. The branch occupation (denoted by thick blue lines)
changes at the points $-x_{\scriptscriptstyle -}$ and $x_{\scriptscriptstyle
+}$ that are associated with the pinning and depinning processes for a vortex
passing the defect. The sum of the associated energy jumps $\Delta e_{\pin}^1$
and $\Delta e_{\pin}^2$ provide a quantitative result for the pinning force
density.} \label{fig:strong}
\end{figure}

The single-defect \textit{Ansatz} is thus meaningful only in the strong
pinning regime where the solution for the on-site displacement is non-unique.
In this case, different values (branches) for the total pinning energy
$e_\pin(\vec{x})$ describe pinned and unpinned vortex states, see Fig.\
\ref{fig:strong}. Proper averaging accounts for the occupation of these
branches which is unsymmetric, resulting in a non-vanishing average
pinning-force density.  We perform this analysis for a radially symmetric
potential with a force $\vec{f}_p(\vec{r}) = \hat{\vec{r}} f_p(r)$.

The condition for the appearance of multiple solutions
is provided by the Labusch criterion \cite{Labusch1969, Blatter2004,
Brazovskii2004} 
\begin{align}\label{eq:Labusch_crit}
   \kappa = \frac{\max f_p'(r)}{\C} > 1
\end{align}
(we note that $\max f_p'(r) = f_p'(r_m)$ with the inflection point $r_m$
obtained from $f_p''(r_m) =0$).  Furthermore, we assume that the vortices are
driven in the positive $x$-direction and we parametrize their trajectories
$\vec{x} = (x,b)$ through the longitudinal vortex position $x$ and an impact
parameter $b$ in the transverse direction (the distinction between the
`longitudinal' field direction along $z$ and the `longitudinal' direction of
motion along $x$ should be clear from the context).  The resulting pinning
force averaged over positions $\vec{x}$ then points in the negative
$x$-direction and is evaluated in two steps: first, we perform an averaging
over $x$ at vanishing impact parameter $b = 0$ and then we average over
contributions from vortex trajectories with finite impact parameters $b \neq
0$.

For the case of a vanishing impact parameter $b = 0$, Eq.~\eqref{eq:u_on_site}
can be reduced to one dimension, $\C u = f_p(x + u)$ (we have dropped the
index $i$).  This is equivalent to minimizing the total pinning energy
$e_\pin(x,u) = e_p(x + u) + \tfrac{1}{2}\C u^2$ with respect to $u$. Provided
$x$ falls into the bistability region, $|x|\in [x_\-,x_\+]$, there exist
multiple solutions $u_\t{f}(x)$, $u_\t{p}(x)$ for the vortex tip displacement
(denoting \textit{free} and \textit{pinned} vortex states) \cite{Blatter2008,
Willa2016, Thomann2017}. Substituting these solutions to the total pinning
energy provides multiple branches $e_\pin^{\t{f},\t{p}}(x)\equiv
e_\pin[x,u_{\t{f},\t{p}}(x)]$, see Fig.~\ref{fig:strong}. When going from
large negative to large positive $x$, the branch occupation first undergoes a
transition from the free to the pinned branch at the pinning point $-x_\-$ and
then another transition from the pinned to the free branch at the depinning
point $x_\+$. Averaging the pinning force $f_p[x + u_o(x)]$ over the occupied
branches (as marked by the index $o\in\lbrace \t{f},\t{p} \rbrace$), the
resulting integral over $x$ can be expressed as (see Eq.\
\eqref{eq:F_pin_weak} and \cite{Thomann2017, Willa2016, Buchacek2019}),
\begin{align}\label{eq:f_pin_av_strong_1D}
   \langle f_p[x + u_o(x)]\rangle_{x} = -\frac{\Delta e_\pin^1 + \Delta e_\pin^2}{a_0},
\end{align}
with the jumps in energy $\Delta e_\pin^1 = [e_\pin^\t{f} - e_\pin^\t{p}]_{x =
-x_\-}$ and $\Delta e_\pin^2 = [e_\pin^\t{p} - e_\pin^\t{f}]_{x = x_\+}$
occurring at the pinning ($-x_\-$) and depinning ($x_\+$) points,
respectively. 

The result \eqref{eq:f_pin_av_strong_1D} remains unchanged even for a
non-vanishing impact parameter $b\neq 0$ \cite{Thomann2017, Buchacek2019},
provided the vortex passes the defect within the pinning distance $y_p$ along
the $y$-direction; for the radially symmetric case, it turns out that $y_p =
x_\-$ and hence pinning occurs for impacts with $|b| < x_\-$. For $|b|> x_\-$,
the pinning forces are small and multiple branches no longer exist, implying a
vanishing average over $x$ \cite{Thomann2017, Buchacek2019}.  Finally,
averaging the result \eqref{eq:f_pin_av_strong_1D} over $y$ contributes a
factor $2x_\-/a_0$ and thus
\begin{align}\label{eq:f_pin_av_strong}
   \langle \vec{f}_p[\vec{x} + \vec{u}(\vec{x})] \rangle_\vec{x} 
   = (-\vec{e}_x)\frac{2 x_\-}{a_0}\frac{\Delta e_\pin^1 + \Delta e_\pin^2}{a_0},
\end{align}
with $-\vec{e}_x$ denoting the unit vector pointing in the negative $x$-direction.

Eq.~\eqref{eq:f_pin_av_strong} assumes different scaling forms for the regime
of very strong pinning $\kappa \gg 1$ and for moderately strong pinning
$\kappa - 1\ll 1$. In the first case, the jump sizes are related to the
pinning potential depth via $\Delta e_\pin^{1}\sim e_p$ and $\Delta
e_\pin^2\sim \kappa e_p$ \cite{Thomann2017, Buchacek2019}, together providing
the estimate for the magnitude of the position-averaged pinning force
$\langle f_p \rangle_{\vec{x}}\sim (\kappa\xi^2/a_0^2)\,f_p$ and a pinning force density
\begin{align}\label{eq:F_pin_av_scaling_strong}
   F_\pin \sim \frac{\kappa \xi^2}{a_0^2}\,n_p f_p 
   = \frac{S_\mathrm{trap}}{a_0^2}\,n_p f_p.
\end{align}
This result is interpreted as a pinning force $f_p\sim e_p/\xi$ [see
Eq.~\eqref{eq:f_p_fun}] due to a single defect exerted within the trapping
area \cite{IvlevOvchinnikov1991,Blatter2004} $S_\mathrm{trap} = 2 y_p
(x_\+ + x_\-) \sim \kappa \xi^2$; $S_\mathrm{trap}/a_0^2$ denotes the fraction
of area occupied by trapped vortices.

For moderately strong pinning with $\kappa$ close to unity (that is
particularly relevant for the pinning by rare events), the energy jumps are
evaluated by expanding the pinning force around the inflection point at $r_m$,
$f_p''(r_m) = 0$, where $f_p'(r_m) = \kappa \C$ is maximally positive
\cite{Blatter2004, Koopmann2004, Buchacek2019, Willa2016},
\begin{align}
   f_p(r_m + \delta r) \approx f_p(r_m) + \kappa \C \delta r 
   + \tfrac{1}{6}f_p'''(r_m) \delta r^3.
\end{align}
In this situation, both jumps are identical and given by the expression
\cite{Blatter2004, Buchacek2019} (note that $f_p'''(r_m) < 0$)
\begin{align}\label{eq:jump_marginal}
   \Delta e_\pin^1 = \Delta e_\pin^2 = \frac{9 \C^2}{2[-f_p'''(r_m)]}(\kappa - 1)^2.
\end{align}
Using the scaling formulas $\C = f_p'(r_m)/\kappa\sim f_p\xi$ (provided
$\kappa \sim \mathcal{O}(1)$), $f_p'''(r_m) \sim f_p/\xi^3$ in
Eq.~\eqref{eq:jump_marginal} then gives
\begin{align}\label{eq:f_pin_av_scaling_weak}
\langle f_p\rangle_\vec{x} \sim \frac{\xi^2}{a_0^2}\, f_p (\kappa - 1)^2.
\end{align}
The pinning force density follows trivially, 
\begin{align}\label{eq:F_strong}
   F_\mathrm{pin} \sim \frac{\xi^2}{a_0^2}\, (\kappa - 1)^2 n_p f_p,
\end{align}
and vanishes at the Labusch point $\kappa = 1$, in accordance with the strong
pinning criterion \eqref{eq:Labusch_crit}.

\subsection{Weak collective pinning}\label{sect:wcp}

When pinning is weak, $\kappa < 1$, individual defects fail to produce
multi-valued solutions for the vortex displacement and the mean pinning force
in Eq.\ \eqref{eq:F_pin_weak} vanishes. Pinning then arises through the random
action of defects within the collective pinning volume $V_c$ defined as the
region where the spatial fluctuations of the vortex displacement $\langle
u^2(\vrh)\rangle=\langle [\vec{u}(\vrh) - \vec{u}(\vec{0})]\rangle^2$ remains
bounded by the pinning scale, $\langle u^2(\vrh)\rangle \leq \xi^2$. The
displacement correlation function can be systematically evaluated from
Eq.~\eqref{eq:u_general} using the disorder-averaged correlator of the pinning
energy density Eq.\ \eqref{eq:pinning_potential} \cite{Blatter1994, Larkin1970},
\begin{align}\label{eq:disorder_correlator}
   \langle \varepsilon_\pin(\vec{\vrh},\vec{u})\varepsilon_\pin(\vrh',\vec{u}') 
   \rangle \! = \! \frac{e_p^2 n_p}{a_0^2}
   \delta^3(\vrh\! -\! \vrh')k(\vec{u}\! -\!  \vec{u}'),
\end{align}
with the correlation function $k(\vec{u} - \vec{u}') = \int d^2\vec{R} \,
p(\vec{R} - \vec{u})p(\vec{R} - \vec{u}')$ related to the vortex form factor $p(R)$.

A qualitative estimate for the displacement correlator is provided by summing
up distortions originating from all defects within a finite volume. In the
vicinity of a reference defect characterized by the pinning force $f_p\sim
e_p/\xi$, the distortion scale $u_0$ is given by the on-site Green's function,
$u_0 \sim G(\vec{0})^{-1}f_p$. Expressing the on-site displacement through the
effective vortex lattice stiffness $\C = G(\vec{0})^{-1}$ and estimating the
Labusch parameter as $\kappa\sim f_p/\C \xi$ provides us with $u_0\sim \kappa
\xi$. Assuming small defect densities and hence large typical inter-defect
separations, the extension of the collective pinning volume falls into the
non-dispersive regime of the Green's function, see Eq.~\eqref{eq:G_non_disp}.
Defects located a distance $\tilde\rho^2 = R^2 + (a_0^2/16\pi\lambda^2)
z^2$ away from the reference defect contribute with the displacement
$u(\tilde\rho) \sim u_0\> G(\tilde\vrh)/G(\vec{0})\sim u_0(a_0^2/\lambda
\tilde\rho)$. Within the collective pinning volume $V_c = R_c^2 L_c \sim
(\lambda/a_0)R_c^3$, these displacements add up with a random sign, as the
forces from different defects are randomly directed; furthermore, only the
fraction $\xi^2/a_0^2$ of defects that reside inside the vortex cores are
directly attacking the vortices, resulting in a total squared displacement
$\langle u^2(R_c) \rangle \sim (\kappa\xi\, a_0^2/\lambda R_c)^2
(\xi^2/a_0^2)\, n_p V_c$ on the scale $R_c$. Finally, the condition $\langle
u^2(R_c)\rangle \sim \xi^2$ provides us with the collective pinning radius
\begin{align}\label{eq:R_c}
   R_c \sim \lambda \frac{1}{\kappa^2\, n_p \xi^2 a_0}.
\end{align}

For small defect densities, as specified by the condition $\kappa^2 n_p \xi^2
a_0 \ll 1$, the pinning radius $R_c\gg \lambda$ indeed falls into the
non-dispersive regime (note that $\kappa \lesssim 1$).  Finally, summing up
the random force-contributions due to the active defects within the bundle
volume $V_c = (\lambda/a_0)R_c^3$, $F_\mathrm{coll} \sim [f_p^2 n_p
(\xi^2/a_0^2) V_c]^{1/2}/V_c$, we find the collective pinning-force density
\begin{align}\label{eq:F_coll}
   F_\mathrm{coll} \sim \frac{\xi^2}{\lambda^2}\, \kappa^3 (n_p a_0 \xi^2) n_p f_p.
\end{align}

\subsection{Pinning by rare events}\label{sect:rare_events}
The collective pinning scenario described above sums up small competing
contributions to the vortex lattice distortions arising from {\it typical}
fluctuations in the defect distribution, involving defects that lie far away
from each other within the collective pinning volume.  However, it does not
account for the presence of \textit{rare} clusters, where two (or more) weak
defects act cooperatively, giving rise to an effectively strong pinning
center; the latter then is supposed to produce a distortion exceeding the
scale $\xi$ of the pinning potential. In looking for promising candidate
pairs, we consider Eqs.~\eqref{eq:G_non_disp} and \eqref{eq:G_interp} that
describe the decay of the Green's function; these imply that the vortex
displacement is substantially suppressed beyond a distance $\sim a_0$ away
from the defect. Hence, two weak defects with $\tfrac{1}{2} < \kappa < 1$ can
be combined into a strongly-pinning object characterized by $\kappa > 1$ and
producing a displacement $u$ of order $\xi$ only if they are at most a
longitudinal distance $z\sim a_0$ apart and pinning the same vortex core,
i.e., they are separated by at most $R\sim \xi$ in the transverse dimension.
This consideration then provides us with the density $(n_p a_0\xi^2) \, n_p$
of strongly-pinning pairs. With only the fraction $\xi^2/a_0^2$ of those
clusters being located within the vortex core area and each cluster exerting a
pinning force $\sim f_p$, we arrive at the following estimate for the pinning
force density due to defect pairs (with $1/2 < \kappa < 1$ still close to
unity),
\begin{align}\label{eq:F_clust_intr}
   F_\mathrm{clust}\sim \frac{\xi^2}{a_0^2} (n_p a_0 \xi^2)\,  n_p f_p.
\end{align}
Assuming a magnetic field sufficiently above $H_{c1}$ such that $a_0 <
\lambda$, the pinning force due to such clusters dominates over the collective
pinning contribution in Eq.~\eqref{eq:F_coll} by a factor of
$(\lambda/a_0)^2$.

This factor in fact arises due to the dispersion of the tilt elastic modulus;
in order to trace its origin, we need to understand the explicit dependence of
the quantities contributing to both pinning mechanisms on the elastic
properties of the vortex lattice, a dispersive tilt modulus $c_{44}(\vec{k})$
and non-dispersive shear modulus $c_{66}$.  The collective pinning radius in
Eq.~\eqref{eq:R_c} can be obtained by comparing the elastic energy
$\mathcal{E}_{\mathrm{el}}(V_c)\sim c_{66}(\xi/R_c)^2 V_c$ and the pinning
energy $\mathcal{E}_\pin(V_c)\sim [f_p^2 n_p (\xi^2/a_0^2) V_c]^{1/2}\xi$
accumulated within the pinning domain of volume $V_c = R_c^2 L_c$. Assuming a
large collective pinning volume where the dispersion of the tilt modulus is
not relevant, the extensions $R_c$ and $L_c$ in the longitudinal and the
transverse directions are related via $c_{66}(\xi/R_c)^2\sim
c_{44}(\vec{0})(\xi/L_c)^2$, that provides us with (we write
$c_{44}(\vec{0}) = c_{44}$)
\begin{align}
   R_c \sim \frac{c_{66}^{3/2} c_{44}^{1/2} a_0^2}{f_p^2 n_p},
\end{align}
and the pinning force density is estimated as
\begin{align}\label{eq:F_coll_reexpressed_0}
    F_\mathrm{coll} \sim \frac{\mathcal{E}_\mathrm{el}(V_c)}{V_c\,\xi}
    \sim \frac{\xi}{a_0^4}\frac{f_p^4 n_p^2}{c_{66}^2 c_{44}}.
\end{align}
We replace a factor $f_p^3$ in \eqref{eq:F_coll_reexpressed_0} with
$\kappa^3(\C\xi)^3$,
\begin{align}\label{eq:F_coll_reexpressed_1}
    F_\mathrm{coll} 
    \sim \frac{\xi^4}{a_0^4}n_p^2 f_p \frac{\kappa^3\C^3}{c_{66}^2 c_{44}},
\end{align}
and using $\kappa \sim 1$, $\C\sim a_0 [c_{44}(K_\BZ)c_{66}]^{1/2}$, and
$[c_{44}(K_\BZ)/c_{66}]^{1/2} \sim 1$, we obtain the desired result
\begin{align} \label{eq:F_coll_reexpressed}
    F_\mathrm{coll} \sim \frac{c_{44}(K_\BZ)}{c_{44}(0)} \>
    \frac{\xi^2}{a_0^2}(n_p a_0 \xi^2)n_p f_p.
\end{align}

On the other hand, our strongly-pinning pairs are small and the associated
elastic scales $R_p$ and $L_p$ in the longitudinal and transverse directions
are related by $c_{66}(u/R_p)^2\sim c_{44}(K_{\rm\scriptscriptstyle
BZ})(u/L_p)^2$ with the short scale elasticity $c_{44} (K_{\rm
\scriptscriptstyle BZ})$; hence $L_p \sim a_0 \sqrt{c_{44}(K_\BZ)/c_{66}} \sim
a_0$, where we have chosen the smallest transverse scale $R_p \sim a_0$ of the
lattice. The density of pairs then is given by $n_p (n_p \xi^2 a_0)$ and the
resulting pair pinning-force density is (assuming again $\kappa \sim
\mathcal{O}(1)$, cf.\ Eq.\ \eqref{eq:F_clust_intr})
\begin{align}\label{eq:F_clust_reexpressed}
   F_\mathrm{clust} \sim \frac{\xi^2}{a_0^2} (n_p a_0 \xi^2) n_p f_p.
\end{align}
Finally, comparing the weak-collective- and cluster-pinning force densities in
Eqs.~\eqref{eq:F_coll_reexpressed} and \eqref{eq:F_clust_reexpressed} provides
us with
\begin{align}\label{eq:pinning_force_ratio}
   \frac{F_\mathrm{clust}}{F_\mathrm{coll}} \sim \frac{c_{44}(0)}{c_{44}(K_\BZ)}
   \sim \frac{\lambda^2}{a_0^2},
\end{align}
that demonstrates that the cluster pinning dominates over the weak-collective
pinning contributions due to the dispersion in the tilt modulus with its
reduction $\propto (a_0/\lambda^2)$ at the Brillouin zone boundary.

The concept of pair-pinning described above can be extended to larger
clusters, pushing the domain of pinning by rare events further down to smaller
values of $\kappa$.  Within the interval $\kappa \in [1/n,1/(n-1)]$, $n\geq 2$
and integer, $n$ neighboring defects are required to form a strong-pinning
cluster with $n\kappa > 1$; the density of such clusters is given by $(n_p
a_0\xi^2)^{n-1} n_p$ and the resulting pinning force density becomes
$F_\mathrm{clust}\sim (\xi/a_0)^2 (n_p a_0 \xi^2)^{n-1} n_p f_p$. However, for
pinning strengths $\kappa \leq 1/n$ with $n\approx 2 + 2[\ln
(\lambda/a_0)]/[\ln (1/n_p a_0 \xi^2)]$, the collective pinning dominates;
given a low density of defects such that $n_p a_0\xi^2 \ll (a_0/\lambda)^2$,
this crossover lies close to $n = 2$, $\kappa = \tfrac{1}{2}$.

The idea of pinning due to rare events has been previously touched upon in the
context of charge density wave pinning in high dimensions, see Ref.\
\cite{Fisher1985}. In this case, the disorder-induced distortions accumulated
over a finite-sized domain are not sufficient to induce pinning.  This can be
easily seen by considering the elastic Green's function $G(\vrh)\propto
\rho^{2-D}$ in $D$ dimensions,  yielding a total displacement accumulated
within a pinning domain of size $R$ that scales as $\langle u^2(R) \rangle\sim
R^{4-D}$, see Eq.~\eqref{eq:u_general}.  While for $D<4$, the accumulated
displacement will eventually exceed the threshold required for the existence
of bistabilities at large domain sizes $R$, this is not the case for
dimensions $D\geq 4$. As noted by Fisher \cite{Fisher1985}, this does not
render the weak disorder irrelevant, since, although with exponentially small
probability, one will always find rare domains with anomalously coherent
pinning.  The manifold is then pinned by such rare fluctuations rather than by
the collective action of the disorder landscape. In our $D=3$ vortex lattice,
the situation is somewhat different: for $D=3$, weak pinning is still active
and competes with the pinning by rare events, which take the specific form of
close-by defect pairs making up for a strongly-pinning object.  The latter
mechanism dominates for defect strengths $\tfrac{1}{2} < \kappa < 1$ and a
small density of defects. The dominance of pinning by such rare events is,
however, not an inherent property of the {\it pinning} mechanism, but rather
appears as a result of the specific, i.e., dispersive, {\it elastic} response
of the vortex lattice.

\section{Two-defect problem}\label{sect:two_defect_problem}
We have seen in Sect.~\ref{sect:sp} that the strong pinning paradigm assuming
independent action of defects is meaningful only provided the Labusch
parameter \eqref{eq:Labusch_crit} satisfies $\kappa > 1$; in this case, the
single-defect Ansatz gives rise to multi-valued solutions for the vortex
displacement, what results in a finite averaged pinning force. Here, we
consider the range of pinning strength $\tfrac{1}{2}<\kappa < 1$ and go a step
beyond the single-defect ansatz by considering pairs of defects. Given $N$
defects, of all possible $N(N-1)/2$ pairings there will be a finite set of
pairs that reach the strong pinning criterion, thus generating multi-valued
solutions of vortex states; given that defects are dilute, these strong
pinning pairs will be dilute as well and hence act independently. 

\subsection{Geometry}

To find the relevant pairs, we consider two defects labelled by $i = 1,\,2$ at
positions $(\vec{R}_i,z_i)$ and the associated vortices $\vec{R}_{\mu(i)}$
separated from the pins by $\vec{x}_i = \vec{R}_{\mu(i)} - \vec{R}_i$; the
displacement fields $\vec{u}_i$ at $z_i$ are solutions of the coupled
equations (see Eq.\ \eqref{eq:u_i_VL}),
\begin{align}\label{eq:two_defects}
\begin{split}
   \C\vec{u}_1 = \vec{f}_p(\vec{x}_1 + \vec{u}_1) + g\vec{f}_p(\vec{x}_2 + \vec{u}_2),\\
   \C\vec{u}_2 = \vec{f}_p(\vec{x}_2 + \vec{u}_2) + g\vec{f}_p(\vec{x}_1 + \vec{u}_1).
\end{split}
\end{align}
The coupling $g\in (0,1]$ renormalizes the force at the site of the first
impurity due to the action of the second impurity (and vice-versa) and reads
\begin{align}\label{eq:g}
   g = G(\vec{R}_{\mu(1)} - \vec{R}_{\mu(2)}, z_1 - z_2)/G(\vec{0}).
\end{align}
While the impurity positions $\vec{R}_i$ and displacements
$\vec{u}_{1,2}$ in \eqref{eq:two_defects} are continuous variables (with the
small scale set by $\xi$), the vortex positions $\vec{R}_{\mu(1,2)}$ in
\eqref{eq:g} are restricted to the vortex lattice involving the scale $a_0$.
A coupling $g$ of order unity implies that both impurities act with their
maximal pinning force on the {\it same vortex}$\,$; typical separations of such
defects are below $\xi$ in the transverse and below $a_0$ in the longitudinal
direction.  Hence, large couplings $g$ are associated with close defect pairs
lying within a volume $a_0 \xi^2$. Small couplings $g \ll 1$ refer to the
situation where the impurities are separated far away from one another, of
order several lattice constants $a_0$, typically; in this situation, the
defects act on different vortices and their mutual effect on the vortex pair
is small.

\begin{figure}[t]
\includegraphics[scale=1]{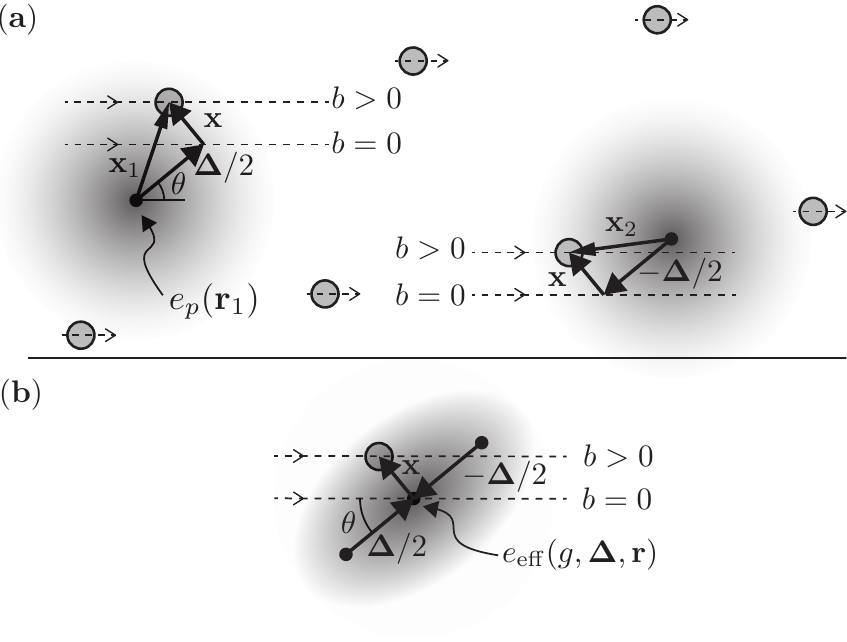}
\caption{Explanation of the mean vortex position $\vec{x}$ and the mismatch
vector $\vec{\Delta} = (\Delta \cos\theta, \Delta \sin\theta)$ for a vortex
lattice driven in the positive $x$-direction. (a) Two defects (black points)
are pinning two vortices (grey circles) at the unperturbed relative (to the
defects) positions $\vec{x}_1 = \vec{\Delta}/2 + \vec{x}$ and $\vec{x}_2 =
-\vec{\Delta}/2 + \vec{x}$. The areas shaded in grey represent the defects'
pinning potentials $e_p(\vr_i)$ with $\vr_i = \vec{x}_i + \vec{u}_i$,
$i\in\lbrace 1,2 \rbrace$, the positions of the vortex tips displaced by
$\vec{u}_i$ (not shown on the picture). The other vortices of the lattice are
not affected by any of the defects but their asymptotic positions are
co-moving with the vortex lattice. The pinning-force averaging is performed
over the trajectories $\vec{x} = (x,b)$; shown is an example of a trajectory
with impact parameter $b > 0$. Shifting the whole lattice downwards produces
the special trajectory of maximal symmetry characterized by $b = 0$. In this
case, both vortices are separated from the defect by the same transverse
distance $|\Delta \sin\theta|$ (along $y$). When passing through the point
$\vec{x} = 0$, their position relative to the defect is given by opposite
vectors $\vec{\Delta}/2$ and $-\vec{\Delta}/2$. (b) Reduction of the
two-vortex problem to the pinning of an effective vortex at the position
$\vec{x} = \tfrac{1}{2}(\vec{x}_1 + \vec{x}_2)$ relative to an effective
defect characterized by an anisotropic (or non-radial) pinning potential
$e_\eff(g,\vec{\Delta};\vr)$, with $\vr = \vec{x} + \vec{u}$ and $\vec{u} =
\tfrac{1}{2}(\vec{u}_1 + \vec{u}_2)$ the displacement of the effective vortex.
For $b=0$ the effective vortex passes through the center of $e_\eff(\vr)$.
}\label{fig:mismatch}
\end{figure}

As in Sect.~\ref{sect:sp}, we consider a driving force applied in the positive
$x$-direction and assume that the vortex lattice structure is preserved; under
application of the drive, the vortices are displaced from their initial
positions $\vec{R}_{\mu(i)}^0$ by a constant shift of magnitude $X$ along $x$,
i.e, $\vec{R}_{\mu(i)} = \vec{R}_{\mu(i)}^0 + X\, \vec{e}_x$.  It is
convenient to reformulate the problem in terms of the \textit{mean}
vortex position $\vec{x}$ (relative to the defects)
\begin{align}
   \vec{x} &= \tfrac{1}{2}(\vec{x}_1 + \vec{x}_2)\nonumber\\
   &=\tfrac{1}{2}(\vec{R}_{\mu(1)}^0 + \vec{R}_{\mu(2)}^0) 
   - \tfrac{1}{2}(\vec{R}_1 + \vec{R}_2) + X\,\vec{e}_x,\label{eq:def_x}
\end{align}
and the \textit{mismatch} vector $\vec{\Delta}$,
\begin{align}
   \vec{\Delta} &= \vec{x}_1 - \vec{x}_2 = \vec{R}_{\mu(1)}^0 - \vec{R}_1 -
   (\vec{R}_{\mu(2)}^0 - \vec{R}_2),\label{eq:def_Delta}
\end{align}
see Fig.\ \ref{fig:mismatch}. Note that the vortex positions $\vec{x}_1$ and
$\vec{x}_2$ (relative to the defects) as well as the mismatch vector
$\vec{\Delta}$ are restricted to the unit cell of the vortex lattice.

Pushing the vortex lattice a distance $X$ along the $x$-direction, the mean vortex
position $\vec{x}$ is parametrized as $\vec{x} = [x(X),b]$ with a fixed
impact parameter $b$, while the vector $\vec{\Delta}$ remains constant. Since
$\vec{x}_1 = \vec{x} + \vec{\Delta}/2$ and $\vec{x}_2 = \vec{x} -
\vec{\Delta}/2$ (see Fig.~\ref{fig:mismatch}), the vector $\vec{\Delta}$ is
interpreted as the \textit{mismatch} in the pinning by the two defects. If
$\vec{\Delta} = \vec{0}$, the defects are perfectly synchronized: for any
$X$, the position of both vortices relative to the defects is the same,
$\vec{x}_1 = \vec{x}_2$, the pinning forces acting on both vortices are
identical, and pinning by the defect pair is maximal. For a finite
$\vec{\Delta}\neq\vec{0}$, the two vortices are subject to different pinning
forces, which reduces the total pinning strength. As shown in Fig.\
\ref{fig:mismatch}(b), the geometry can be reduced to one where an effective
vortex at the position $\vec{x}$ impacts on an effective defect with a
non-radial pinning potential $e_\eff(g,\vec{\Delta};\vr)$, with $\vr = \vec{x}
+ \vec{u}$ and $\vec{u} = \tfrac{1}{2}(\vec{u}_1 + \vec{u}_2)$ the displacement of
the effective vortex.

Fig.~\ref{fig:mismatch} also clarifies the meaning of the \textit{head-on}
vortex trajectory $\vec{x} = [x(X),0]$ with vanishing impact parameter $b$ in the
context of the two-defect problem. When $\vec{\Delta}\neq \vec{0}$, it is not
possible for both trajectories $\vec{x}_1$, $\vec{x}_2$ to simultaneously pass
through the defect centers. The special value $b = 0$ then describes the
situation where both vortices are separated by the same transverse distance from
the defects and the vortex trajectory passes through the special point
$\vec{x} = 0$ when the vortices are located at opposite positions
$\vec{\Delta}/2$, $-\vec{\Delta}/2$ with respect to the defects. Translated to
the effective geometry, for $b=0$ the effective vortex passes through the
center of $e_\eff(\vr)$.

\subsection{Averaging}

Given the geometric layout of the strong pinning problem with two vortices and
two defects, we have to find the associated pinning force density $F_\pin$ by
proper averaging.  This averaging involves i) the averaging over trajectories
$\vec{x} = (x,b)$ of vortex pairs with fixed mismatch $\vec{\Delta}$ and fixed
coupling $g$, ii) the averaging over all possible mismatch vectors
$\vec{\Delta}$, and iii) the averaging over couplings $g$ in the pair pinning
Eq.\ \eqref{eq:two_defects} that involves the relative distances between
vortices $\vec{R}_{\mu(1)} - \vec{R}_{\mu(2)}$ and the distance in elevation
$z_1 -z_2$ of defects, see Eq.\ \eqref{eq:g}. The final result will
provide us with a formula, Eq.\ \eqref{eq:F_pin_av}, that expresses the
pinning-force density $F_\pin$ due to defect pairs in terms of the individual
pair forces $f_\mathrm{pair} [g(\vrh),\vec\Delta]$ for defects separated by
$\vrh$ and with a mismatch $\vec\Delta$ between them and the vortices. While
this expression can be evaluated precisely using numerical techniques, here,
we will discuss analytic results that are necessarily of approximate nature.

In the first step i), we fix $\vec{\Delta}$ and average over the vector
$\vec{x} = [x(X),b]$ in a similar fashion as in Sect.~\ref{sect:sp}. For each
$b$, we average the aggregated pinning force exerted by the two defects while
pushing $X$ from large negative to large positive values and then take the
average over the impact parameter $b$. This procedure provides us with the
pinning force $\vec{f}_\mathrm{pair}$ of a defect pair at fixed mismatch vector
$\vec{\Delta}$ and coupling $g$,
\begin{align}\label{eq:x_average}
   &\vec{f}_\mathrm{pair}(g,\vec{\Delta}) = \nonumber\\
   &\,=\Bigl\langle \vec{f}_p\Bigl[\vec{x} + \frac{\vec{\Delta}}{2} 
   + \vec{u}_1(\vec{x})\Bigr] + \vec{f}_p\Bigl[\vec{x} 
   - \frac{\vec{\Delta}}{2} + \vec{u}_2(\vec{x})\Bigr]\Bigr\rangle_{\vec{x}}
\end{align}
where $\vec{u}_1(\vec{x}), \vec{u}_2(\vec{x})$ are the solutions to the
coupled system, Eq.~\eqref{eq:two_defects}. As before, this average involves
the jumps in energy between bistable solutions for the pair-pinning problem
defined by Eq.~\eqref{eq:two_defects}.

In the next step ii), we average over mismatch vectors $\vec{\Delta}$ (the
normalization $a_0^2$ follows from $\vec{\Delta}$ being restricted to the unit
cell of the vortex lattice),
\begin{align}\label{eq:Delta_average}
   \langle f_\mathrm{pair}(g,\vec{\Delta}) \rangle_{\vec{\Delta}} 
   = \int \frac{d^2\Delta}{a_0^2}\, f_\mathrm{pair}(g,\vec{\Delta}),
\end{align}
where $f_\mathrm{pair} = \vec{f}_\mathrm{pair}\cdot (-\vec{e}_x)$ denotes the
(negative) $x$-component of $\vec{f}_\mathrm{pair}$; note that for a vortex
lattice pushed along the positive $x$-direction, the $\vec{x}$-averaged
pinning force points in the negative $x$-direction.  The $y$-component of
$\vec{f}_\mathrm{pair}$ vanishes after the $\vec{\Delta}$-averaging since it
is compensated by the configuration with $\vec{\Delta}\to -\vec{\Delta}$ and
$b\to - b$.

Third, we determine the pinning force density $F_\pin$ exerted by all defect
pairs within a volume $V$ by summing over pairs that are pinned at different
separation $\vec{\vrh} = (\vec{R}_{\mu(1)} - \vec{R}_{\mu(2)}, z_1 - z_2)$,
where $\vec{R}_{\mu(1)} - \vec{R}_{\mu(2)}$ refers to the separation between
the vortices and $z_1 - z_2$ is the distance between the defects along the
$z$-axis. This final sum (or average) accounts for the dependence of the
pair-force $\vec{f}_\mathrm{pair}(g,\vec{\Delta})$ on the coupling $g(\vrh)$.
Approximating the sum by an integral gives (note the factor of $\tfrac{1}{2}$
to avoid double-counting of the defects)
\begin{align}\label{eq:F_pin_av_0}
   F_\pin = \frac{n_p^2}{2V} &\int_V d^2\vec{R}_{\mu(1)}\,d^2 
   \vec{R}_{\mu(2)}\,dz_1\,dz_2 \\ \nonumber
   &\times \langle f_\mathrm{pair}[g(\vec{R}_{\mu(1)} - \vec{R}_{\mu(2)}, z_1 -
   z_2),\vec{\Delta}] \rangle_{\vec{\Delta}}.
\end{align}
Carrying out one volume integral, we arrive at the final expression
\begin{align}\label{eq:F_pin_av}
   F_\mathrm{pin} = \frac{n_p^2}{2}\int_V d^3\vrh\,
   \langle f_\mathrm{pair}[g(\vrh),\vec{\Delta}] \rangle_{\vec{\Delta}}.
\end{align}

It remains to solve the coupled equations \eqref{eq:two_defects} and determine
the resulting pinning force $\vec{f}_\mathrm{pair}(g,\vec{\Delta})$ of defect
pairs that enters the final expression \eqref{eq:F_pin_av} for the
pinning-force density $F_\mathrm{pin}$. We first provide a qualitative
overview of the results, before presenting the detailed derivations.

\subsection{Overview of results}\label{sect:overview}
\begin{figure}
\begin{center}
\includegraphics[scale=1]{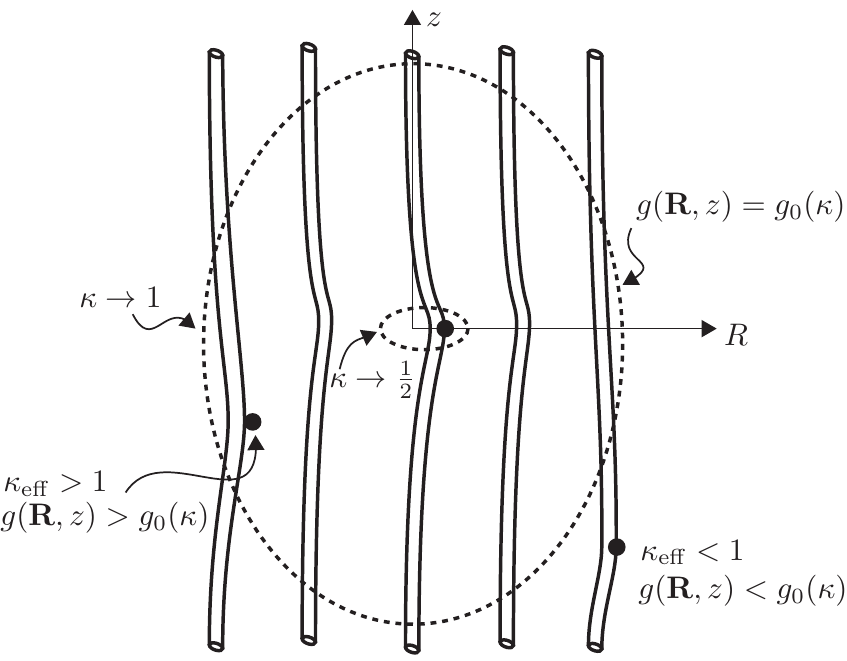}
\caption{Illustration of domains with different effective pair-pinning
strengths. A reference vortex with undisturbed transverse position
$\vec{R}_\mu = \vec{0}$ is pinned by a defect at height $z = 0$, see center of
the figure. The surface $g(\vec{R},z) = g_0(\kappa)$ (large ellipse) where
the coupling is critical determines the boundary between strong and weak
pinning by defect pairs. A defect located inside this region combines with the
central defect to form a strongly pinning pair with an effective Labusch
parameter (or pinning strength) $\kappa_\mathrm{eff} > 1$, while a defect
outside the region does not contribute to the strong pinning by rare events.
For $\kappa\to 1$, the ellipse diverges to infinity as $R_0, z_0 \propto
(1-\kappa)^{-1}$, see Eq.\ \eqref{eq:maximal_separation}.  For $\kappa\to
\tfrac{1}{2}$, the ellipse shrinks to $R\sim \xi$ and $z \sim (\kappa-
\tfrac{1}{2}) a_0$.} \label{fig:g0_explanation}
\end{center}
\end{figure}

Pinning is maximally strong if both defects are synchronized, i.e.,
$\vec{\Delta} = 0$. In this case, $\vec{u}_1 = \vec{u}_2 = \vec{u}$ and
Eq.~\eqref{eq:two_defects} reduces to a single equation
\begin{align}
   \C\vec{u} = (1 + g)\vec{f}_p(\vec{x} + \vec{u}),
\end{align}
that is equivalent to the single-defect problem with renormalized pinning
strength 
\begin{align}\label{eq:def_kappa_eff} 
   \kappa_\eff(g,\vec{\Delta} = \vec{0}) = \kappa(1+g).
\end{align}
The condition $\kappa_\eff(g,\vec{0}) > 1$ for a strong-pinning pair then
requires $g(\vrh) > g_0(\kappa)$ with
\begin{align}\label{eq:g0}
   g_0(\kappa) = \frac{1}{\kappa} - 1
\end{align}
restricting the maximal separation between the defects, see
Fig.~\ref{fig:g0_explanation}. 

In order to arrive at an expression for $f_\pair(g,\vec{\Delta} = 0)$, it is
convenient to express $\kappa_\eff$ in terms of $g$ and the critical value
$g_0$,
\begin{align}\label{eq:kappa_eff_simple}
   \kappa_\eff(g,\vec{0}) = 1 + \frac{g-g_0(\kappa)}{1 + g_0(\kappa)}.
\end{align}
With $\kappa_\eff$ above but close to unity, we can make use of Eq.\
\eqref{eq:f_pin_av_scaling_weak} and find that 
\begin{align}\nonumber
   f_{\pair}(g,\vec{0})&\sim (\xi/a_0)^2 [\kappa_\eff(g,\vec{0}) - 1]^2 f_p\\
   \label{eq:f2dg0}
   &\sim (\xi/a_0)^2 (g-g_0)^2 f_p
\end{align}
scales with $(g-g_0)^2 \leq 1$.

Defect pairs satisfying $g(\vrh)>g_0(\kappa)$ can pin strongly at a finite
mismatch vector $\vec{\Delta}$ as well. We then have to generalize the
effective pair-pinning strength $\kappa_\eff(g,\vec{\Delta})$ to finite
$\vec{\Delta}$ and the strong-pinning condition $\kappa_\eff(g,\vec{\Delta})
=1$ will provide us with the $\vec{\Delta}(g)$-domain where pair-pinning is
strong. The latter will allow us to determine the pair-pinning force
$f_\pair(g,\vec{\Delta})$. However, a quantitative evaluation of
$\kappa_\mathrm{eff}(g,\vec{\Delta})$ for general $g$ and $\vec{\Delta}$ is
quite cumbersome, given the complex geometry of the problem.

\begin{figure}[b]
\begin{center}
\includegraphics[scale=1]{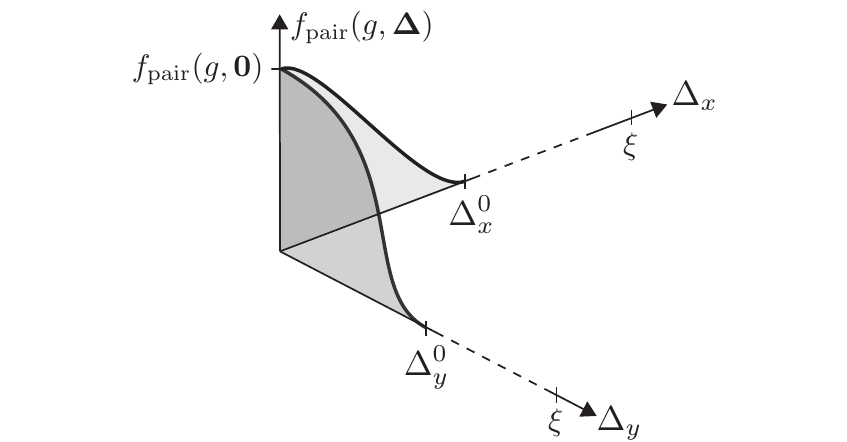}
\caption{Sketch of the pinning force $f_\mathrm{pair}(g,\vec{\Delta})$ for a
pair of defects on the verge of strong pair-pinning, $g-g_0\ll 1$ as a
function of the mismatch vector $\vec{\Delta} = (\Delta_x, \Delta_y)$. The
pinning force for the pair-defect decays from its maximal value at
$\vec{\Delta} = 0$, Eq.\ \eqref{eq:f2dg0}, on the scales $\Delta_x^0,
\Delta_y^0 \ll \xi$ given by Eqs.~\eqref{eq:mismatch_scaling} that are small
compared to the vortex core size $\xi$.} \label{fig:f_2d_sketch}
\end{center}
\end{figure}

Progress can be made by carrying out a perturbative analysis in $\Delta \ll
\xi$, which requires defect pairs to be at the verge of strong pinning, i.e.,
$g(\vrh)-g_0(\kappa)\ll 1$. Such a calculation for the effective pinning strength
$\kappa_\eff(g,\vec{\Delta})$ is carried out in Sec.\ \ref{sect:kappa_eff};
furthermore, it is shown there, that strong pinning with $\kappa_\eff
(g,\vec{\Delta})>1$ is limited to small mismatches $\Delta_x<\Delta_x^0$ and
$\Delta_y<\Delta_y^0$ with (see Eqs.\ \eqref{eq:Delta_x_0_param} and
\eqref{eq:Delta_y_0_param})
\begin{align}\label{eq:mismatch_scaling} \begin{split}
   \Delta_x^0&\sim \xi (g-g_0)^{1/2}(g+g_0),\\ \Delta_y^0&\sim \xi
   (g-g_0)^{1/2}(g+g_0)^{-1/2},
\end{split} \end{align}
see Fig.~\ref{fig:f_2d_sketch}. An estimate for the $\vec{\Delta}$-averaged
pair force can be obtained by combining the maximal pair force
\eqref{eq:f2dg0} at $\vec\Delta = 0$ and the region in $\vec\Delta$ where
pinning is strong, see Fig.~\ref{fig:f_2d_sketch},
\begin{align}\nonumber
   \langle f_{\pair}(g,\vec{\Delta})\rangle_{\vec{\Delta}} 
   &\sim \frac{\Delta_x^0\,\Delta_y^0}{a_0^2}\,f_{\pair}(g,\vec{0})\\
   \label{eq:f_pin_Delta_av}
   &\sim \frac{\xi^4}{a_0^4} (g-g_0)^3(g+g_0)^{1/2}\,f_p.
\end{align}
A more precise result for the average pair-pinning force $\langle
f_{\pair}(g,\vec{\Delta})\rangle_{\vec{\Delta}}$ is derived in Sec.\
\ref{sect:average_f_pin}, see Eq.\ \eqref{eq:f_pair_final}.

For $g(\vrh) - g_0(\kappa)\sim \mathcal{O}(1)$, the action of a defect-pair
can be analyzed only on a qualitative level. Such large coupling $g$ implies
that defects are closeby and the mismatch vector $\vec{\Delta}$ can go up to
the vortex-core radius $\xi$.  The averaging of the pair force
$f_\mathrm{pair} \sim (\xi/a_0)^2 f_p$, see Eq.\
\eqref{eq:f_pin_av_scaling_weak} for $\kappa -1$ of order unity, over
$\vec{\Delta}$ then gives
\begin{align}
\label{eq:f_pair_qual}
   \langle f_{\pair}(g,\vec{\Delta}) \rangle_{\vec{\Delta}} \sim (\xi/a_0)^4 f_p.
\end{align}
Unlike in the previous case, a precise form for the dependencies of the
effective pinning strength $\kappa_\mathrm{eff}(g,\vec{\Delta})$ and the
pair-pinning force $f_\pair(g,\vec{\Delta})$ on the mismatch vector
$\vec{\Delta}$ cannot be derived analytically. Quantitative insights can be
made by a numerical treatment of the problem (see Sec.\
\ref{sect:average_f_pin} and Appendix \ref{APP:num_pin_force}).  Such proper
averaging over the vector $\vec{\Delta}$ will only provide a numerical
prefactor to the result in Eq.~\eqref{eq:f_pair_qual}, that we do not consider
here.

Remarkably, Eq.~\eqref{eq:f_pin_Delta_av} as originally derived from a
perturbative analysis in small $g(\vrh) - g_0(\kappa) \ll 1$ produces the
correct scaling result Eq.\ \eqref{eq:f_pair_qual} also for $g(\vrh) -
g_0(\kappa)\sim\mathcal{O}(1)$. On a qualitative level, we can thus extend the
regime of applicability of the expression Eq.~\eqref{eq:f_pin_Delta_av} to any
coupling $g$ within the interval $[g_0,1]$. 

Substituting $\langle f_{\pair}(g,\vec{\Delta})\rangle_{\vec{\Delta}}$ from
\eqref{eq:f_pin_Delta_av} into Eq.~\eqref{eq:F_pin_av} yields the pinning-force
density
\begin{align}\label{eq:F_pin_integral1}
\begin{split}
   F_\pin &\sim \Bigl(\frac{\xi}{a_0}\Bigr)^2 (n_p \xi^2 a_0)n_p f_p\\
   &\times \!\!\!\!\! \int\limits_{g(\vrh)>g_0} \!\!\!\!\! \frac{d^3\vrh}{a_0^3}\, 
   [g(\vrh)-g_0]^3 [g(\vrh)+g_0]^{1/2},
\end{split}
\end{align}
with the integration over distances $\vrh$ restricted through the condition
$g(\vrh) > g_0$.

We now proceed to discuss the resulting pinning-force density $F_\pin$ due to
the defect pairs. Of crucial importance in this discussion is the behavior of
the coupling $g(\vec{R},z)$ that is of order unity at distances $R,z < a_0$
and rapidly decays further away, see Sec.\ \ref{sect:Greens_fun}.  We
distinguish three cases, i) the limit $\kappa \to \tfrac{1}{2}$, where $g_0
\to 1$, implying that the coupling $g > g_0$ has to be close to unity. This
condition demands that relevant distances $R$ are far below $a_0$, from what
follows that both defects pin the same vortex.  This regime, where defect
pairs act on the same vortex, extends throughout all of the regime ii), where
$\tfrac{1}{2} < \kappa < 1/[1+g(R=a_0)]$; in this regime, the coupling $g$ is
never large enough to produce strong pinning of different vortices that are
always further apart than $a_0$.  Finally, in region iii), $\kappa \to 1$ and
$g_0 \to 0$, hence, even a small coupling $g$ is sufficient to establish a
strong-pinning defect pair. In this case, distances $R > a_0$ become relevant
and different vortices can get strongly pinned to separated defects.

Starting with i), we consider pinning strengths of individual defects close to
the threshold $\kappa = \tfrac{1}{2}$.  The expansion of Eq.~\eqref{eq:g0} for
$\kappa - \tfrac{1}{2} \ll 1$ gives the critical coupling $g_0(\kappa) \approx
1 - 4(\kappa-\tfrac{1}{2})$ close to unity, and the condition $g(\vrh) >
g_0(\kappa)$ requires both defects to act on the same vortex. Using
Eq.~\eqref{eq:G_vertical_small_z} further implies that $g(\vec{0},z)-g_0\approx
4(\kappa-\tfrac{1}{2})-z/a_h > 0$, hence the maximal longitudinal separation is
limited by $z_0\approx 4(\kappa-\tfrac{1}{2})a_h$.  Since for such range of
coordinates we have $g(\vrh)+g_0 \sim \mathcal{O}(1)$, the pinning-force
density in Eq.~\eqref{eq:F_pin_integral1} can be cast into the form
\begin{align}\label{eq:F_clust_marginal}
   F_\pin &\sim \Bigl(\frac{\xi}{a_0}\Bigr)^2 (n_p \xi^2 a_0) n_p f_p
   \int_{0}^{z_0} \frac{dz}{a_0}\, 
   \Bigl[4(\kappa - \tfrac{1}{2}) - \frac{z}{a_h}\Bigr]^3\nonumber\\
   &\sim \Bigl(\frac{\xi}{a_0}\Bigr)^2 (\kappa-\tfrac{1}{2})^4 
   \, (n_p \xi^2 a_0) n_p f_p,
\end{align}
where we have ignored the logarithmic factor in the expression for the healing
length [see the definition of $a_h$ below Eq.~\eqref{eq:G_vertical_small_z}]
and replaced $a_h\sim a_0$.  The result in Eq.~\eqref{eq:F_clust_marginal}
defines the onset of the pinning force density due to defect pairs for
$\kappa$ rising above $\kappa = \tfrac{1}{2}$; it starts dominating over the
collective pinning result, Eq.~\eqref{eq:F_coll}, as soon as the pinning
strength surpasses the threshold $\kappa = \tfrac{1}{2}$ by a small amount
$\sim (a_0/\lambda)^{1/2}$.

Increasing $\kappa$ through region ii), i.e., staying below the threshold
$\kappa = 1/[1+g(R=a_0)]$, the critical coupling $g_0(\kappa)$ will remain
finite, of order unity.  Referring to Figs.~\ref{fig:G_schematic} and
\ref{fig:g_Rz}, we note that in this situation, the separation $\vrh =
(\vec{R},z)$ between defects must remain within the peak region III, since
otherwise $g$ rapidly decays to a value $\sim a_0^2/\lambda^2\ll 1$ and the
criterion $g(\vrh) > g_0(\kappa)$ cannot be met.  The integral in
Eq.~\eqref{eq:F_pin_integral1} is of order unity and the pinning-force density
assumes the small-pair or {\it cluster} value
\begin{align}\label{eq:F_clust}
   F_\pin = F_\mathrm{clust} &\sim \frac{\xi^2}{a_0^2} (n_p \xi^2 a_0) \, n_p f_p.
\end{align}

Finally, when $\kappa$ resides within the small interval $1-a_0^2/ \lambda^2
\lesssim \kappa\lesssim 1$, the critical coupling becomes small, $g_0\lesssim
a_0^2/\lambda^2$, and the separation between the defects producing strong
pair-pinning extends beyond region III into the regions I, II, and even the
(yellow) non-dispersive region of Fig.~\ref{fig:G_schematic}. In this case,
that includes region iii), we drop the small value $g_0(\kappa)$ against
$g(\vrh)$ in Eq.~\eqref{eq:F_pin_integral1} and rewrite the result for the
pinning force density as
\begin{align}\label{eq:F_pin_kappa_1}
   F_\pin &\sim \Bigl(\frac{\xi}{a_0}\Bigr)^2 (n_p \xi^2 a_0) n_p f_p
   \!\!\!\! \int\limits_{g(\vrh)>g_0}\!\!\!\! \frac{d^2\vec{R}\, dz}{a_0^3}\, 
   g(\vec{R}, z)^{7/2}.
\end{align}
The integration is dominated by small distances: the small-pair or
\textit{cluster} region (region III in Fig.~\ref{fig:G_schematic}) contributes
the same estimate as in Eq.~\eqref{eq:F_clust},
\begin{align}\label{eq:F_III}
   F_\mathrm{III} &\sim F_\mathrm{clust}.
\end{align}
A similar contribution arises from region I: we integrate over the transverse
coordinate $R < \sqrt{a_0z}$ and find the expression
\begin{align}\label{eq:F_I}
   F_\mathrm{I}\sim F_\mathrm{clust}\int_{a_0}^{\lambda^2/a_0} 
   \frac{d z}{a_0}\frac{a_0z}{a_0^2}\Bigl(\frac{a_0}{z}\Bigr)^{7/2}\sim 
   F_\mathrm{clust},
\end{align}
where the main contribution originates from the lower bound $z\sim a_0$ (and
hence also $R\sim a_0$). 

The contributions of region II and the non-dispersive regime are smaller by a
factor $(a_0/\lambda)^3$: In region II with transverse and longitudinal
extension $R\sim \lambda$ and $z\sim \lambda^2/a_0$, the Green's function
assumes a constant value with $g^{7/2}\sim (a_0/\lambda)^{7}$ and the 
integration gives a result
\begin{align}\label{eq:F_II}
   F_\mathrm{II}\sim \Bigl(\frac{a_0}{\lambda}\Bigr)^3 F_\mathrm{clust}.
\end{align}
In dealing with the non-dispersive region, we introduce the rescaled
distance $\tilde{\vrh} = [\vec{R},(a_0/ 4\sqrt{\pi}\lambda)\,z]$; the
non-dispersive region is bounded below by $|\tilde{\vrh}|\gtrsim \lambda$ and the
condition $g(\tilde\vrh)>g_0(\kappa)$ translates to the upper boundary 
$|\tilde{\vrh}|\lesssim a_0^2/\lambda g_0$. The integral over the
non-dispersive region then takes the form
\begin{align}\label{eq:F_non_disp}
   F_\mathrm{non-disp} &\sim \Bigl(\frac{\xi}{a_0}\Bigr)^2 n_p 
   (n_p \xi^2 a_0) f_p \!\!\!\! \int\limits_{\tilde{\rho}
   \sim\lambda}^{\tilde{\rho}\sim a_0^2/\lambda g_0} \!\!\!\! 
   \frac{d^3 \tilde{\vrh}}{a_0^3} \frac{\lambda}{a_0}
   \Bigl(\frac{a_0^2}{\lambda \tilde{\rho}}\Bigr)^{7/2}\nonumber\\
   &\sim \Bigl(\frac{a_0}{\lambda}\Bigr)^3 F_{\mathrm{clust}}
\end{align}
that, given the large exponent $7/2$, is determined by the lower bound
$\tilde\vrh \sim \lambda$. The total pinning-force density then sums up to
\begin{align}\label{eq:F_pin_I_II_III}
   F_\pin \sim F_\mathrm{I} + F_\mathrm{III} \sim F_\mathrm{clust}.
\end{align}

In the limit $\kappa\to 1$, we have $g_0(\kappa)\to 0$ and pairs of
arbitrarily distant defects induce a finite pinning force. Indeed, the maximal
distance between defects [providing the upper bound of the integral in
Eq.~\eqref{eq:F_non_disp}] diverges as $\tilde{\rho}\sim a_0^2/ \lambda
(1-\kappa)$; this translates into maximal longitudinal and transverse
separations $z\leq z_0$ and $R\leq R_0$ between defects with
\begin{align}\label{eq:maximal_separation}
  z_0\sim \frac{a_0}{1-\kappa},\qquad R_0\sim \frac{a_0}{\lambda}
  \frac{a_0}{1-\kappa},
\end{align}
see also Fig.~\ref{fig:g0_explanation}. However, the pair-pinning force in
Eq.~\eqref{eq:F_non_disp} is dominated by the lower $\tilde\rho$ bound and
contributions from distant defects are irrelevant, implying a finite integral
even in the limit $\kappa\to 1$.

The origin of the power $\alpha = 7/2$ can be traced back to the maximal size
of the mismatch vector $\vec{\Delta}$ ensuring strong pinning, see
Eqs.~\eqref{eq:mismatch_scaling}.  The derivation of the latter requires a
detailed quantitative understanding of the pinning mechanism due to defect
pairs at the verge of strong pinning that is presented in sections
\ref{sect:effective_potential}, \ref{sect:kappa_eff}, and
\ref{sect:average_f_pin} below.

\subsection{Effective pinning potential $e_\mathrm{eff}$}\label{sect:effective_potential}

In the following sections, we provide a systematic derivation of the results
presented above. In a first step, we reduce the two-defect equation
\eqref{eq:two_defects} to a single equation describing the interaction of a
fictitious vortex with an effective pinning potential, see
Fig.~\ref{fig:mismatch}(b). We have seen that for vanishing mismatch
$\vec{\Delta}$, the action of both defects is synchronized and the
displacement of both vortices is identical, $\vec{u}_1 = \vec{u}_2$.  For a
finite but small mismatch $\vec\Delta$, we reformulate the problem in terms of
the \textit{mean} position $\vr$ of the displaced vortices relative to the
defects and the \textit{internal} relative position $\delta\vr$ (we remind
that $\vec{x}_i = \vec{R}_{\mu(i)} - \vec{R}_i$ is the unperturbed
defect--vortex distance),
\begin{align}
\vr &= \tfrac{1}{2}(\vec{x}_1 + \vec{u}_1 + \vec{x}_2 + \vec{u}_2),\\
\delta\vr &= \tfrac{1}{2}(\vec{x}_1 + \vec{u}_1 - \vec{x}_2 - \vec{u}_2).
\end{align}
Solving perturbatively for the internal coordinate $\delta\vr$ will allow us
to reformulate the two-defect problem in terms of a single equation for the
mean `fictitious' vortex tip position $\vr$. This reformulated problem then
will involve an effective pinning potential $e_\eff(g,\vec{\Delta};\vr)$
exerting the pinning force $\vec{f}_\eff(\vr) = -\nabla_{\vr} e_\eff(g,
\vec{\Delta}; \vr)$ on the `fictitious' vortex.

In the above new coordinates, the two-defect problem of Eqs.~\eqref{eq:two_defects}
takes the form
\begin{align}
\begin{split}\label{eq:two_defects2}
   \C(\vec{r} + \delta\vec{r} - \vec{x} - \vec{\Delta}/2) 
   &= \vec{f}_p(\vec{r} + \delta \vec{r}) + g\,\vec{f}_p(\vec{r} - \delta \vec{r}),\\
   \C(\vec{r} - \delta\vec{r} - \vec{x} + \vec{\Delta}/2) 
   &= \vec{f}_p(\vec{r} - \delta \vec{r}) + g\,\vec{f}_p(\vec{r} + \delta \vec{r}).
\end{split}
\end{align}
Expanding in $\delta \vec{r}$ to second order and subtracting one equation
from the other provides us with an expression for $\delta \vec{r}$,
\begin{align}\label{eq:vec_delta_r}
   \Bigl[\delta_{ij} - \frac{1-g}{\C}\partial_j f_{p,i}(\vec{r})\Bigr]
   \delta r_j = \frac{\Delta_i}{2} + \mathcal{O}(\vec{\Delta}^3).
\end{align}

We write the gradient of the radial pinning force ${\vec{f}_p(r) = f_p(r)\,
\hbr}$ (with $\hbr = \vec{r}/r$ the unit vector in radial direction) in terms
of the projectors $\mathcal{P}^\parallel_{ij} = \hr_i\hr_j$ and
${\mathcal{P}^\perp_{ij} = \delta_{ij} - \hr_i\hr_j}$,
\begin{align}\label{eq:grad_central_force}
   \partial_j f_{p,i}(\vec{r}) = \mathcal{P}^\parallel_{ij}f_p'(r) 
   + \mathcal{P}^\perp_{ij}\frac{f_p(r)}{r}.
\end{align}
Using $\delta_{ij} = \mathcal{P}^\parallel_{ij} + \mathcal{P}^\perp_{ij}$,
Eq.~\eqref{eq:vec_delta_r} is rewritten as
\begin{align}\label{eq:vec_delta_r_proj}
   \bigl[\alpha_\parallel (r) \mathcal{P}^\parallel_{ij} + \alpha_\perp (r) 
   \mathcal{P}^\perp_{ij} \bigr]\delta r_j = \frac{\Delta_i}{2} 
   + \mathcal{O}(\vec{\Delta}^3)
\end{align}
with
\begin{align}\label{eq:projectors}
   \alpha_\parallel (r) = 1 - \frac{1\!-\!g}{\C}f_p'(r),\quad 
   \alpha_\perp (r) = 1 - \frac{1\!-\!g}{\C}\frac{f_p(r)}{r}.
\end{align}
Making use of the relation $\mathcal{P}^\alpha_{ij}\mathcal{P}^\beta_{jk} =
\delta_{ik}\delta_{\alpha\beta}$ for the projectors with
$\alpha,\beta\in \{\parallel, \perp\})$, Eq.~\eqref{eq:vec_delta_r_proj} is easily
inverted and provides a relation between the internal coordinate
$\delta \vec{r}$ and the mean coordinate $\vec{r}$ of the vortex
pair,
\begin{align}\label{eq:delta_r}
   \delta \vec{r} &=\! \Bigl[\frac{\mathcal{P}^\parallel}{\alpha_\parallel (r)} 
   + \frac{\mathcal{P}^\perp}{\alpha_{\perp}}\Bigr]\frac{\vec{\Delta}}{2} 
   = \frac{(\hbr\!\cdot\!\vec{\Delta})\hbr}{2\alpha_\parallel (r)} 
   + \frac{\vec{\Delta} - (\hbr\!\cdot\!\vec{\Delta})\hbr}{2\alpha_\perp (r)}.
\end{align}

Adding the two Eqs.~\eqref{eq:two_defects2} provides us with an equation for
the mean vortex tip position,
\begin{align}\label{eq:force_balance_eff}
   \C(\vec{r} - \vec{x}) = \vec{f}_\eff(\vec{r}),
\end{align}
reminiscent of the single-defect case but with an \textit{effective} pinning
force
\begin{align}\label{eq:f_eff}
   \vec{f}_\eff(\vec{r}) = \tfrac{1}{2}(1+g)\bigl[\vec{f}_p(\vec{r} 
   + \delta\vec{r}) + \vec{f}_p(\vec{r} - \delta\vec{r})\bigr].
\end{align}
Expanding to 2-{nd} order in $\delta \vec{r}$ (or $\vec{\Delta}$), the
effective pinning force becomes
\begin{align}\label{eq:f_eff_expansion}
   f_{\eff,k} = (1\!+\!g)f_{p}(r)\hr_k + \tfrac{1}{2}(1\!+\!g)\partial_i\partial_j 
   f_{p,k}(\vec{r})\delta r_i \delta r_j,
\end{align}
where the matrix of second derivatives of the pinning force can be expressed
as $\partial_i\partial_j f_{p,k}(\vec{r}) = \gamma \hr_i\hr_j\hr_k +
\mu(\hr_i\delta_{jk} + \hr_j\delta_{ki} + \hr_k\delta_{ij})$ with
\begin{align}
   \gamma = f_p''(r) - 3\partial_r[f_p(r)/r]~~\mathrm{and}~~
   \mu = \partial_r[f_p(r)/r].
\end{align}
The sums $\partial_i\partial_j f_{p,k} (\vec{r}) \delta r_i \delta r_j$ involve the 
expressions 
\begin{align} 
   &\gamma\hr_i\hr_j\hr_k\delta r_i \delta r_j 
   = \gamma \hr_k\frac{(\hbr\cdot\vec{\Delta})^2}{4\alpha^2_\parallel (r)},\\
   &\mu\delta_{ij}\hr_k\delta r_i \delta r_j = \mu \hr_k\Bigl[\frac{(\hbr \cdot
   \vec{\Delta})^2}{4\alpha^2_\parallel (r)} + \frac{\vec{\Delta}^2 - (\hbr \cdot
   \vec{\Delta})^2}{4\alpha_\perp (r)}\Bigr],\\ 
   &\mu(\delta_{jk}\hr_i + \delta_{ki}\hr_j)\delta r_i \delta r_j = \nonumber\\ 
   &\qquad= \mu\frac{\hbr\cdot \vec{\Delta}}{2\alpha_\parallel (r)}\Bigl[\frac{(\hbr
   \cdot\vec{\Delta})\hr_k}{\alpha_\parallel (r)} + \frac{\Delta_k - (\hbr\cdot
   \vec{\Delta})\hr_k}{2\alpha_\parallel (r)}\Bigr].\label{eq:f_eff_calculations}
\end{align}
Combining Eqs.~\eqref{eq:f_eff_expansion}--\eqref{eq:f_eff_calculations} gives
the effective pinning force up to second order in the mismatch $\vec{\Delta}$,
\begin{align}\label{eq:f_eff_result} 
\begin{split}
   \vec{f}_\eff(\vec{r}) &= (1+g)\Bigl\lbrace f_p(r)\hbr +
   f_p''(r)\frac{(\hbr\cdot\vec{\Delta})^2}{8\alpha^2_\parallel (r)}\hbr \\ &+
   \partial_r[f_p(r)/r]\frac{\Delta^2 - (\hbr\cdot \vec{\Delta})^2}
   {8\alpha^2_\perp (r)}\hbr\\ 
   &+ \partial_r[f_p(r)/r]\frac{\hbr\cdot\vec{\Delta}}{4\alpha_\parallel (r)
   \alpha_\perp (r)}[\vec{\Delta}-(\hbr\cdot\vec{\Delta})\hbr]\Bigr\rbrace,
\end{split}
\end{align}
with the first three terms producing a radial force (along the
vector $\hbr$), while the last term contributes a transverse force
(along the vector $\vec{\Delta} - (\hbr\cdot\vec{\Delta})\hbr$
perpendicular to $\hbr$). 

This effective pinning force can be written as the gradient of the effective
pinning potential $e_\eff(\vec{r})$, $\vec{f}_\eff(\vec{r}) = -\nabla
e_\eff(\vec{r})$, that has a much simpler form. Indeed, using
$\nabla(\hbr\cdot\vD) = [\vD - (\hbr\cdot\vD)\hbr]/r$ and fixing the
integration constant by requiring $e_\eff(\br)\to 0$ as $r\to\infty$, we find
the effective pinning potential in the form
\begin{align}\label{eq:e_eff}
   e_\eff(g,\vec{\Delta}; \vec{r}) &= (1\!+\!g)e_p(\vec{r})
   -\frac{\C(1\!+\!g)}{8(1\!-\!g)}\\ \nonumber &\times
   \Bigl[\frac{(\hbr\cdot\vec{\Delta})^2}
   {\alpha_\parallel (r)} + \frac{\vec{\Delta}^2 
   - (\hbr\cdot\vec{\Delta})^2}{\alpha_\perp (r)} - \vec{\Delta}^2\Bigr].
\end{align}
Using polar coordinates $\vec{\Delta} = \Delta(\cos\theta,
\sin\theta)$ and $\vec{r} = r(\cos\varphi,\sin\varphi)$, the effective
potential can be described in terms of the magnitude $\Delta$ of the mismatch
and the angle $\varphi-\theta$ enclosed by $\vec{r}$ and $\vec{\Delta}$,
\begin{align}\label{eq:e_eff_angular}
   e_\eff(g,\vec{\Delta}; \vr) &= (1\!+\!g)e_p(r)
   -\frac{\C\Delta^2(1\!+\!g)}{8(1\!-\!g)}
   \\ \nonumber &\times \Bigl[\frac{\cos^2(\varphi -\theta)}
   {\alpha_\parallel(r)} + \frac{\sin^2(\varphi - \theta)}{\alpha_\perp(r)} - 1 \Bigr].
\end{align}
Inserting the expressions for $\alpha_{\parallel,\perp}$, see
Eq.~\eqref{eq:projectors}, and expanding for large distances $r\gg \xi$
(i.e., small values of $f_p'(r)$, $f_p(r)/r$), we find that
\begin{align}\label{eq:e_eff_asymp}
   e_\eff(g,\vec{\Delta}; \vr) & \approx (1\!+\!g)e_p(r) -
   (1\!+\!g) \frac{\Delta^2}{8} \\ \nonumber & \times
   \Bigl[f_p'(r)\cos^2(\varphi-\theta) +
   \frac{f_p(r)}{r}\sin^2(\varphi-\theta)\Bigr],
\end{align}
with the anisotropic terms appearing at finite mismatch $\Delta$ proportional
to $f_p'(r),f_p(r)/r\propto (\xi/r)^4$ vanishing faster than the isotropic term
$\propto e_p(r)\propto (\xi/r)^2$.

In the limit $g = 1$ (with defects acting on the same vortex at the same height
$z$), the effective potential becomes
\begin{align}\label{eq:e_eff_g_1}
\begin{split}
   e_\eff(g=1,\vec{\Delta}; \vr) &= 2e_p(\vr) 
   + \frac{1}{4}(\hbr\cdot\vec{\Delta})^2 f_p'(r)\\
   &+\frac{1}{4}\Bigl[\vec{\Delta}^2 - (\hbr\cdot\vec{\Delta})^2\Bigr]
   \frac{f_p(r)}{r},
\end{split}
\end{align}
that corresponds, up to order $\mathcal{O}(\vec{\Delta}^2)$, to the simple
superposition of two mutually shifted pinning potentials, ${e_\eff(g=1,
\vec{\Delta}; \vec{r})} \approx e_p(\vec{r} + \vec{\Delta}/2) + e_p(\vec{r} -
\vec{\Delta}/2)$. In the limit $g = 0$, the two-defect problem
\eqref{eq:two_defects2} decouples and we can obtain independently each
perturbed vortex tip position $\vec{r}_i = \vec{x}_i + \vec{u}_i$,
$i\in\lbrace 1,2\rbrace$.

\subsection{Effective Labusch parameter $\kappa_\mathrm{eff}$ and
strong-pinning range $\vec\Delta^0$}\label{sect:kappa_eff}

We proceed with the calculation of the effective Labusch parameter
$\kappa_\eff$ (or pinning strength) for the anisotropic pinning potential of
Eq.\ \eqref{eq:e_eff_angular}. For a single isotropic defect, the Labusch
parameter $\kappa$ is defined in Eq.\ \eqref{eq:Labusch_crit} and involves the
(maximal) potential curvature $f_p' = -e_p''$ and the effective elasticity
$\C$. Going to the defect pair, the anisotropic potential
\eqref{eq:e_eff_angular} depends on the distance and arrangement of defects
through the parameters $g$ and $\vec\Delta = \Delta(\cos\theta,\sin\theta)$.
In the following, we consider a vortex with an asymptotic trajectory $\vec{x}
= (x,0)$ and determine the angular dependence (on $\theta$) of the effective
Labusch parameter $\kappa_\eff(g,\vec\Delta)  = \kappa_\eff(g,\Delta,\theta)$,
i.e., we consider defect pairs with different angular arrangement relative to
an asymptotically fixed vortex motion.  Once the function $\kappa_\eff(g,
\vec\Delta)$ is known, the condition $\kappa_\eff(g,\vec\Delta^0) = 1$ will
provide us with the maximal misfit $\vec\Delta^0(g)$ limiting the
strong-pinning range. Trajectories with different angle of incidence and/or
finite impact parameter $b$ will be discussed later.  Finally, the pinning
strength in the vortex--defect system can be tuned by either changing the
effective elasticity $\C$ or the energy scale $e_p$ of the defect
potential---in the present discussion, we will tune $\kappa_\eff$ via changing
$\C$.

We start our derivation of the pinning strength $\kappa_\eff(g,
\vec\Delta)$ by going back to its defining equation. For an isotropic pinning
potential, this is given by \eqref{eq:Labusch_crit}, that in turn derives from
the self-consistency equation \cite{Labusch1969,Larkin1979,Blatter2004}
\begin{align}\label{eq:force_balance_f_p}
   \C(r - x) = f_p(r).
\end{align}
Equation \eqref{eq:force_balance_f_p} allows us to connect incremental changes
in the asymptotic and tip positions, $\delta x = [1-f_p'(r)/\C] \delta r$,
with jumps in $\delta r$ occurring when $1-f_p'(r)/\C = 0$. Combining this
relation with the condition of its first appearance, $f_p'(r) \to
\max[f'(r)]$, leads to \eqref{eq:Labusch_crit}. Finally, the maximum force
derivative $\max[f'(r)]$ is achieved at the inflection point $r_m$ defined via
$f_p''|_{r_m} = 0$.

In the present anisotropic situation, Eq.\ \eqref{eq:force_balance_f_p}
has to be generalized to its vectorial form \eqref{eq:force_balance_eff}
and incremental changes in asymptotic and tip positions of vortices are
related via
\begin{align}\label{eq:dx_dr}
   \delta x_i = [\delta_{ij} + H_{ij}(\vec{r})/\C]\,\delta r_{j},
\end{align}
where 
\begin{align}\label{eq:Hessian}
   H_{ij}(\vec{r})
   = \partial_{r_i} \partial_{r_j} e_\eff(g,\vec\Delta;\vec{r})
\end{align}
is the Hessian matrix associated with the pinning energy landscape
$e_\eff(g,\vec\Delta;\vec{r})$. The Labusch criterion again marks the first
appearance of an instability in the vortex tip position $\delta\vec{r}$.
We thus have to invert Eq.\ \eqref{eq:dx_dr} and find the solution
$\delta\vec{r}(\delta\vec{x})$---a diverging result for $\delta\vec{r}$ then
signals the presence of a jump in the vortex tip position. Approaching this
divergence from the weak pinning domain, i.e., starting with a large $\C$ and
decreasing its value, the jump appears when the determinant in the matrix
relation \eqref{eq:dx_dr} vanishes,
\begin{align}\label{eq:det_Hessian}
   \det[\C\delta_{ij} + H_{ij}(\vec{r})] = 0.
\end{align}

Evaluating the Hessian in cylindrical coordinates $(r,\varphi)$, we obtain the
matrix
\begin{align}\label{eq:H_cyl}
   H = (1\!+\!g)\!\begin{bmatrix}
   -f_p' + \beta \, \Delta^2 & \!\!\! \gamma\,\Delta^2 \\
   \gamma\, \Delta^2 & \!\!\!  -f_p(r)/r + \delta\,\Delta^2
\end{bmatrix},
\end{align}
with 
\begin{align}\nonumber 
   \beta(r,\theta) = \frac{-1}{8}\biggl\lbrace 
   \biggl[\frac{f_p''(r)}{\alpha_\parallel^2(r)}\biggr]'\!\cos^2\!\theta
   + \biggl[\frac{(f_p(r)/r)'}{\alpha_\perp^2(r)}\biggr]'\!
   \sin^2\!\theta \biggr\rbrace
\end{align}
and functions $\gamma(r,\theta),\,\delta(r,\theta)$ that we do not need to
calculate explicitly. Above, we have made use of the fact that the tip
trajectory stays always close to the $x$-axis (up to corrections of order
$\Delta^2$) and hence, we have set the angle $\varphi$ in
$\beta(\vec{r},\theta)$ to zero, $\beta(\vec{r},\theta) \to \beta(r,\theta)$.

The condition of vanishing determinant \eqref{eq:det_Hessian} is equivalent to
matching up the lower eigenvalue $\lambda_\-(r,\theta) < 0$ of $H$ with $\C$,
$\lambda_\-(r,\theta) + \C = 0$; furthermore, we need to find the location
where this happens first, i.e., we have to determine the distance
$r_m^\eff(\theta)$ that generalizes $r_m$ to the anisotropic situation.  Once
this program is executed, the generalized Labusch parameter is given by
\begin{align}\label{eq:gen_Lab}
   \kappa_\eff(\theta) = \frac{-\lambda_\-[r_m^\eff(\theta),\theta]}{\C}
\end{align}
that assumes unity at the weak-to-strong pinning transition and larger values
on decreasing $\C$ further into the strong pinning region.

Let us first consider the above generalized formulation of the Labusch
parameter for the isotropic situation with $\Delta = 0$. Then $H_{ij}$ is
already diagonal, with eigenvalues $\lambda_\-(r)  = -f_p'(r) < 0$ and
$\lambda_\+(r) = -f_p(r)/r > 0$ close to the inflection point $r_m$, where
$f_p''(r_m) = 0$ and the maximum in $-\lambda_\- = f'(r)$ is realized. These
results are fully in line with the previous discussion of the Labusch
criterion \eqref{eq:Labusch_crit}.

The perturbative analysis of the anisotropic situation contributes
corrections to order $\Delta^2$ that introduce an angular dependence of the
results on $\theta$. The eigenvalues $\lambda_\ppm(r,\theta)$ of $H$, see
\eqref{eq:H_cyl}, coincide, to order $\Delta^2$, with its diagonal entries,
since the off-diagonal terms only add a correction $\gamma^2\Delta^4$ to the
determinant appearing in their calculation.  In particular, the lower
eigenvalue assumes the form
\begin{align} \label{eq:lm}
   \lambda_\-(r,\theta) 
   \approx (1\!+\!g)\bigl[-f_p'(r) + \beta(r,\theta)\,\Delta^2 \bigr].
\end{align}
Following the definition in Eq.\ \eqref{eq:gen_Lab}, we have to evaluate this
expression at the generalized inflection point $r_m^\eff(\theta)$. The latter
remains close to $r_m$, $r_m^\eff(\theta) = r_m + \mathcal{O}(\Delta^2)$, and
using this Ansatz in \eqref{eq:lm}, we find that the correction to $r_m$ is
irrelevant since $f_p'[r_m^\eff(\theta)] = f_p'[r_m + \mathcal{O}(\Delta^2)]
\approx f_p'(r_m) + \mathcal{O}(\Delta^4)$ as $f_p''(r_m) = 0$. We thus arrive
at the formal expression for the effective Labusch parameter
\begin{align} \label{eq:kap_eff_th}
   \kappa_\eff(g,\vec\Delta)
   \approx \frac{1+g}{\C}
   \Bigl[f_p'(r_m) -\beta(r_m,\theta)\Delta^2\Bigr].
\end{align}
Inserting the expression for $\beta(r,\theta)$ from above, we can rewrite
this result into the convenient form
\begin{align} \label{eq:kap_eff_fin}
   \kappa_\eff(g,\vec\Delta) \approx \frac{-\partial_r^2
   e_\eff(g,\vec\Delta;r)}{\C}\bigg|_{r_m}.
\end{align}
When the vortex trajectory is oriented at a finite angle $\phi$ with respect
to the $x$-axis, the angular dependence in \eqref{eq:kap_eff_fin} has to be
replaced according to $\theta \to \theta - \phi$.  Finally, the discussion for
a finite impact parameter $b$ can be easily reduced to the situation where the
vortex approaches the defect center from an angle, see the discussion in Sec.\
\ref{sect:average_f_pin} below.

Note that, while $r_m^\eff \approx r_m$ does not depend on angle to order
$\Delta^2$, the effective pinning strength $\kappa_\eff(g,\Delta,\theta -
\phi)$ experienced by a vortex incident at an angle $\phi$ does. This implies
that pinning is not equally strong when approaching the same defect from
different direction. Rather, $\kappa_\eff$ may be larger (or smaller) than
unity when changing $\phi \in [0, 2\pi]$.  As a result, vortex trajectories
crossing the same defect may undergo pinning and depinning jumps in some
directions but not in other.

Next, we return to our vortex incident along $x$, substitute the anisotropic
defect-pair potential Eq.\ \eqref{eq:e_eff_angular} into the expression
\eqref{eq:kap_eff_fin}, and use $\kappa = f_p'(r_m)/\C$ to find the explicit
result
\begin{align}\label{eq:kappa_eff}
   \kappa_\eff(g,\Delta,\theta) &\approx (1+g)\kappa
   +(1+g)\frac{\Delta^2}{8\C} \\ \nonumber
   & \!\!\!\! \times \biggl\lbrace\frac{f_p'''}{\alpha_\parallel^2}
   \cos^2\theta+ \partial_r\biggl[\frac{\partial_r(f_p/r)}{\alpha_\perp^2}\biggr]
   \sin^2\theta \biggr\rbrace_{r = r_m} \! .
\end{align}
Setting the pinning strength to its critical value, $\kappa_\eff(g,
\vec\Delta) = 1$, we now can determine the misfit parameter $\vec\Delta^0(g)$
below which pinning is strong.  We first analyze the two special cases $\theta
= 0$ and $\theta = \pi/2$, where the mismatch $\vec{\Delta}$ is parallel and
perpendicular to the vortex trajectory, before generalizing the result to
other angles $\theta$.

For $\theta = 0$, we obtain
\begin{align}\label{eq:kappa_eff_x}
   \kappa_\eff(g,\Delta,0) \approx (1\!+\!g)\kappa + (1\!+\!g)\frac{\Delta^2}{8\C}
   \frac{f_p'''(r_m)}{[1 - \kappa(1\!-\!g)]^2}.
\end{align}
Since $f_p'''(r_m) < 0$, the pinning strength decreases below its maximal
value $(1+g)\kappa$ as the mismatch $\Delta$ is increased. By setting
$\kappa_\eff(g,\Delta,0) = 1$ in Eq.~\eqref{eq:kappa_eff_x}, we can obtain the
maximum longitudinal mismatch $\Delta_x^0$ below which the pinning by the two
defects is strong,
\begin{align}\label{eq:Delta_x_0}
   \Delta_x^0 &= \frac{(8\C\kappa^3)^{1/2}}{[-f_p'''(r_m)]^{1/2}}
   \frac{(g-g_0)^{1/2}(g+g_0)}{(1+g)^{1/2}}.
\end{align}
Estimating $f_p''' \sim f_p/\xi^3$ and $f_p/\C\xi\sim \mathcal{O}(1)$ and
dropping the factor $1 + g$ in the denominator, we can express the relevant
$g$-dependence of $\Delta_x^0$ in the parametric form
\begin{align}\label{eq:Delta_x_0_param}
   \Delta_x^0 \sim \xi ( g-g_0)^{1/2}(g+g_0).
\end{align}

For $\theta = \pi/2$, we first rewrite the prefactor of the $\sin^2\theta$ term in
Eq.~\eqref{eq:kappa_eff}. Introducing the notation $\beta(r) = f_p(r)/r$, we
find that
\begin{align} \nonumber
   &\partial_r(\alpha_\perp^{-2}\beta') 
   = \alpha_\perp^{-3}(\alpha_\perp \beta'' - 2\alpha_\perp'\beta') \\
   \label{eq:tech}
   &=\alpha_\perp^{-3}\C^{-1}[\C\beta'' - (1-g)\beta\beta'' + 2(1-g)(\beta')^2].
\end{align}
Expressing the effective elasticity $\C$ through $\kappa$ and $\beta$, $\C =
\kappa^{-1}(\beta r)'_{r = r_m}$, the second factor in this expression
simplifies to
\begin{align}\nonumber
   &\kappa^{-1}(\beta r)'\beta'' - (1\!-\!g)\beta\beta'' + 2(1\!-\!g)(\beta')^2
   \\ \nonumber
   &=(\kappa^{-1}\!\! +\! g\! -\! 1)(\beta r)'\beta'' \!+ (1\!-\!g)[2(\beta')^2 
   \!+ (\beta r)'\beta'' \!- \beta\beta'']
   \\ \nonumber
   &=(g+g_0)(\beta r)'\beta'' + (1\!-\!g)\beta'(r\beta)''
   \\ \label{eq:manip}
   &= (g+g_0)(\beta r)'\beta'' = -(2\kappa \C/r_m)(g+g_0)\beta',
\end{align}
where in the last two steps we used $(r\beta)'' = f_p''(r_m) = 0$ and $\beta''
= -(2/r)\beta'$. Since $f_p'(r_m) > 0$, we have $\beta'> 0$ at $r = r_m$ and
hence the effective Labusch parameter 
\begin{align}\label{eq:keff_y}
   \kappa_\eff(g,\Delta,\pi/2) &= (1\!+\!g)\kappa \\ \nonumber
   &- (1\!+\!g) \frac{\Delta^2}{4 \C}
   \frac{\kappa(g+g_0)\beta'(r_m)/r_m}{[1 - \beta(r_m)(1\!-\!g)/C]^{3}}
\end{align}
again decreases with increasing $\Delta$.  Setting $\kappa_{\eff}(\pi/2) = 1$ then
defines the maximal transverse mismatch for strong pinning,
\begin{align}\label{eq:Delta_y_0}
   \Delta_y^0 = \frac{2\C^{1/2}[1\!-\!\beta(r_m)(1\!-\!g)/\C]^{3/2}}
   {[\beta'(r_m)/r_m]^{1/2}} \frac{(g-g_0)^{1/2}}{[(1\!+\!g)(g\!+\!g_0)]^{1/2}}.
\end{align}
Using similar estimates as above, we obtain the result in parametric form
\begin{align}\label{eq:Delta_y_0_param}
   \Delta_y^0 \sim \xi \frac{(g-g_0)^{1/2}}{(g+g_0)^{1/2}}.
\end{align}
Finally, expressing the effective Labusch parameter in Eq.~\eqref{eq:kappa_eff}
in terms of the maximum longitudinal and transverse mismatches $\Delta_x^0$,
$\Delta_y^0$ in Eqs.~\eqref{eq:Delta_x_0} and \eqref{eq:Delta_y_0} yields the
angular dependence
\begin{align}\label{eq:kappa_eff_angular}
   \kappa_\eff(\theta) = 1 \!+ \kappa(g\!-\!g_0)
   \Bigl[1 \!-\! \frac{\Delta^2\cos^2\!\theta}{(\Delta_x^0)^2} 
   \!-\! \frac{\Delta^2\sin^2\!\theta}{(\Delta_y^0)^2}\Bigr].
\end{align}

It is interesting to compare the pinning strengths for different
vortex--defect configurations where the vortex trajectories are either
parallel or perpendicular to the mismatch vector $\vec{\Delta}$. Combining
Eqs.~\eqref{eq:Delta_x_0} and \eqref{eq:Delta_y_0} and using $g_0=1/\kappa
-1$, we obtain the following ratio for the maximal longitudinal and transverse
mismatches
\begin{align}\label{eq:mismatches_ratio}
   \frac{\Delta_y^0}{\Delta_x^0} = 
   \frac{1}{\sqrt{2}}\frac{[-f_p'''(r_m)]^{1/2}}{[\beta'(r_m)/r_m]^{1/2}}
   \Bigl[\frac{1\!-\!\beta(r_m)(1\!-\!g)/\C}{1-\kappa(1\!-\!g)}\Bigr]^{3/2}\!\!\!\!.
\end{align}
For a maximal coupling between defects, $g = 1$, such that the defect
potentials directly add up [cf.~Eq.~\eqref{eq:e_eff_g_1}], the factor
$[\cdots]^{3/2}$ in the above expression is unity.  The ratio of the two
scales then depends on the specific form of the pinning potential in the
vicinity of the inflection point $r_m$. For the Lorentzian pinning potential
producing the pinning force in Eq.~\eqref{eq:f_p_fun}, the ratio of the two
scales reads $\Delta_y^0/\Delta_x^0|_{g=1} = \sqrt{3/2}$ and pinning is always
stronger, i.e., $\kappa_\eff(\theta)$ is larger along the direction
perpendicular to the mismatch vector.  Furthermore, since $\beta(r_m) < 0$,
the ratio $\Delta_y^0/\Delta_x^0$ grows as $g$ is decreased away from unity
towards its minimum value $g_0=1/\kappa -1$. When $g = g_0$, the term
$1-\kappa(1-g)$ in the denominator of Eq.~\eqref{eq:mismatches_ratio} takes
the value $2(1-\kappa)$, hence as $\kappa\to 1$, pinning due to pairs of {\it
distant} defects is always stronger in the direction perpendicular to the
mismatch, regardless of the specific form of the pinning potential.

The maximal longitudinal and transverse mismatches $\Delta_x^0$ and
$\Delta_y^0$ allow us to identify the region of applicability of the
perturbative approach that we have used to derive the effective pinning
potential. For a longitudinal mismatch with $\theta = 0$,
Eq.~\eqref{eq:delta_r} tells us that $\delta r \leq
\Delta_x^0/2\alpha_\parallel$ and using the bound $\alpha_\parallel \geq 1 -
\kappa(1-g) = \kappa(g+g_0)$, we find that $\delta r \lesssim \xi
(g-g_0)^{1/2}$. The perturbative approach is valid provided that
$\delta r \ll \xi$ which is the case for $g - g_0 \ll 1$.  Similarly, for a
transverse mismatch, we use $\alpha_\perp\sim\mathcal{O}(1)$ and therefore
$\delta r \leq \Delta_y^0/2\alpha_\perp\sim \Delta_y^0$. The perturbation
$\delta r$ then again remains small for $g - g_0 \ll 1$, except for the
crossover to strong pinning where $\kappa = 1$, $g_0 = 0$, and $\Delta_y^0
\sim \xi$. In this case, the validity of the perturbative approach requires
$\Delta_y \ll \xi$ (see Eq.~\eqref{eq:delta_r}), however, a comparison with
the numerical results in Fig.~\ref{fig:f_comparison} demonstrates that the
perturbative approach still provides excellent agreement even for a mismatch
$\Delta_y$ comparable with $\xi$.

\subsection{Average pinning force}\label{sect:average_f_pin}
In this section, we use the findings on the effective Labusch parameter to
calculate the average pinning force due to two defects coupled by $g$.
Restricting first to vortex trajectories $\vec{x} = [x(X),0]$ passing through
the center of the effective pinning potential, we evaluate the $x$-averaged
pinning force for the longitudinal component (along the $x$-direction), see
Eq.~\eqref{eq:f_pin_av_strong_1D},
\begin{align} \nonumber
   f_\mathrm{pair}(&g,\Delta,\theta,b=0)
   \\ \nonumber
   &= (-\vec{e}_x)\cdot\langle \vec{f}_p[\vec{r}(x) + \delta \vec{r}(x)] + \vec{f}_p
   [\vec{r}(x) - \delta \vec{r}(x)] \rangle_x
   \\ \nonumber
   &= \frac{2}{1+g}\langle -f_{\eff,x}[\vec{r}(x)]\rangle_x
   \\ \label{eq:f_2d_b0_result}
   &= \frac{2}{1+g}\frac{\Delta e_\pin^{\eff,1} + \Delta e_\pin^{\eff,2}}{a_0},
\end{align}
where $\Delta e_\pin^{\eff,1}$ and $\Delta e_\pin^{\eff,2}$ denote the jumps
in the effective pinning potential defined as in Eq.~\eqref{eq:e_pin_def},
i.e., $e_\pin^\eff(\vec{x}) = e_\eff[\vec{r}(\vec{x})] + \tfrac{1}{2}\C
[\vec{x} - \vec{r}(x)]^2$.

Provided that $\kappa_\eff(g,\Delta,\theta) > 1$, the jumps $\Delta
e_\pin^{\eff, i}$, $i\in\lbrace 1,2\rbrace$ in the pinning energy are given by
Eq.~\eqref{eq:jump_marginal} with the replacements $f_p'''(r_m)\to
f_\eff'''(r_m)$ and $\kappa\to \kappa_\eff$; otherwise, for
$\kappa_\eff(g,\Delta,\theta) < 1$, the effective pinning force vanishes after
the averaging. This gives the two-defect pinning force
\begin{figure}[t]
\begin{center}
\includegraphics[scale=1]{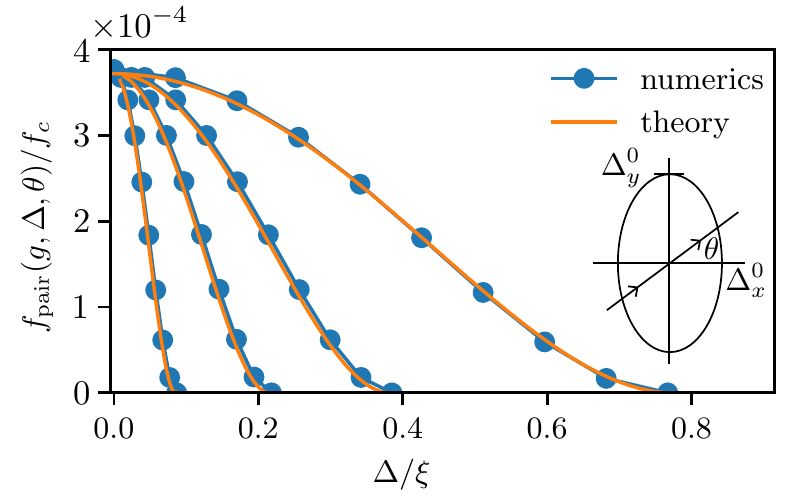}
\caption{Comparison of numerical (see Appendix \ref{APP:num_pin_force}) and
analytical (see Eq.\ \eqref{eq:fdp_gDth0_ex}) results of the effective
pinning force for a `fictitious' vortex trajectory $\vec{x} = [x(X),0]$
passing through the center of the effective pinning potential. Results are
shown for $\kappa = 0.8$ and a small coupling $g = g_0(\kappa) + 2. \>
10^{-2}$, producing a $\kappa_\eff(g,\vec{0})$ (Eq.\
\eqref{eq:kappa_eff_simple}) slightly above unity. In this regime, the maximal
longitudinal and transverse mismatches are related by $\Delta_x^0 \ll
\Delta_y^0$.  The mismatch vector $\vec{\Delta} = (\Delta\cos\theta,
\Delta\sin\theta)$ encloses an angle $\theta$ with the $x$-direction: the four
curves correspond to (from left to right, see also the inset) $\theta =
0,\,3\pi/8,\,7\pi/16$, and $\pi/2$. The pinning force is normalized by the
scale $f_c = (\xi/a_0)\,f_p$.}\label{fig:f_comparison}
\end{center}
\end{figure}
\begin{align}\nonumber
   f_\mathrm{pair}(g,\Delta,\theta,0) &= \frac{18\C^2}{(1\!+\!g)\,a_0[-f_\eff'''(r_m)]}\\
   &\times\lbrace\max[0,\kappa_\eff(\theta) - 1]\rbrace^2.\label{eq:fdp_gDth0}
\end{align}
Using Eq.~\eqref{eq:kappa_eff_angular} for the effective Labusch parameter and
noting that $f_\eff'''(r_m) = (1+g)f_p'''(r_m) + \mathcal{O}(\vec{\Delta}^2)$,
we obtain
\begin{align} \label{eq:fdp_gDth0_ex}
   f_\mathrm{pair}(g,\Delta,\theta,0) &\approx \frac{18(\kappa\C)^2}{(1\!+\!g)^2a_0
   [-f_p'''(r_m)]}(g-g_0)^2 \\ \nonumber
   \times&\Bigl\lbrace\max\Bigl[0,1 - \frac{\Delta^2\cos^2\theta}{(\Delta_x^0)^2} 
   - \frac{\Delta^2\sin^2\theta}{(\Delta_y^0)^2}\Bigr]\Bigr\rbrace^2.
\end{align}
The pinning force thus decays with $\Delta$ from its maximal value $\sim
f_p(g-g_0)^2$ at $\vec{\Delta} = 0$ to zero as the mismatch increases to
$\Delta_x^0$ and $\Delta_y^0$ along and perpendicular to the vortex
trajectory, respectively, see Fig.\ \ref{fig:f_2d_sketch}. In
Fig.~\ref{fig:f_comparison}, we compare this analytic formula to the numerical
results (see Sec.\ \ref{APP:num_pin_force}) in the regime $g - g_0\ll 1$ for
different angles $\theta$. Note that, while the (perturbative) analytic
results assume a small mismatch $\Delta\ll \xi$, they remain applicable for
angles close to $\theta = \pi/2$, where $\Delta$ becomes comparable to $\xi$.

\begin{figure}[b]
\begin{center}
\includegraphics[scale=1]{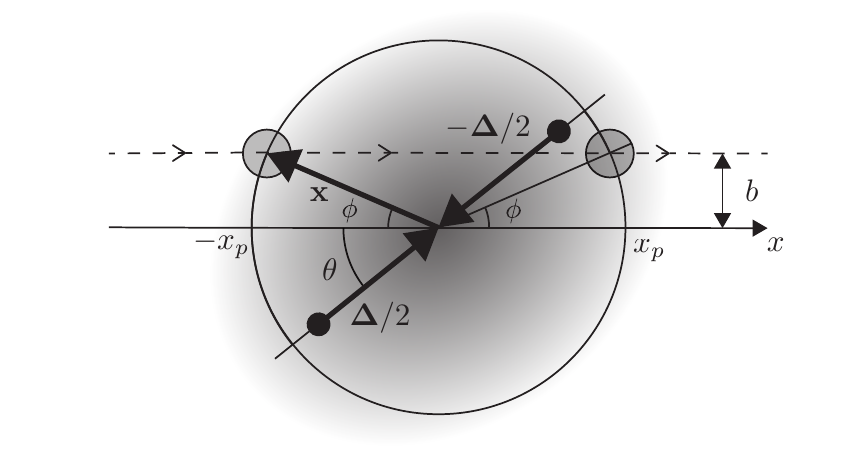}
\caption{Pinning and depinning of a `fictitious' vortex moving along the
trajectory $\vec{x} = (x,b)$ with a non-vanishing impact parameter $b > 0$
across a defect characterized by an effective pinning potential $e_\eff(g,
\vec{\Delta}, \vec{r})$ (shaded grey). The vortex is pinned and depinned from
the defect at the intersection points of the trajectory with the circle (up to
corrections of order $\Delta^2$) of radius $x_p$. The angular distance of the
vortex from the mismatch vector $\vec{\Delta} = (\Delta\cos\theta,
\Delta\sin\theta)$ at these points is $\theta + \phi$ and $\theta -
\phi$, with $\phi = \arcsin(b/x_p)$.}\label{fig:jumps_angles}
\end{center}
\end{figure}

The geometric complexity arising at finite impact $b$, see Fig.\
\ref{fig:jumps_angles},  produces interesting new features, e.g., asymmetric
pinning and depinning jumps or even trajectories with only one of the
pinning/depinning jumps realized.  In the generic situation, the vortex tip
associated with the trajectory $\vec{x} = [x(X),b]$ undergoes a jump every
time the position $\vec{r}$ hits the distance $r_m + \mathcal{O}
(\vec{\Delta}^2)$ from the center of the effective pinning potential.  At the
instance  of the jump, the vortex asymptotic position is $\vec{x} = \vec{r} -
\vec{f}_\eff(\vec{r})/\C$ [see Eq.~\eqref{eq:force_balance_eff}], i.e., at a
distance $x_p = r_m - (1+g)f_p(r_m)/\C + \mathcal{O}(\vec{\Delta}^2)$ from the
center of the pinning potential. Pinning and depinning then occur at the
asymptotic positions $\vec{x} = (\pm x_p\cos\phi, x_p \sin\phi)$, with
$\phi \approx \arcsin (b/x_p)$. Note that for the single-defect pinning in
Fig.\ \ref{fig:strong} corresponding to a large $\kappa$, the energy jumps
associated with pinning and depinning appear at \textit{different} asymptotic
distances $x_\-$, $x_\+$ from the defect center; for marginally strong
effective pinning with $\kappa_\eff(g,\vec{\Delta})-1\ll 1$, we can neglect
the difference $x_\+-x_\-\sim \xi(\kappa_\eff - 1)^{3/2}$ (see Refs.\
\cite{Buchacek2019, Willa2016}) and set $x_\-\approx x_\+\approx x_p$.

The angles enclosed between $\vec{x}$ and $\vec{\Delta}$ at the pinning and
depinning events are $\theta_p = \theta + \phi$ and $\theta_\dep = \theta -
\phi$, see Fig.~\ref{fig:jumps_angles}.  The jump size at the pinning
transition does not depend on the direction of the vortex motion and thus can
be evaluated from the vortex trajectory passing directly through the center of
the effective pinning potential but at an angle $\theta + \phi$, i.e.,
$\Delta e_\eff^1(g,\Delta,\theta,b) = \Delta e_\eff^1(g, \Delta,
\theta+\phi, 0)$.  Similarly, $\Delta e_\eff^2(g, \Delta, \theta,b) =
\Delta e_\eff^2(g, \Delta, \theta-\phi,0)$ at depinning. Hence, the pinning
and depinning jumps assume different values at finite impact $b$.  The pinning
force is then expressed as $f_\mathrm{pair}(g, \Delta, \theta,
b)=\tfrac{1}{2}[\Delta e_\eff^1(g, \Delta, \theta+\phi,0) + \Delta
e_\eff^2(g, \Delta, \theta-\phi,0)]/a_0$.  Furthermore, since for
marginally strong pinning and $b = 0$ trajectories, the pinning and depinning
jumps in energy are equally-sized, we express the resulting pinning force in
terms of the forces exerted on the $b = 0$ trajectories,
\begin{align}
\begin{split}
   f_\mathrm{pair}(g,\Delta,\theta,b) = \tfrac{1}{2}\bigl[
   &f_\mathrm{pair}(g,\Delta,\theta + \phi,0)\\
   &+ f_\mathrm{pair}(g,\Delta,\theta - \phi,0)\bigr].
\end{split}
\end{align}

\begin{figure}[h!]
\begin{center}
\includegraphics[scale=1]{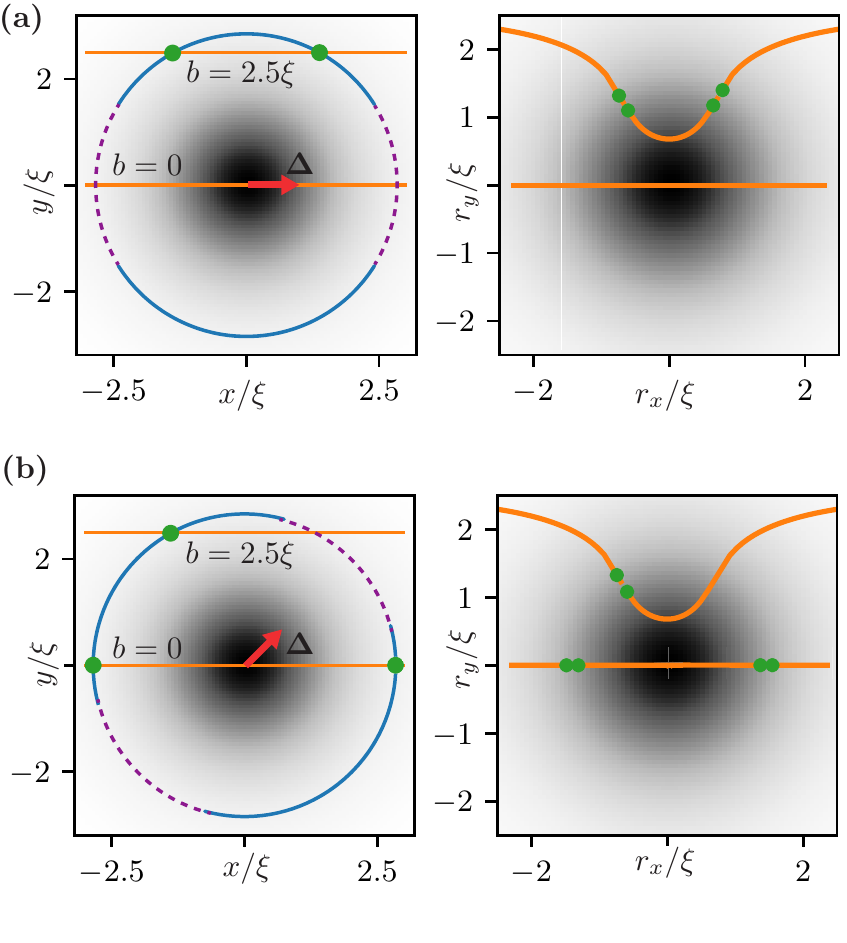}
\end{center}
\caption{Examples of vortex trajectories for an effective pinning potential
due to two weak defects with $\kappa=0.8$, $g = g_0(\kappa) + 2. \> 10^{-2}$,
$\Delta = 0.1 \, \xi$. (a) Longitudinal mismatch vector $\vec{\Delta} =
(\Delta,0)$ (red arrow, not to scale). For this setting, $\kappa_\eff(g,
\Delta,0) = 0.994$ at $b=0$ and the vortex does not undergo any jumps in
energy when the asymptotic position crosses the circle of radius $x_p$ (dashed
purple line).  The plot on the right shows the vortex tip position $\vec{r}$
which remains continuous. A vortex passing at the transverse distance $b =
2.5\, \xi$ experiences an effective Labusch parameter
$\kappa_\eff(g,\Delta,\phi) = 1.011$ (with $\phi = \arcsin (b/x_p)$) and
the vortex undergoes pinning and depinning jumps (green dots) when its
asymptotic position crosses the circle of radius $x_p$. The blue arcs denote
the sections on the circle of radius $x_p$ where the effective pinning is
strong; for $g-g_0\ll 1$ and $\kappa$ approaching unity, pinning is stronger
along the direction perpendicular to the mismatch vector $\vec{\Delta}$ (that
is for $\phi = \pi/2$) than along the direction parallel to the mismatch,
see the discussion below Eq.~\eqref{eq:kappa_eff_angular}. The tip position
(right plot) jumps at the pinning and depinning events, as illustrated by the
two pairs of green dots. (b) Mismatch vector enclosing an angle $\theta =
\pi/4$ with the vortex trajectory.  At this angle, pinning is strong,
$\kappa_\eff(g, \Delta,\theta) = 1.005$, and a vortex with vanishing impact
parameter $b = 0$ undergoes both pinning and depinning jumps.  However, for
the vortex passing at $b = 2.5\,\xi$, the corresponding effective Labusch
parameters read $\kappa_\eff(g,\Delta,\theta + \phi) = 1.014$ and
$\kappa_\eff(g, \Delta,\theta - \phi) = 0.996$ and the vortex undergoes a
jump only upon pinning. The right plot shows the corresponding tip
trajectories with its jumps.} \label{fig:trajectories_e_eff}
\end{figure}

With the pinning and depinning jumps no longer equal, we may encounter
situations where one of the jumps is absent. This is the case for misfits
$\vec\Delta$ with $\theta \neq 0$ and $b \neq 0$; e.g., when
$\kappa_\eff(g,\Delta,\theta + \phi) > 1$ but $\kappa_\eff(g,\Delta,\theta
-\phi) < 1$, the vortex undergoes a pinning jump in energy when its
asymptotic trajectory passes the circle of radius $x_p$ for the first time but
does not undergo any depinning jump when the asymptotic trajectory crosses the
circle a second time, see Fig.~\ref{fig:trajectories_e_eff}(b) (note that
the corresponding trajectory with opposite impact parameter $y = -b$ will
undergo a depinning jump but will not jump upon pinning).

The pinning force averaging is done through integration over the mismatch
$\vec{\Delta}$ and the impact parameter $b$,
\begin{align}\nonumber
   &\langle f_\mathrm{pair}(g,\vec{\Delta})\rangle_{\vec{\Delta}} =
   \int_{-x_p}^{x_p}\frac{d b}{a_0}\,\int_{-\pi}^\pi \!\!\!\! d\theta \int
   \frac{\Delta\,d\Delta}{a_0^2}\, f_\mathrm{pair}(g,\Delta,\theta,b) \\
   \nonumber &=\frac{x_p}{2 a_0}\!\!\int_{-\pi/2}^{\pi/2} \!\!\! d\sin
   \phi\!  \int_{-\pi}^{\pi} \!\!\!\! d\theta \int
   \frac{\Delta\,d\Delta}{a_0^2} \bigl[f_\mathrm{pair}(g,\Delta,\theta +
   \phi,0) \\ \nonumber &\qquad\qquad\qquad\qquad\qquad\qquad 
   + f_\mathrm{pair}(g,\Delta,\theta - \phi,0)\bigr]\\
   \label{eq:f_pair_integration} &= \frac{2x_p}{a_0}\int_{-\pi}^{\pi} \!\!\!\!
   d\theta \int \frac{\Delta\, d\Delta}{a_0^2}\,
   f_\mathrm{pair}(g,\Delta,\theta,0).
\end{align}
The integration over $\Delta$ is restricted to the strongly-pinning region,
i.e.  $\kappa_\eff(\theta) > 1$. Rewriting the integration in terms of
$\vec{\Delta} = (\Delta_x,\Delta_y)$, we compute the factor
\begin{align}\nonumber
   &\int_0^{\Delta_x^0} \! \frac{d\Delta_x}{a_0}\,\int_0^{\Delta_y^0}\! 
   \frac{d \Delta_y}{a_0}\, \Bigl[1 - \frac{(\Delta_x)^2}{(\Delta_x^0)^2} 
   - \frac{(\Delta_y)^2}{(\Delta_y^0)^2}\Bigr]^2
	= \frac{\pi}{3}\frac{\Delta_x^0 \Delta_y^0}{a_0^2}
\end{align}
and obtain the averaged pinning force
\begin{align}\label{eq:f_pair_final}
   \langle f_\mathrm{pair}(g,\vec{\Delta})\rangle_{\vec{\Delta}} 
   = \frac{2x_p}{a_0^2} \frac{6\pi(\kappa\C)^2}{[-f_p'''(r_m)]}
   \frac{(g\!-\!g_0)^2}{(1\!+\!g)^4}
   \frac{\Delta_x^0\Delta_y^0}{a_0^2}.
\end{align}
Keeping systematically the corrections in the jump position $r_m +
\mathcal{O}(\vec{\Delta}^2)$ is equivalent to replacing $b = x_p\sin\phi +
\mathcal{O}(\vec{\Delta}^2)$ in the integration leading to the
Eq.~\eqref{eq:f_pair_integration}. Carrying out the integration of the
additional $\mathcal{O}(\vec{\Delta}^2)$ term would contribute with a quartic
correction $\propto \Delta^4$ to Eq.~\eqref{eq:f_pair_final} which we ignore
here. With $x_p\sim \xi$ and the results for the maximum mismatches
$\Delta_x^0$, $\Delta_y^0$ in Eqs.~\eqref{eq:Delta_x_0_param} and
\eqref{eq:Delta_y_0_param}, we obtain a parametric estimate for the pinning
force originating from defect pairs in the form,
\begin{align}\label{eq:f_pair_final_parametric}
   \langle f_\mathrm{pair}(g,\vec{\Delta})\rangle_{\vec{\Delta}} 
   \sim \frac{\xi^4}{a_0^4}(g-g_0)^3(g+g_0)^{1/2}
   \, f_p,
\end{align}
with the following interpretation: the defect pair induces the pinning force
$\sim f_p$ rescaled by the factor $(g-g_0)^2$ that accounts for the distance
between defects, a factor $\xi^2/a_0^2 \sim S_\mathrm{trap}/a_0^2$ that
represents the areal fraction where vortices are trapped, and the additional
factor $\xi^2/a_0^2$ together with the distance-dependent factor $(g-g_0)
(g+g_0)^{1/2}$ that derive from the constraint on the mismatch $\vec\Delta$.
The result Eq.~\eqref{eq:f_pair_final_parametric} is the basis for the
evaluation of the pinning force due to defect pairs at different separations
in Sec.~\ref{sect:overview}.

\section{Summary and conclusion}\label{sect:conclusion} 

We have extended the strong pinning paradigm into the weak pinning domain by
accounting for correlations between defects. The most relevant correlations
arise from defect pairs---they reduce the critical Labusch parameter (or
pinning strength) $\kappa \sim -e_p^{\prime\prime}/\bar{C}$ for strong pinning
from its standard value $\kappa_c =1$ to $\kappa_{c, \mathrm{pairs}} = 1/2$.
When decreasing the individual defect's pinning strength $\kappa$ towards the
critical value $\kappa_c$, the strong pinning-force density vanishes as
$F_\mathrm{pin} \propto n_p (\kappa - 1)^2$ and strong pair-pinning takes
over. Upon a further decrease of $\kappa$, the pair-induced strong
pinning-force density $F_\mathrm{pin}$ scales with $n_p^2$ and vanishes at
$\kappa = 1/2$ according to $F_\mathrm{pin} \propto n_p^2 (\kappa - 1/2)^4$.
The contributions to $F_\pin$ from higher-order correlations between $n$
defects scale as $(n_p)^n$ and quickly become irrelevant, with the dominant
contribution to the pinning force density being taken over by weak collective
pinning with $F_\pin \propto n_p^2 \kappa^3$ as $\kappa$ drops below $1/2$.

The origin of the pair-induced strong pinning condition $\kappa > 1/2$ is
easily understood---for two defects that overlap in position, their joint
pinning strength doubles and they reach the strong pinning criterion $2\kappa
> 1$. The substantial enhancement in pinning strength remains in place for
{\it small} defect pairs that are separated at most by $\sim \xi$ and $\sim
a_0$ in transverse and longitudinal {(field-)} directions and act on the same
vortex.  However, this is not the full story: with $\kappa > 1/2$ approaching unity
from below, pairs separated by transverse distances beyond $a_0$ can
constitute a strong-pinning pair as well. These pairs, rather than pinning the
{\it same} vortex, will pin {\it different} vortices.  The interaction between
these two defect--vortex entities is transmitted by the elastic properties of
the vortex lattice, specifically, the static non-local Green's function
$G_{\alpha\beta}({\bf R},z)$, and determines the effective pinning strength of
the extended pair which is smaller than the one of a small pair.

The Green's function $G_{\alpha\beta}({\bf R},z)$ describes the displacement
field ${\bf u}({\bf R},z)$ for a $\delta$-force acting at the origin and hence
the distortion at the site of a second defect that is positioned a distance
$({\bf R},z)$ away from the first defect. In our analysis, we have simplified
the expression for the Green's function and considered its diagonal,
transverse part $G({\bf R},z)$; the result, shown in Fig.\ \ref{fig:g_Rz},
exhibits a sharp asymmetric and structured peak in the shape of a dumbbell.
This complex real-space structure has not been considered before and is
expected to be present in the full expression for the response matrix
$G_{\alpha\beta}({\bf R},z)$ as well.

For our {\it extended} pairs, the effective Labusch parameter or pinning
strength $\kappa_\mathrm{eff}$, rather than simply doubling $\kappa$, scales
as $\kappa_\mathrm{eff} \sim (1 + g)\,\kappa$, with $g = g({\bf R},z) = G({\bf
R},z)/G(0,0) < 1$; hence, the partner defect contributes to the strong pinning
with a reduced weight. Extended pairs within a distance determined by the
condition $g(({\bf R},z) > 1/\kappa -1 \equiv g_0(\kappa)$ thus potentially
contribute to strong pair-pinning; geometric considerations refine this
analysis and produce an effective Labusch parameter $\kappa_\mathrm{eff} (g,
\vec\Delta)$ that depends on distance (through $g$) and on the misfit
$\vec\Delta$ between the defect-pair and the vortex lattice, with a finite
$\Delta$ further reducing the effective pinning strength
$\kappa_\mathrm{eff}$, see Eq.\ \eqref{eq:kappa_eff_angular}.

The effective Labusch parameter $\kappa_\mathrm{eff} (g, \vec\Delta)$
exhibits a non-trivial angular dependence encoded in the direction of
$\vec\Delta$.  While for isotropic single-defect pinning, strong-pinning jumps
appear near the inflection points arranged in a circle of radius $r_m$, for an
anisotropic potential as in Eq.\ \eqref{eq:e_eff_asymp}, strong-pinning jumps
appear on arcs that grow with decreasing elasticity $\C$ or increasing pinning
strength $e_p$ as illustrated in Figs.\ \ref{fig:trajectories_e_eff}.  The
direction away from the defect center where these arcs make their first
appearance depends, besides the direction of $\vec\Delta$, on the detailed
shape of the pinning potential, see Eq.\ \eqref{eq:mismatches_ratio}.

With contributions to strong pair-pinning arising both from small pairs
pinning one vortex and extended pairs pinning two separated (and relatively
misfitted) vortices, the question arises about their relative total weight. It
turns out, that the extended-pair force decreases with the scaled distance
$\tilde\rho = [R^2 + (a_0^2/16\pi\lambda^2) z^2]^{1/2}$ as $\propto
\tilde\rho^{-7/2}$, that makes the small-pair contribution (originating from
pairs in a small volume $\xi^2 a_0$) dominate the strong-pair pinning-force
density $F_\mathrm{clust}$ in Eq.\ \eqref{eq:F_clust_intr}.

As follows from the above discussion, the elastic properties of the vortex
lattice take an important role in the calculation of the strong-pair pinning
force.  Furthermore, they also define the dominance of strong-pair pinning
over weak-collective pinning that is reduced by the factor $(a_0/\lambda)^2$.
This reduction is a consequence of the non-local interaction between vortices
producing a dispersion in $c_{44}(\vec{k})$. While collective pinning in the
non-dispersive regime (with a Larkin length $R_c > \lambda$) involves a
`stiff' lattice with $c_{44}(0) = B^2/4 \pi$, the small defect pairs involve
the soft lattice with $c_{44}(k) = B^2/4 \pi \lambda^2 k^2$; with the relevant
$k \sim K_{\rm\scriptscriptstyle BZ} \approx \sqrt{4\pi}/a_0$, the lattice is
softer by a factor $\sim a_0^2/\lambda^2$ and hence pair-pinning is stronger. The
large factor $\lambda^2/a_0^2$ also guarantees, that strong-pair pinning is
larger than weak-collective pinning deep in the dispersive regime where the
lattice becomes softer.

It is interesting to compare the situation described above with the one
studied by Fisher \cite{Fisher1985}: Focusing on weak defects and large
dimensions $D > 4$, it turns out that weak-collective pinning is ineffective
due to the fast spatial relaxation of the manifold's distortions. Pinning then
is exclusively due to rare configurations appearing in the random pinning
landscape. In our analysis, we start from the opposite limit, strong defects
that pin the elastic manifold (here, vortices) individually. Upon decreasing
the pinning strength $\kappa$ below unity, we loose the strong pinning of
individual defects and would expect weak-collective pinning to take over in
$D=3$, where distortions decay slowly, proportional to the inverse distance.
Instead, due to the non-local interaction between vortices producing a
dispersive elastic response, we find that specific rare events, small defect
pairs, take over and produce the leading pinning mechanism.  In a
non-dispersive elastic medium, the stiffening at large scales is absent and the
two types of pinning, rare and collective, come with equal (parametric)
weight.

While weak collective pinning arises from {\it typical} fluctuations in the
defect distribution, strong pair- or cluster-pinning arises from {\it rare}
fluctuations.  In reality, both types of fluctuations coexist and hence
simultaneously contribute to the pinning force density $F_\pin$. Similar to
the addition of resistivities arising from different scattering mechanisms in
the Matthiessen rule describing metallic transport, the pinning-force
densities from different pinning mechanisms should be added up to the total
pinning force $F_\pin \approx F_\mathrm{coll} + F_\mathrm{clust}$ when
describing the transport in a superconductor. However, given the inductive
response of a superconductor, this corresponds to an addition of (critical)
currents rather than voltages.  Microscopically, comparing the distance
between small pairs, $d_\mathrm{pairs} \sim [n_p(n_p a_0 \xi^2)]^{-1/3}$, with
the size of the collective pinning volume $V_c \sim L_c R_c^2 \sim
(\lambda/a_0) R_c^3$, one notes that $V_c$ contains many pairs.  Hence, when
dragging a vortex system slowly over the pinning landscape, one should observe
a complex stick-slip type motion where small slips of individual vortices
depinning from defect pairs combine with large slips of collectively pinned
vortex bundles.  It would be interesting to observe the motion of such a
pinned vortex system in a numerical simulation. Another future topic of
interest is the further investigation of the real-space structure of the
Green's function $G_{\alpha\beta}(\vec{R},z)$, both theoretically as well as
experimentally.  In particular, it would be interesting to come up with a
proposal for an experiment that is sensitive to the non-trivial dumbbell
structure of the peak in the response function.

\acknowledgements
We thank Alexei Koshelev and Roland Willa for inspiring discussions.
M.B.\ acknowledges financial support from the Swiss
National Science Foundation, Division II.

\bibliography{bib_correlations}

%merlin.mbs apsrev4-1.bst 2010-07-25 4.21a (PWD, AO, DPC) hacked
%Control: key (0)
%Control: author (8) initials jnrlst
%Control: editor formatted (1) identically to author
%Control: production of article title (-1) disabled
%Control: page (0) single
%Control: year (1) truncated
%Control: production of eprint (0) enabled
\begin{thebibliography}{37}%
\makeatletter
\providecommand \@ifxundefined [1]{%
 \@ifx{#1\undefined}
}%
\providecommand \@ifnum [1]{%
 \ifnum #1\expandafter \@firstoftwo
 \else \expandafter \@secondoftwo
 \fi
}%
\providecommand \@ifx [1]{%
 \ifx #1\expandafter \@firstoftwo
 \else \expandafter \@secondoftwo
 \fi
}%
\providecommand \natexlab [1]{#1}%
\providecommand \enquote  [1]{``#1''}%
\providecommand \bibnamefont  [1]{#1}%
\providecommand \bibfnamefont [1]{#1}%
\providecommand \citenamefont [1]{#1}%
\providecommand \href@noop [0]{\@secondoftwo}%
\providecommand \href [0]{\begingroup \@sanitize@url \@href}%
\providecommand \@href[1]{\@@startlink{#1}\@@href}%
\providecommand \@@href[1]{\endgroup#1\@@endlink}%
\providecommand \@sanitize@url [0]{\catcode `\\12\catcode `\$12\catcode
  `\&12\catcode `\#12\catcode `\^12\catcode `\_12\catcode `\%12\relax}%
\providecommand \@@startlink[1]{}%
\providecommand \@@endlink[0]{}%
\providecommand \url  [0]{\begingroup\@sanitize@url \@url }%
\providecommand \@url [1]{\endgroup\@href {#1}{\urlprefix }}%
\providecommand \urlprefix  [0]{URL }%
\providecommand \Eprint [0]{\href }%
\providecommand \doibase [0]{http://dx.doi.org/}%
\providecommand \selectlanguage [0]{\@gobble}%
\providecommand \bibinfo  [0]{\@secondoftwo}%
\providecommand \bibfield  [0]{\@secondoftwo}%
\providecommand \translation [1]{[#1]}%
\providecommand \BibitemOpen [0]{}%
\providecommand \bibitemStop [0]{}%
\providecommand \bibitemNoStop [0]{.\EOS\space}%
\providecommand \EOS [0]{\spacefactor3000\relax}%
\providecommand \BibitemShut  [1]{\csname bibitem#1\endcsname}%
\let\auto@bib@innerbib\@empty
%</preamble>
\bibitem [{\citenamefont {Abrikosov}(1957)}]{Abrikosov1957}%
  \BibitemOpen
  \bibfield  {author} {\bibinfo {author} {\bibfnamefont {A.~A.}\ \bibnamefont
  {Abrikosov}},\ }\href {{http://www.jetp.ac.ru/cgi-bin/dn/e_005_06_1174.pdf}}
  {\bibfield  {journal} {\bibinfo  {journal} {{[Zh. Eksp. Teor. Fiz.
  \textbf{32}, 1442 (1957)]} JETP}\ }\textbf {\bibinfo {volume} {5}},\ \bibinfo
  {pages} {1174} (\bibinfo {year} {1957})}\BibitemShut {NoStop}%
\bibitem [{\citenamefont {Bloch}(1932)}]{Bloch1932}%
  \BibitemOpen
  \bibfield  {author} {\bibinfo {author} {\bibfnamefont {F.}~\bibnamefont
  {Bloch}},\ }\href {\doibase doi.org/10.1007/BF01337791} {\bibfield  {journal}
  {\bibinfo  {journal} {Z. Physik}\ }\textbf {\bibinfo {volume} {74}},\
  \bibinfo {pages} {295} (\bibinfo {year} {1932})}\BibitemShut {NoStop}%
\bibitem [{\citenamefont {Landau}\ and\ \citenamefont
  {Lifshitz}(1935)}]{LandauLifshitz1935}%
  \BibitemOpen
  \bibfield  {author} {\bibinfo {author} {\bibfnamefont {L.~D.}\ \bibnamefont
  {Landau}}\ and\ \bibinfo {author} {\bibfnamefont {E.}~\bibnamefont
  {Lifshitz}},\ }\href {\doibase 10.1016/B978-0-08-036364-6.50008-9} {\bibfield
   {journal} {\bibinfo  {journal} {Phys. Z. Sowjet}\ }\textbf {\bibinfo
  {volume} {8}},\ \bibinfo {pages} {153} (\bibinfo {year} {1935})}\BibitemShut
  {NoStop}%
\bibitem [{\citenamefont {Labusch}(1969)}]{Labusch1969}%
  \BibitemOpen
  \bibfield  {author} {\bibinfo {author} {\bibfnamefont {R.}~\bibnamefont
  {Labusch}},\ }\href@noop {} {\bibfield  {journal} {\bibinfo  {journal}
  {Crystal Lattice Defects}\ }\textbf {\bibinfo {volume} {1}},\ \bibinfo
  {pages} {1} (\bibinfo {year} {1969})}\BibitemShut {NoStop}%
\bibitem [{\citenamefont {Larkin}\ and\ \citenamefont
  {Ovchinnikov}(1979)}]{Larkin1979}%
  \BibitemOpen
  \bibfield  {author} {\bibinfo {author} {\bibfnamefont {A.~I.}\ \bibnamefont
  {Larkin}}\ and\ \bibinfo {author} {\bibfnamefont {Y.~N.}\ \bibnamefont
  {Ovchinnikov}},\ }\href {\doibase 10.1007/BF00117160} {\bibfield  {journal}
  {\bibinfo  {journal} {Journal of Low Temperature Physics}\ }\textbf {\bibinfo
  {volume} {34}},\ \bibinfo {pages} {409} (\bibinfo {year} {1979})}\BibitemShut
  {NoStop}%
\bibitem [{\citenamefont {Kittel}(1949)}]{Kittel1949}%
  \BibitemOpen
  \bibfield  {author} {\bibinfo {author} {\bibfnamefont {C.}~\bibnamefont
  {Kittel}},\ }\href {\doibase 10.1103/RevModPhys.21.541} {\bibfield  {journal}
  {\bibinfo  {journal} {Rev. Mod. Phys.}\ }\textbf {\bibinfo {volume} {21}},\
  \bibinfo {pages} {541} (\bibinfo {year} {1949})}\BibitemShut {NoStop}%
\bibitem [{\citenamefont {Blatter}\ \emph {et~al.}(1994)\citenamefont
  {Blatter}, \citenamefont {Feigel'man}, \citenamefont {Geshkenbein},
  \citenamefont {Larkin},\ and\ \citenamefont {Vinokur}}]{Blatter1994}%
  \BibitemOpen
  \bibfield  {author} {\bibinfo {author} {\bibfnamefont {G.}~\bibnamefont
  {Blatter}}, \bibinfo {author} {\bibfnamefont {M.~V.}\ \bibnamefont
  {Feigel'man}}, \bibinfo {author} {\bibfnamefont {V.~B.}\ \bibnamefont
  {Geshkenbein}}, \bibinfo {author} {\bibfnamefont {A.~I.}\ \bibnamefont
  {Larkin}}, \ and\ \bibinfo {author} {\bibfnamefont {V.~M.}\ \bibnamefont
  {Vinokur}},\ }\href {\doibase 10.1103/RevModPhys.66.1125} {\bibfield
  {journal} {\bibinfo  {journal} {Rev. Mod. Phys.}\ }\textbf {\bibinfo {volume}
  {66}},\ \bibinfo {pages} {1125} (\bibinfo {year} {1994})}\BibitemShut
  {NoStop}%
\bibitem [{\citenamefont {Nattermann}\ and\ \citenamefont
  {Scheidl}(2000)}]{NattermannScheidl2000}%
  \BibitemOpen
  \bibfield  {author} {\bibinfo {author} {\bibfnamefont {T.}~\bibnamefont
  {Nattermann}}\ and\ \bibinfo {author} {\bibfnamefont {S.}~\bibnamefont
  {Scheidl}},\ }\href {\doibase 10.1080/000187300412257} {\bibfield  {journal}
  {\bibinfo  {journal} {Advances in Physics}\ }\textbf {\bibinfo {volume}
  {49}},\ \bibinfo {pages} {607} (\bibinfo {year} {2000})}\BibitemShut
  {NoStop}%
\bibitem [{\citenamefont {Larkin}\ and\ \citenamefont
  {Ovchinnikov}(1974)}]{Larkin1974}%
  \BibitemOpen
  \bibfield  {author} {\bibinfo {author} {\bibfnamefont {A.~I.}\ \bibnamefont
  {Larkin}}\ and\ \bibinfo {author} {\bibfnamefont {Y.~N.}\ \bibnamefont
  {Ovchinnikov}},\ }\href
  {{http://www.jetp.ac.ru/cgi-bin/dn/e_038_04_0854.pdf}} {\bibfield  {journal}
  {\bibinfo  {journal} {[Zh. Eksp. Teor. Fiz. \textbf{65}, 1704 (1974)] JETP}\
  }\textbf {\bibinfo {volume} {38}},\ \bibinfo {pages} {854} (\bibinfo {year}
  {1974})}\BibitemShut {NoStop}%
\bibitem [{\citenamefont {Schmid}\ and\ \citenamefont
  {Hauger}(1973)}]{Schmid1973}%
  \BibitemOpen
  \bibfield  {author} {\bibinfo {author} {\bibfnamefont {A.}~\bibnamefont
  {Schmid}}\ and\ \bibinfo {author} {\bibfnamefont {W.}~\bibnamefont
  {Hauger}},\ }\href {\doibase 10.1007/BF00654452} {\bibfield  {journal}
  {\bibinfo  {journal} {Journal of Low Temperature Physics}\ }\textbf {\bibinfo
  {volume} {11}},\ \bibinfo {pages} {667} (\bibinfo {year} {1973})}\BibitemShut
  {NoStop}%
\bibitem [{\citenamefont {Larkin}(1970)}]{Larkin1970}%
  \BibitemOpen
  \bibfield  {author} {\bibinfo {author} {\bibfnamefont {A.~I.}\ \bibnamefont
  {Larkin}},\ }\href@noop {} {\bibfield  {journal} {\bibinfo  {journal} {[Zh.
  Eksp. Teor. Fiz. \textbf{58}, 1466 (1970)] JETP}\ }\textbf {\bibinfo {volume}
  {31}},\ \bibinfo {pages} {784} (\bibinfo {year} {1970})}\BibitemShut
  {NoStop}%
\bibitem [{\citenamefont {Giamarchi}\ and\ \citenamefont
  {Le~Doussal}(1994)}]{Giamarchi1994}%
  \BibitemOpen
  \bibfield  {author} {\bibinfo {author} {\bibfnamefont {T.}~\bibnamefont
  {Giamarchi}}\ and\ \bibinfo {author} {\bibfnamefont {P.}~\bibnamefont
  {Le~Doussal}},\ }\href {\doibase 10.1103/PhysRevLett.72.1530} {\bibfield
  {journal} {\bibinfo  {journal} {Phys. Rev. Lett.}\ }\textbf {\bibinfo
  {volume} {72}},\ \bibinfo {pages} {1530} (\bibinfo {year}
  {1994})}\BibitemShut {NoStop}%
\bibitem [{\citenamefont {Giamarchi}\ and\ \citenamefont
  {Le~Doussal}(1995)}]{Giamarchi1995}%
  \BibitemOpen
  \bibfield  {author} {\bibinfo {author} {\bibfnamefont {T.}~\bibnamefont
  {Giamarchi}}\ and\ \bibinfo {author} {\bibfnamefont {P.}~\bibnamefont
  {Le~Doussal}},\ }\href {\doibase 10.1103/PhysRevB.52.1242} {\bibfield
  {journal} {\bibinfo  {journal} {Phys. Rev. B}\ }\textbf {\bibinfo {volume}
  {52}},\ \bibinfo {pages} {1242} (\bibinfo {year} {1995})}\BibitemShut
  {NoStop}%
\bibitem [{\citenamefont {Korshunov}(1993)}]{Korshunov1993}%
  \BibitemOpen
  \bibfield  {author} {\bibinfo {author} {\bibfnamefont {S.~E.}\ \bibnamefont
  {Korshunov}},\ }\href {\doibase 10.1103/PhysRevB.48.3969} {\bibfield
  {journal} {\bibinfo  {journal} {Phys. Rev. B}\ }\textbf {\bibinfo {volume}
  {48}},\ \bibinfo {pages} {3969} (\bibinfo {year} {1993})}\BibitemShut
  {NoStop}%
\bibitem [{\citenamefont {Kleemann}(2007)}]{Kleemann2007}%
  \BibitemOpen
  \bibfield  {author} {\bibinfo {author} {\bibfnamefont {W.}~\bibnamefont
  {Kleemann}},\ }\href
  {https://doi.org/10.1146/annurev.matsci.37.052506.084243} {\bibfield
  {journal} {\bibinfo  {journal} {Annual Review of Materials Research}\
  }\textbf {\bibinfo {volume} {37}},\ \bibinfo {pages} {415} (\bibinfo {year}
  {2007})}\BibitemShut {NoStop}%
\bibitem [{\citenamefont {Gorchon}\ \emph {et~al.}(2014)\citenamefont
  {Gorchon}, \citenamefont {Bustingorry}, \citenamefont {Ferr\'e},
  \citenamefont {Jeudy}, \citenamefont {Kolton},\ and\ \citenamefont
  {Giamarchi}}]{Gorchon2014}%
  \BibitemOpen
  \bibfield  {author} {\bibinfo {author} {\bibfnamefont {J.}~\bibnamefont
  {Gorchon}}, \bibinfo {author} {\bibfnamefont {S.}~\bibnamefont
  {Bustingorry}}, \bibinfo {author} {\bibfnamefont {J.}~\bibnamefont
  {Ferr\'e}}, \bibinfo {author} {\bibfnamefont {V.}~\bibnamefont {Jeudy}},
  \bibinfo {author} {\bibfnamefont {A.~B.}\ \bibnamefont {Kolton}}, \ and\
  \bibinfo {author} {\bibfnamefont {T.}~\bibnamefont {Giamarchi}},\ }\href
  {\doibase 10.1103/PhysRevLett.113.027205} {\bibfield  {journal} {\bibinfo
  {journal} {Phys. Rev. Lett.}\ }\textbf {\bibinfo {volume} {113}},\ \bibinfo
  {pages} {027205} (\bibinfo {year} {2014})}\BibitemShut {NoStop}%
\bibitem [{\citenamefont {Jeudy}\ \emph {et~al.}(2016)\citenamefont {Jeudy},
  \citenamefont {Mougin}, \citenamefont {Bustingorry}, \citenamefont
  {Savero~Torres}, \citenamefont {Gorchon}, \citenamefont {Kolton},
  \citenamefont {Lema\^{\i}tre},\ and\ \citenamefont {Jamet}}]{Jeudy2016}%
  \BibitemOpen
  \bibfield  {author} {\bibinfo {author} {\bibfnamefont {V.}~\bibnamefont
  {Jeudy}}, \bibinfo {author} {\bibfnamefont {A.}~\bibnamefont {Mougin}},
  \bibinfo {author} {\bibfnamefont {S.}~\bibnamefont {Bustingorry}}, \bibinfo
  {author} {\bibfnamefont {W.}~\bibnamefont {Savero~Torres}}, \bibinfo {author}
  {\bibfnamefont {J.}~\bibnamefont {Gorchon}}, \bibinfo {author} {\bibfnamefont
  {A.~B.}\ \bibnamefont {Kolton}}, \bibinfo {author} {\bibfnamefont
  {A.}~\bibnamefont {Lema\^{\i}tre}}, \ and\ \bibinfo {author} {\bibfnamefont
  {J.-P.}\ \bibnamefont {Jamet}},\ }\href {\doibase
  10.1103/PhysRevLett.117.057201} {\bibfield  {journal} {\bibinfo  {journal}
  {Phys. Rev. Lett.}\ }\textbf {\bibinfo {volume} {117}},\ \bibinfo {pages}
  {057201} (\bibinfo {year} {2016})}\BibitemShut {NoStop}%
\bibitem [{\citenamefont {Lee}\ and\ \citenamefont {Rice}(1979)}]{Lee1979}%
  \BibitemOpen
  \bibfield  {author} {\bibinfo {author} {\bibfnamefont {P.~A.}\ \bibnamefont
  {Lee}}\ and\ \bibinfo {author} {\bibfnamefont {T.~M.}\ \bibnamefont {Rice}},\
  }\href {\doibase 10.1103/PhysRevB.19.3970} {\bibfield  {journal} {\bibinfo
  {journal} {Phys. Rev. B}\ }\textbf {\bibinfo {volume} {19}},\ \bibinfo
  {pages} {3970} (\bibinfo {year} {1979})}\BibitemShut {NoStop}%
\bibitem [{\citenamefont {Brazovskii}\ and\ \citenamefont
  {Nattermann}(2004)}]{Brazovskii2004}%
  \BibitemOpen
  \bibfield  {author} {\bibinfo {author} {\bibfnamefont {S.}~\bibnamefont
  {Brazovskii}}\ and\ \bibinfo {author} {\bibfnamefont {T.}~\bibnamefont
  {Nattermann}},\ }\href {\doibase 10.1080/00018730410001684197} {\bibfield
  {journal} {\bibinfo  {journal} {Advances in Physics}\ }\textbf {\bibinfo
  {volume} {53}},\ \bibinfo {pages} {177} (\bibinfo {year} {2004})}\BibitemShut
  {NoStop}%
\bibitem [{\citenamefont {Chauve}\ \emph {et~al.}(2000)\citenamefont {Chauve},
  \citenamefont {Giamarchi},\ and\ \citenamefont {Le~Doussal}}]{Chauve2000}%
  \BibitemOpen
  \bibfield  {author} {\bibinfo {author} {\bibfnamefont {P.}~\bibnamefont
  {Chauve}}, \bibinfo {author} {\bibfnamefont {T.}~\bibnamefont {Giamarchi}}, \
  and\ \bibinfo {author} {\bibfnamefont {P.}~\bibnamefont {Le~Doussal}},\
  }\href {\doibase 10.1103/PhysRevB.62.6241} {\bibfield  {journal} {\bibinfo
  {journal} {Phys. Rev. B}\ }\textbf {\bibinfo {volume} {62}},\ \bibinfo
  {pages} {6241} (\bibinfo {year} {2000})}\BibitemShut {NoStop}%
\bibitem [{\citenamefont {Brazovskii}(1996)}]{Brazovskii1996}%
  \BibitemOpen
  \bibfield  {author} {\bibinfo {author} {\bibfnamefont {S.}~\bibnamefont
  {Brazovskii}},\ }\href@noop {} {\bibfield  {journal} {\bibinfo  {journal} {in
  \emph{Physics and Chemistry of Low-Dimensional Inorganic Conductors}, Plenum
  Press, New York}\ }\textbf {\bibinfo {volume} {354}},\ \bibinfo {pages} {465}
  (\bibinfo {year} {1996})}\BibitemShut {NoStop}%
\bibitem [{\citenamefont {Brazovskii}\ and\ \citenamefont
  {Larkin}(1999)}]{Brazovskii1999}%
  \BibitemOpen
  \bibfield  {author} {\bibinfo {author} {\bibfnamefont {S.}~\bibnamefont
  {Brazovskii}}\ and\ \bibinfo {author} {\bibfnamefont {A.}~\bibnamefont
  {Larkin}},\ }\href {https://hal.archives-ouvertes.fr/hal-00003959} {\bibfield
   {journal} {\bibinfo  {journal} {{Journal de Physique IV Colloque}}\ }\textbf
  {\bibinfo {volume} {9}},\ \bibinfo {pages} {Pr10} (\bibinfo {year} {1999})},\
  \bibinfo {note} {to be published in Proceedings of ECRYS-99, J. de Physique,
  Coll., December 1999}\BibitemShut {NoStop}%
\bibitem [{\citenamefont {Thomann}\ \emph {et~al.}(2017)\citenamefont
  {Thomann}, \citenamefont {Geshkenbein},\ and\ \citenamefont
  {Blatter}}]{Thomann2017}%
  \BibitemOpen
  \bibfield  {author} {\bibinfo {author} {\bibfnamefont {A.~U.}\ \bibnamefont
  {Thomann}}, \bibinfo {author} {\bibfnamefont {V.~B.}\ \bibnamefont
  {Geshkenbein}}, \ and\ \bibinfo {author} {\bibfnamefont {G.}~\bibnamefont
  {Blatter}},\ }\href {\doibase 10.1103/PhysRevB.96.144516} {\bibfield
  {journal} {\bibinfo  {journal} {Phys. Rev. B}\ }\textbf {\bibinfo {volume}
  {96}},\ \bibinfo {pages} {144516} (\bibinfo {year} {2017})}\BibitemShut
  {NoStop}%
\bibitem [{\citenamefont {Willa}\ \emph
  {et~al.}(2018{\natexlab{a}})\citenamefont {Willa}, \citenamefont
  {Marziali~Berm\'udez},\ and\ \citenamefont {Pasquini}}]{Willa2018a}%
  \BibitemOpen
  \bibfield  {author} {\bibinfo {author} {\bibfnamefont {R.}~\bibnamefont
  {Willa}}, \bibinfo {author} {\bibfnamefont {M.}~\bibnamefont
  {Marziali~Berm\'udez}}, \ and\ \bibinfo {author} {\bibfnamefont
  {G.}~\bibnamefont {Pasquini}},\ }\href {\doibase 10.1103/PhysRevB.98.184520}
  {\bibfield  {journal} {\bibinfo  {journal} {Phys. Rev. B}\ }\textbf {\bibinfo
  {volume} {98}},\ \bibinfo {pages} {184520} (\bibinfo {year}
  {2018}{\natexlab{a}})}\BibitemShut {NoStop}%
\bibitem [{\citenamefont {Willa}\ \emph
  {et~al.}(2018{\natexlab{b}})\citenamefont {Willa}, \citenamefont {Koshelev},
  \citenamefont {Sadovskyy},\ and\ \citenamefont {Glatz}}]{Willa2018b}%
  \BibitemOpen
  \bibfield  {author} {\bibinfo {author} {\bibfnamefont {R.}~\bibnamefont
  {Willa}}, \bibinfo {author} {\bibfnamefont {A.~E.}\ \bibnamefont {Koshelev}},
  \bibinfo {author} {\bibfnamefont {I.~A.}\ \bibnamefont {Sadovskyy}}, \ and\
  \bibinfo {author} {\bibfnamefont {A.}~\bibnamefont {Glatz}},\ }\href
  {\doibase 10.1103/PhysRevB.98.054517} {\bibfield  {journal} {\bibinfo
  {journal} {Phys. Rev. B}\ }\textbf {\bibinfo {volume} {98}},\ \bibinfo
  {pages} {054517} (\bibinfo {year} {2018}{\natexlab{b}})}\BibitemShut
  {NoStop}%
\bibitem [{\citenamefont {Buchacek}\ \emph {et~al.}(2018)\citenamefont
  {Buchacek}, \citenamefont {Willa}, \citenamefont {Geshkenbein},\ and\
  \citenamefont {Blatter}}]{Buchacek2018}%
  \BibitemOpen
  \bibfield  {author} {\bibinfo {author} {\bibfnamefont {M.}~\bibnamefont
  {Buchacek}}, \bibinfo {author} {\bibfnamefont {R.}~\bibnamefont {Willa}},
  \bibinfo {author} {\bibfnamefont {V.~B.}\ \bibnamefont {Geshkenbein}}, \ and\
  \bibinfo {author} {\bibfnamefont {G.}~\bibnamefont {Blatter}},\ }\href
  {\doibase 10.1103/PhysRevB.98.094510} {\bibfield  {journal} {\bibinfo
  {journal} {Phys. Rev. B}\ }\textbf {\bibinfo {volume} {98}},\ \bibinfo
  {pages} {094510} (\bibinfo {year} {2018})}\BibitemShut {NoStop}%
\bibitem [{\citenamefont {Buchacek}\ \emph {et~al.}(2019)\citenamefont
  {Buchacek}, \citenamefont {Willa}, \citenamefont {Geshkenbein},\ and\
  \citenamefont {Blatter}}]{Buchacek2019}%
  \BibitemOpen
  \bibfield  {author} {\bibinfo {author} {\bibfnamefont {M.}~\bibnamefont
  {Buchacek}}, \bibinfo {author} {\bibfnamefont {R.}~\bibnamefont {Willa}},
  \bibinfo {author} {\bibfnamefont {V.~B.}\ \bibnamefont {Geshkenbein}}, \ and\
  \bibinfo {author} {\bibfnamefont {G.}~\bibnamefont {Blatter}},\ }\href
  {\doibase 10.1103/PhysRevB.100.014501} {\bibfield  {journal} {\bibinfo
  {journal} {Phys. Rev. B}\ }\textbf {\bibinfo {volume} {100}},\ \bibinfo
  {pages} {014501} (\bibinfo {year} {2019})}\BibitemShut {NoStop}%
\bibitem [{\citenamefont {Blatter}\ \emph {et~al.}(2004)\citenamefont
  {Blatter}, \citenamefont {Geshkenbein},\ and\ \citenamefont
  {Koopmann}}]{Blatter2004}%
  \BibitemOpen
  \bibfield  {author} {\bibinfo {author} {\bibfnamefont {G.}~\bibnamefont
  {Blatter}}, \bibinfo {author} {\bibfnamefont {V.~B.}\ \bibnamefont
  {Geshkenbein}}, \ and\ \bibinfo {author} {\bibfnamefont {J.~A.~G.}\
  \bibnamefont {Koopmann}},\ }\href {\doibase 10.1103/PhysRevLett.92.067009}
  {\bibfield  {journal} {\bibinfo  {journal} {Phys. Rev. Lett.}\ }\textbf
  {\bibinfo {volume} {92}},\ \bibinfo {pages} {067009} (\bibinfo {year}
  {2004})}\BibitemShut {NoStop}%
\bibitem [{\citenamefont {Koopmann}\ \emph {et~al.}(2004)\citenamefont
  {Koopmann}, \citenamefont {Geshkenbein},\ and\ \citenamefont
  {Blatter}}]{Koopmann2004}%
  \BibitemOpen
  \bibfield  {author} {\bibinfo {author} {\bibfnamefont {J.}~\bibnamefont
  {Koopmann}}, \bibinfo {author} {\bibfnamefont {V.}~\bibnamefont
  {Geshkenbein}}, \ and\ \bibinfo {author} {\bibfnamefont {G.}~\bibnamefont
  {Blatter}},\ }\href {\doibase https://doi.org/10.1016/j.physc.2003.11.046}
  {\bibfield  {journal} {\bibinfo  {journal} {Physica C: Superconductivity}\
  }\textbf {\bibinfo {volume} {404}},\ \bibinfo {pages} {209 } (\bibinfo {year}
  {2004})},\ \bibinfo {note} {proceedings of the Third European Conference on
  Vortex Matter in Superconductors at Extreme Scales and
  Conditions}\BibitemShut {NoStop}%
\bibitem [{\citenamefont {Blatter}\ and\ \citenamefont
  {Geshkenbein}(2008)}]{Blatter2008}%
  \BibitemOpen
  \bibfield  {author} {\bibinfo {author} {\bibfnamefont {G.}~\bibnamefont
  {Blatter}}\ and\ \bibinfo {author} {\bibfnamefont {V.~B.}\ \bibnamefont
  {Geshkenbein}},\ }\enquote {\bibinfo {title} {Vortex matter},}\ in\ \href
  {\doibase 10.1007/978-3-540-73253-2_12} {\emph {\bibinfo {booktitle}
  {Superconductivity: Conventional and Unconventional Superconductors}}},\
  \bibinfo {editor} {edited by\ \bibinfo {editor} {\bibfnamefont {K.~H.}\
  \bibnamefont {Bennemann}}\ and\ \bibinfo {editor} {\bibfnamefont {J.~B.}\
  \bibnamefont {Ketterson}}}\ (\bibinfo  {publisher} {Springer Berlin
  Heidelberg},\ \bibinfo {address} {Berlin, Heidelberg},\ \bibinfo {year}
  {2008})\ pp.\ \bibinfo {pages} {495--637}\BibitemShut {NoStop}%
\bibitem [{\citenamefont {Ovchinnikov}\ and\ \citenamefont
  {Ivlev}(1991)}]{IvlevOvchinnikov1991}%
  \BibitemOpen
  \bibfield  {author} {\bibinfo {author} {\bibfnamefont {Y.~N.}\ \bibnamefont
  {Ovchinnikov}}\ and\ \bibinfo {author} {\bibfnamefont {B.~I.}\ \bibnamefont
  {Ivlev}},\ }\href {\doibase 10.1103/PhysRevB.43.8024} {\bibfield  {journal}
  {\bibinfo  {journal} {Phys. Rev. B}\ }\textbf {\bibinfo {volume} {43}},\
  \bibinfo {pages} {8024} (\bibinfo {year} {1991})}\BibitemShut {NoStop}%
\bibitem [{\citenamefont {Fisher}(1985)}]{Fisher1985}%
  \BibitemOpen
  \bibfield  {author} {\bibinfo {author} {\bibfnamefont {D.~S.}\ \bibnamefont
  {Fisher}},\ }\href {\doibase 10.1103/PhysRevB.31.1396} {\bibfield  {journal}
  {\bibinfo  {journal} {Phys. Rev. B}\ }\textbf {\bibinfo {volume} {31}},\
  \bibinfo {pages} {1396} (\bibinfo {year} {1985})}\BibitemShut {NoStop}%
\bibitem [{\citenamefont {Willa}\ \emph {et~al.}(2016)\citenamefont {Willa},
  \citenamefont {Geshkenbein},\ and\ \citenamefont {Blatter}}]{Willa2016}%
  \BibitemOpen
  \bibfield  {author} {\bibinfo {author} {\bibfnamefont {R.}~\bibnamefont
  {Willa}}, \bibinfo {author} {\bibfnamefont {V.~B.}\ \bibnamefont
  {Geshkenbein}}, \ and\ \bibinfo {author} {\bibfnamefont {G.}~\bibnamefont
  {Blatter}},\ }\href {\doibase 10.1103/PhysRevB.93.064515} {\bibfield
  {journal} {\bibinfo  {journal} {Phys. Rev. B}\ }\textbf {\bibinfo {volume}
  {93}},\ \bibinfo {pages} {064515} (\bibinfo {year} {2016})}\BibitemShut
  {NoStop}%
\bibitem [{\citenamefont {Koshelev}()}]{Koshelev_priv_comm}%
  \BibitemOpen
  \bibfield  {author} {\bibinfo {author} {\bibfnamefont {A.~E.}\ \bibnamefont
  {Koshelev}},\ }\href@noop {} {}\bibinfo {howpublished} {private
  communication}\BibitemShut {NoStop}%
\bibitem [{\citenamefont {Brandt}(1977)}]{Brandt1977b}%
  \BibitemOpen
  \bibfield  {author} {\bibinfo {author} {\bibfnamefont {E.~H.}\ \bibnamefont
  {Brandt}},\ }\href {\doibase 10.1007/BF00654877} {\bibfield  {journal}
  {\bibinfo  {journal} {Journal of Low Temperature Physics}\ }\textbf {\bibinfo
  {volume} {26}},\ \bibinfo {pages} {735} (\bibinfo {year} {1977})}\BibitemShut
  {NoStop}%
\bibitem [{\citenamefont {Schmid}(1966)}]{Schmid1966}%
  \BibitemOpen
  \bibfield  {author} {\bibinfo {author} {\bibfnamefont {A.}~\bibnamefont
  {Schmid}},\ }\href@noop {} {\bibfield  {journal} {\bibinfo  {journal} {Physik
  der kondensierten Materie}\ }\textbf {\bibinfo {volume} {5}},\ \bibinfo
  {pages} {302} (\bibinfo {year} {1966})}\BibitemShut {NoStop}%
\bibitem [{\citenamefont {Clem}(1975)}]{Clem1975}%
  \BibitemOpen
  \bibfield  {author} {\bibinfo {author} {\bibfnamefont {J.~R.}\ \bibnamefont
  {Clem}},\ }\href@noop {} {\bibfield  {journal} {\bibinfo  {journal} {Journal
  of Low Temperature Physics}\ }\textbf {\bibinfo {volume} {18}},\ \bibinfo
  {pages} {427} (\bibinfo {year} {1975})}\BibitemShut {NoStop}%
\end{thebibliography}%

\appendix

\section{Green's function in the dispersive regime}\label{APP:g_disp} 

We discuss the derivation of the interpolation formula
\eqref{eq:G_interp} for the diagonal Green's function in the dispersive region
(short distances). We consider the simplified model of the vortex lattice
elasticity where we drop the longitudianal part in Eq.~\eqref{eq:G_VL}
containing the large compression modulus $c_{11}(\vec{k})>c_{66}$;
furthermore, we replace the transverse projector by unity and study the
diagonal Green's function $G_{\alpha\beta}(\vrh) = G(\vrh)\,
\delta_{\alpha\beta}$. The integration in reciprocal space then reads
\begin{align}
   G(\vec{R},z) = \!\!\!\!\!\!
   \int\limits_{K<K_{\rm\scriptscriptstyle BZ}}\!\!\!\!\!\!
   \frac{d^2\vec{K}\,d k_z}{(2\pi)^3} \frac{e^{i \vec{K}\cdot \vec{R}}
   e^{i k_z z}}{c_{66}K^2 + c_{44}(\vec{k})k_z^2}
\end{align}
with the non-dispersive shear modulus $c_{66}$, the dispersive tilt modulus
$c_{44}(\vec{k})\approx c_{44}(\vec{0})/(1+\lambda^2 k^2)$, and
$c_{66}/c_{44}(\vec{0}) = a_0^2/16\pi\lambda^2$.  We first perform the complex
integration over $k_z$ (extended to infinity) with a pole at
\begin{align}
   k_z = i K \frac{(1+\lambda^2 K^2)^{1/2}}{[\lambda^2 K^2 
   + c_{44}(\vec{0})/c_{66}]^{1/2}}.
\end{align}
We drop the term $\lambda^2 K^2$ in the denominator that provides a numerical
correction to the large ratio $c_{44}(\vec{0})/c_{66}$ when $K < K_{\rm
\scriptscriptstyle BZ} \approx \sqrt{4\pi}/a_0$ is residing within the
Brillouin zone.  The Green's function then takes the form (we write
$c_{44}(\vec 0) = c_{44}$)
\begin{align}\label{eq:g_2D_integration}
\begin{split}
   G(\vec{R},z)\approx& \frac{\lambda}{2\sqrt{c_{44}c_{66}}}
   \int\frac{d^2\vec{K}}{(2\pi)^2}\, 
   e^{i\vec{K}\cdot\vec{R}}\frac{(1+\lambda^2 K^2)^{1/2}}{\lambda K}\\
   &\qquad\times \exp\Bigl[-\frac{a_0 z}{4\sqrt{\pi}\lambda}K(1+\lambda^2 K^2)^{1/2}\Bigr].
\end{split}
\end{align}
We assume a small distance $\vec\rho = (\vec{R},z)$ within the ellipse $R^2 +
(a_0^2/16\pi\lambda^2)z^2 \ll \lambda^2$ (see Eq.~\eqref{eq:G_non_disp} for
the opposite limit). We first focus on the contribution from $\lambda K
\gg 1$ and approximate $(1+\lambda^2K^2)^{1/2} \to \lambda K$ in
Eq.~\eqref{eq:g_2D_integration}, that provides us with the dispersive
approximation
\begin{align}\label{eq:g_D}
   G_\mathrm{d}(\vec{R},z) &= \frac{\lambda}{2\sqrt{c_{44}c_{66}}}
   \int\frac{d^2\vec{K}}{(2\pi)^2}\, 
   e^{i\vec{K}\cdot\vec{R}}\exp\Bigl[-\frac{a_0 z K^2}{4\sqrt{\pi}}\Bigr]\\
   \label{eq:g_D_result}
   & = \frac{1/\sqrt{4\pi}}{a_0\sqrt{c_{44}c_{66}}}\frac{\lambda}{z} 
   \exp\Bigl[-\frac{\sqrt{\pi} R^2}{a_0 z}\Bigr].
\end{align}
Next, we account for the difference between the full expression
\eqref{eq:g_2D_integration} and the dispersive approximation \eqref{eq:g_D};
we split this difference into two terms $\delta G_<$ and $\delta G_>$ arising
from small ($K< K_0$) and large ($K > K_0$) momenta,
\begin{align}\label{eq:delta_g_low}
   \delta G_< & = \frac{\lambda}{2\sqrt{c_{44}c_{66}}}\int\limits_{K<K_0}
   \frac{d^2\vec{K}}{(2\pi)^2}\, \frac{e^{i\vec{K}\cdot\vec{R}}}{\lambda K}\\
   \nonumber \times
   &\Bigl\lbrace(1+\lambda^2 K^2)^{1/2}\exp\Bigl[-\frac{a_0 z}
   {4\sqrt{\pi}\lambda}K(1+\lambda^2 K^2)^{1/2}\Bigr]\\ \nonumber
   &\qquad\qquad 
   -\lambda K \exp\Bigl[-\frac{a_0 z K^2}{4\sqrt{\pi}}\Bigr]\Bigr\rbrace,
\end{align}
and a corresponding expression for $\delta G_>$ covering the remaining region
$K_0<K<K_{\rm \scriptscriptstyle BZ}$. The scale $K_0$ is chosen such as to
satisfy $\lambda^{-1}\ll K_0\ll R^{-1}$ as well as $\lambda^{-1}\ll K_0\ll
(a_0z)^{-1/2}$, consistent with the assumption that $R\ll \lambda$ and $z\ll
\lambda^2/a_0$.

In carrying out the integration over small momenta $K<K_0$, we note that
the arguments in the exponentials of Eq.~\eqref{eq:delta_g_low} remain small
since $KR < K_0 R \ll 1$ and $K_0\ll (a_0z)^{-1/2}$; performing the
integration over angles then provides us with
\begin{align}\nonumber
   \delta G_< & = \frac{1/4\pi}{\sqrt{c_{44}c_{66}}} \int_0^{K_0} d K\, 
   \bigl[(1+\lambda^2 K^2)^{1/2} - \lambda K \bigr]\\
   &\label{eq:delta_g_low_expansion}
   \approx \frac{1/16\pi}{\lambda \sqrt{c_{44}c_{66}}} \bigl[1 + 2\ln (2K_0\lambda)\bigr],
\end{align}
where we have used that $\lambda K_0\gg 1$ in the last relation.

For large momenta $K_0<K<K_{\rm \scriptscriptstyle BZ}$, we rewrite the
integral in the form 
\begin{align}\label{eq:delta_g_large}
   &\delta G_> = \frac{\lambda}{2\sqrt{c_{44}c_{66}}}\!\!\!\!\!\!\!
   \int\limits_{K_0<K<K_{\rm \scriptscriptstyle BZ}}\!\!\!\!\!\!\!
   \frac{d^2\vec{K}}{(2\pi)^2}\,\frac{e^{i\vec{K}\cdot\vec{R}}}
   {\lambda K}e^{-a_0z K^2/4\sqrt{\pi}}\\ \nonumber
   &\times \Bigl\lbrace \sqrt{1+\lambda^2 K^2}
   e^{-\frac{a_0z} {4\sqrt{\pi}\lambda^2}
   [\lambda K(\sqrt{1+\lambda^2K^2} -\lambda K)]}
   - \lambda K \Bigr\rbrace.
\end{align}
Since $K\lambda > K_0\lambda \gg 1$, we can expand the square roots;
furthermore the $K$-integration is cut off by the exponentials, either via
$K\sim(a_0z)^{-1/2} < K_{\rm \scriptscriptstyle BZ}$ or by $K\sim R^{-1} <
K_{\rm \scriptscriptstyle BZ}$ since $R,z > a_0$, hence the BZ cutoff at
$K_{\rm \scriptscriptstyle BZ}$ can be replaced by infinity,
\begin{align}\label{eq:delta_g_large_expansion}
   \delta G_> \approx \frac{1}{4\lambda\sqrt{c_{44}c_{66}}}\!\!\!\int\limits_{K>K_0}
   \!\!\!\frac{d^2\vec{K}}{(2\pi)^2}\frac{e^{i\vec{K}\cdot\vec{R}}}{K^2}
   e^{-a_0 z K^2/4\sqrt{\pi}}.
\end{align}
\begin{figure}
\begin{center}
\includegraphics[width=8.6cm]{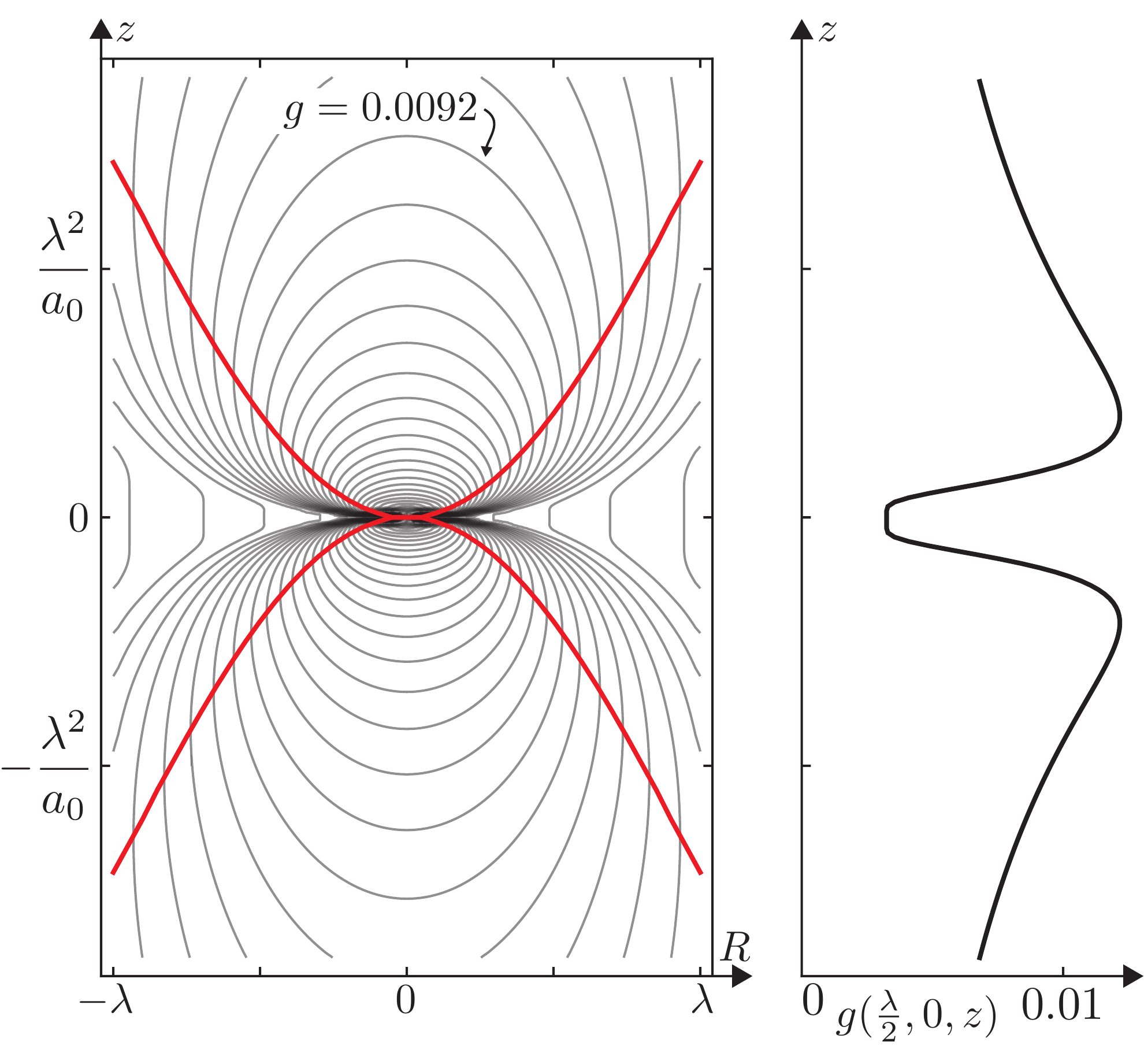}
\caption{Left: contour plot of the rescaled Green's function $g(x=R,0,z) =
G(R, 0, z) / G(\vec{0})$ evaluated for $\lambda = 10\,a_0$ illustrating the
dumbbell structure of the central peak; axes are not to scale.  The ridge
marking the maximum of $g$ when increasing $z$ at fixed $R < \lambda$ has a
parabolic shape (red lines). Subsequent contours are separated by a factor
$2^{1/4}$.  Right: interpreting the Green's function $G$ as providing the
displacement field due to a point-force in the origin, the ratio
$g(x=\lambda/2,0,z)$ follows the profile of a vortex placed at a distance
$\lambda/2$ away from the force center. The ridges in $g$ manifest as
maxima in the vortex displacement away from $z = 0$.}
\label{fig:g_peak}
\end{center}
\end{figure}

In the evaluation along the longitudinal direction, we 
expand the integrand in the small parameter ${(a_0z)^{1/2}K_0\ll 1}$
and obtain the result
\begin{align}\label{eq:delta_g_large_long}
   \delta G_>(\vec{R}=0,z) \approx \frac{1/8\pi}{\lambda \sqrt{c_{44}c_{66}}}
   \bigl[-\gamma/2 + \ln\tfrac{2\pi^{1/4}}{K_0\sqrt{a_0 z}}\bigr]
\end{align}
with the Euler-Mascheroni constant $\gamma$. Along the transverse direction, we 
expand in $K_0R\ll 1$ and find that
\begin{align}\label{eq:delta_g_large_transv}
   \delta G_>(\vec{R},z=0) \approx \frac{1/8\pi}{\lambda \sqrt{c_{44}c_{66}}}
   \bigl[-\gamma + \ln\tfrac{2}{K_0 R}\bigr].
\end{align}
We combine these results to arrive at the interpolation formula for a general
distance $\vec{\rho} = (\vec{R},z)$ within the ellipse
\begin{align}\label{eq:delta_g_large_interp}
   \delta G_>(\vec{R},z)\approx \frac{1/16\pi}{\lambda \sqrt{c_{44}c_{66}}}
   \Bigl[-2\gamma + \ln\tfrac{4/K_0^2}{R^2 + a_0 z/e^\gamma \sqrt{\pi}}\Bigr].
\end{align}
Finally, summing up the contributions $G_\mathrm{d}+ \delta G_< + \delta G_>$
provides us with the result in Eq.~\eqref{eq:G_interp} (note that the momentum
$K_0$ drops out from the final expression); it provides us with the peak
in $G(\vec R, z)$ at small distances with its dumbbell structure that is
illustrated in Fig.\ \ref{fig:g_peak}.

\section{Numerical evaluation of the pinning force}\label{APP:num_pin_force}
Our numerical evaluation of the pinning force $f_\mathrm{pair}(g, \Delta,
\theta,b)$, see Sec.\ \ref{sect:average_f_pin}, makes use of the numerical
solution for the vortex displacements $\vec{u}_1$, $\vec{u}_2$ in the
two-defect problem Eq.~\eqref{eq:two_defects}. We first reformulate the latter
in terms of a minimization problem for the total energy
$e_\pin^{\mathrm{pair}}$ of the two defect system described by the mean
asymptotic vortex position $\vec{x}$, the fixed mismatch $\vec{\Delta}$ and
vortex-tip displacements $\vec{u}_1$, $\vec{u}_2$,
\begin{align}\label{eq:e_pin_minimize}
   &e_\pin^{\mathrm{pair}}(\vec{x},\vec{\Delta};\vec{u}_1,\vec{u}_2)
   \equiv e_p(\vec{x} + \vec{\Delta}/2 + \vec{u}_1)\\ \nonumber
   &+e_p(\vec{x} - \vec{\Delta}/2 + \vec{u}_2)
   + \frac{\C}{2(1\!-\!g^2)}\bigl[\vec{u}_1^2
   + \vec{u}_2^2 -2g\,\vec{u}_1\cdot\vec{u}_2\bigr],
\end{align}
such that setting $\partial e_\pin^{\mathrm{pair}}/\partial\vec{u}_{1,2} = 0$
reproduces Eq.~\eqref{eq:two_defects}. By rewriting $\vec{u}_1^2 + \vec{u}_2^2
-2g\,\vec{u}_1\cdot\vec{u}_2 = (\vec{u}_1-\vec{u}_2)^2 + {2(1-g)\, \vec{u}_1
\cdot \vec{u}_2}$, we note that in the limit $g = 1$ (defects pinning the same
vortex at the same height $z$, such that $\vec{u}_1 = \vec{u}_2$), the elastic
term in Eq.~\eqref{eq:e_pin_minimize} remains regular and reduces to
$\C\vec{u}_1^2/2$; Eq.~\eqref{eq:e_pin_minimize} then describes the
interaction of a single vortex with a defect potential given by the
superposition of two pinning potentials shifted by the mismatch vector
$\vec{\Delta}$ from each other.
\begin{figure}[h]
\begin{center}
\includegraphics[scale=1]{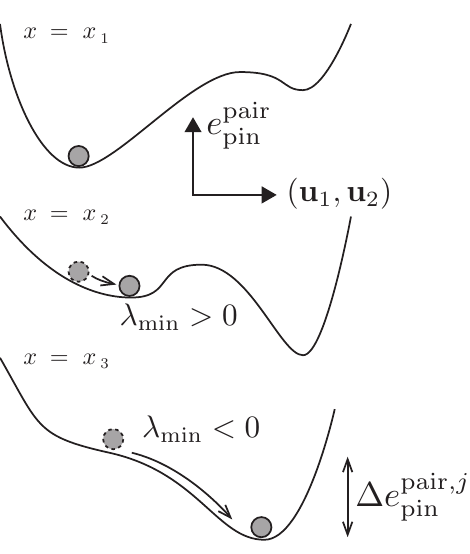}
\end{center}
\caption{Minimization of the pinning energy Eq.\ \eqref{eq:e_pin_minimize}.
Given the position $\vec{x} = (x,b)$ of the vortex system with respect to the
defects, the pinning energy $e_\pin^\mathrm{pair}$ is a function of the
4-dimensional vector $(\vec{u}_1, \vec{u}_2)$. The pictures sketch the
qualitative changes in the shape of $e_\pin^\mathrm{pair}$ when increasing the
mean vortex position through $x_1 < x_2 < x_3$. The neighborhood of each local
minimum is characterized by positive eigenvalues $\lambda$ of the matrix of
second derivatives of $e_\pin^\mathrm{pair}$. The smallest eigenvalue going
negative indicates the disappearance of a local minimum and triggers a jump in
the location of the occupied minimum.}\label{fig:e_pin_minimization}
\end{figure}

We evaluate the critical force for a quasi-static vortex lattice pushed to
the right: For every asymptotic vortex trajectory $\vec{x} = (x,b)$ with fixed
impact parameter $b$, we increase the position $x$ in small steps, and
minimize the pinning energy at each step using the solution from the previous
position as the starting point. Starting far to the left from both defects,
the contribution of the pinning energies $e_p(\vec{x}\pm\Delta/2 +
\vec{u}_{1,2})$ is negligible and the unique solution describes a system of
unperturbed vortices with $\vec{u}_{1,2} \approx \vec{0}$. On approaching the
defects, the vortices deform and multiple minima develop (see Fig.\
\ref{fig:e_pin_minimization}); a small step-wise increase of $x$ ensures that
the occupied minimum develops continuously (see below about details on
the minimization algorithm). At the point where the minimum disappears, the
solution changes abruptly, with the associated jump in the pinning energy
contributing to the average pinning force.

Depending on the specific setting of the problem, the system undergoes a
number of jumps in energy with increasing $x$, see Fig.\
\ref{fig:trajectories_e_eff}, where the fictitious vortex undergoes zero, one,
or two jumps depending on the angle $\theta$ and impact parameter $b$. Going
beyond the transition to strong pinning with $\kappa > 1$, the two-vortex
system may undergo up to four jumps associated with the pinning and depinning
events of two vortices hitting the two strongly-pinning defects. The numerical
procedure then provides a convenient way to evaluate the average pinning force
for the case with an arbitrary number of jumps in energy (even though only the
cases with zero, one, or two jumps have been discussed in the analytic part of
this paper).

As for the single-defect case, it turns out that the pinning force exerted on
the vortices can be expressed as the gradient of the total pinning energy.
Indeed, taking the gradient of $e_\pin^{\mathrm{pair}}(\vec{x}) \equiv
e_\pin^{\mathrm{pair}}[\vec{x},\vec\Delta;\vec{u}_1(\vec{x}),
\vec{u}_2(\vec{x})]$ defined through Eq.~\eqref{eq:e_pin_minimize}, we find
\begin{align}\label{eq:d_e_pin}
   \nabla_{\vec{x}}e_\pin^{\mathrm{pair}}(\vec{x})
   = \frac{\partial e_\pin^{\mathrm{pair}}}{\partial \vec{x}}
   + \sum_{i = {1,2}}\frac{\partial e_\pin^{\mathrm{pair}}}{\partial \vec{u}_i}
   \frac{\partial \vec{u}_i}{\partial \vec{x}}
   = \frac{\partial e_\pin^{\mathrm{pair}}}{\partial \vec{x}}
\end{align}
where the partial derivative $\partial e_\pin^{\mathrm{pair}}/\partial
\vec{x}$ is taken at fixed $\vec{u}_1$, $\vec{u}_2$. The term in
Eq.~\eqref{eq:d_e_pin} involving the $\vec{u}$-derivatives vanishes since
$\partial e_\pin^\mathrm{pair}/\partial \vec{u}_{1,2} = 0$ at the minimum.
Taking the partial $\vec{x}$-derivative in Eq.~\eqref{eq:e_pin_minimize}
provides us with $-\nabla_{\vec{x}} e_\pin^{\mathrm{pair}} = \vec{f}_p[\vec{x}
+ \vec{\Delta}/2 + \vec{u}_1(\vec{x})] + \vec{f}_p[\vec{x} - \vec{\Delta}/2 +
\vec{u}_2(\vec{x})]$, which is precisely the pinning force exerted by the two
defects on the distorted vortices. The $x$-averaged pinning force along the
trajectory $\vec{x} = (x,b)$ is then written as
\begin{align}\label{eq:f_from_jumps}
   f_\mathrm{pair}(g,\vec{\Delta},b) = \!\! \int \! \frac{d x}{a_0}\,
   \vec{e}_x\!\cdot[-\nabla_\vec{x} e_\pin^{\mathrm{pair}}]
   = - \!\sum_j \frac{\Delta e_\pin^{\mathrm{pair},j}}{a_0},
\end{align}
with $\Delta e_\pin^{\mathrm{pair},j} = \lim_{\varepsilon \to 0}
[e_\pin^{\mathrm{pair}}(x_j - \varepsilon,b) - e_\pin^{\mathrm{pair}}(x_j +
\varepsilon,b)]$ quantifying the energy jump at the position $x_j$.
Integrating $\partial e_\pin^\mathrm{pair}/\partial y$ would in general give a
non-vanishing contribution to the pinning force in the $y$-direction; it is
however compensated by the configuration with $\vec{\Delta}\to -\vec{\Delta}$
and $b\to -b$ after averaging. 

The result of this numerical evaluation is compared with the analytic result
Eq.\ \eqref{eq:fdp_gDth0_ex} in Fig.\ \ref{fig:f_comparison} for the case of
marginally-strong pair-pinning $g-g_0\ll 1$ and shows a very good agreement
with the analytic result even at large values of the mismatch $\Delta$ of
order $\xi$ at angles $\theta$ close to $\pi/2$, in which case the theoretical
estimates made in Sec.~\ref{sect:kappa_eff} do not guarantee the validity of
the perturbative approach. For a Lorentzian shape potential and parameters
used in Fig.\ \ref{fig:f_comparison} with $\theta = \pi/2$, we find that the
scaling factor in Eq.\ \eqref{eq:Delta_y_0} assumes a value $[(g-g_0)/ (1+g)
(g+g_0)]^{1/2}\approx 0.17$ and the prefactor contributes a factor $\approx
4.46$, such that $\Delta_y^0 \approx 0.78 \, \xi$.

\subsection{Numerical minimization}\label{APP:minimization}

The main challenge in the minimization of the two-defect pinning energy in
Eq.~\eqref{eq:e_pin_minimize} is to properly track the local minimum
representing the current state of the vortex (the occupied branch) and to
ensure that the minimization algorithm does not overshoot to another minimum.

We define $\vec{u} = (u_{1,x}, u_{1,y}, u_{2,x}, u_{2,y})$ and minimize the
function $e_\pin^\mathrm{pair}(\vec{u},\vec{x})$, see Eq.\
\eqref{eq:e_pin_minimize}, with respect to $\vec{u}$.  For a fixed asymptotic
position $\vec{x}$, we use Newton's method to iterate $\vec{u}$,
\begin{align}\label{eq:Newtons_it}
   \vec{u}_{i+1} = \vec{u}_i - \gamma H(\vec{u}_i)^{-1}
   \nabla_\vec{u}e_\pin^\mathrm{pair}(\vec{u},\vec{x})
\end{align}
as long as the matrix of second derivatives $H_{\alpha \beta} = \partial^2
e_\pin^\mathrm{pair}/\partial u_\alpha\partial u_\beta$ evaluated at
$\vec{u}_i$ remains positive-definite. The parameter $\gamma$ is chosen to
bound the step size ${|\vec{u}_{i+1} - \vec{u}_i| < \delta u_{\max}}$ to a
pre-defined maximum value $\delta u_{\max}$. The method converges if the
initial guess $\vec{u}_0$ lies close to a local minimum and the new minimum is
used as an initial guess for the minimization after changing the parameter
$\vec{x}$.

The appearance of at least one negative eigenvalue of $H(\vec{u})$ during the
minimization signals the disappearance of the local minimum and triggers the
jump to another minimum. Minimisation through this region is performed using
the Nelder-Mead algorithm with the initial simplex size set to $\delta u_{\max}$.
Once the positive-definite region in the neighborhood of the new minimum is
reached, the minimisation procedure switches back to Newton's method.

Obtaining the size of the energy jumps to the desired accuracy requires
precise determination of the jump points $\vec{x}_j = (x_j,b)$ where the
currently occupied local minimum disappears. This is achieved by repeated
interval-halving: assume that for $\vec{x} = (x_0,b)$ the Newton minimization
converged to a local minimum $\vec{u}_0$, that is used as a starting point for
the minimisation at the next position $\vec{x} = (x_0 + \delta x,b)$.  The
appearance of a region with a negative eigenvalue of the Hessian during this
minimisation indicates the presence of the jump point $x_j$ in the interval
$(x_0,x_0 + \delta x)$. Another minimisation is thus performed for $x = x_0 +
\delta x/2$ that reduces the interval either to $(x_0,x_0+\delta x/2)$ (if the
negative eigenvalue region appears during the minimisation) or $(x_0+\delta
x/2,x_0 + \delta x)$. The further iteration of this procedure locates the jump
point to the required precision.

\end{document}